\documentclass[11pt]{article}

\usepackage{amsmath,amsthm,amssymb,ascmac}
\usepackage{bm}
\usepackage[dvipdfmx]{graphicx}	
\usepackage[english]{babel}
\usepackage{cite}

\newcommand{\be}{\beta}
\newcommand{\ka}{\kappa}
\newcommand{\bea}{\begin{eqnarray}}
\newcommand{\eea}{\end{eqnarray}}
\newcommand{\aeq}{\!\!\! &=& \!\!\!}
\newcommand{\aeqe}{\!\!\! & \equiv & \!\!\!}
\newcommand{\aeqd}{\!\!\! & \stackrel{\mathrm{def}}{=} & \!\!\!}

\newcommand{\aeqap}{\!\!\! &\approx & \!\!\!}

\newcommand{\bra}{\langle}
\newcommand{\dbra}{\langle  \hspace{-0.5 mm} \langle}
\newcommand{\ket}{\rangle}
\newcommand{\dket}{\rangle \hspace{-0.5 mm} \rangle }
\newcommand{\mC}{\mathcal{C}}
\newcommand{\mP}{\mathcal{P}}
\newcommand{\mQ}{\mathcal{Q}}
\newcommand{\mO}{\mathcal{O}}
\newcommand{\mL}{\mathcal{L}}

\newcommand{\mN}{\mathcal{N}}

\newcommand{\mD}{\mathcal{D}}
\newcommand{\mU}{\mathcal{U}}
\newcommand{\mJ}{\mathcal{J}}
\newcommand{\mF}{\mathcal{F}}
\newcommand{\mA}{\mathcal{A}}
\newcommand{\mR}{\mathcal{R}}
\newcommand{\mT}{\mathcal{T}}
\newcommand{\mS}{\mathcal{S}}
\newcommand{\mK}{\mathcal{K}}
\newcommand{\mG}{\mathcal{G}}
\newcommand{\tr}{\mbox{Tr}}
\newcommand{\dg}{^\dagger}
\newcommand{\tl}{\tilde}

\newcommand{\defe}{\stackrel{\mathrm{def}}{=}}
\newcommand{\e}{\equiv}
\newcommand{\ep}{\varepsilon}
\newcommand{\ga}{\gamma}
\newcommand{\Ga}{\Gamma}
\newcommand{\up}{\uparrow}
\newcommand{\dw}{\downarrow}
\newcommand{\al}{\alpha}
\newcommand{\bu}{\bullet}
\newcommand{\sig}{\sigma}

\newcommand{\f}{\frac}
\newcommand{\half}{\frac{1}{2}}
\newcommand{\pr}{\prime}
\newcommand{\dl}{\delta}
\newcommand{\Dl}{\Delta}
\newcommand{\lm}{\lambda}
\newcommand{\Lm}{\Lambda}

\newcommand{\om}{\omega}
\newcommand{\Om}{\Omega}
\newcommand{\sinc}{\mbox{sinc}}

\newcommand{\tot}{{\rm{tot}}}

\newcommand{\st}{{\rm{ss}}}

\newcommand{\la}{\label}
\newcommand{\p}{\partial}
\newcommand{\ke}[1]{ \vert #1 \rangle }
\newcommand{\br}[1]{ \langle #1 \vert }
\newcommand{\abs}[1]{\vert #1 \vert }
\newcommand{\no}{\nonumber}
\newcommand{\hn}{{\hat{n}}}
\newcommand{\ha}{{\hat{\al}}}

\newcommand{\hc}{\mbox{h.c.}}

\newcommand{\re}[1]{(\ref{#1})}
\newcommand{\hs}{\hspace}

\newcommand{\dbr}[1]{ \langle  \hspace{-0.5 mm} \langle #1 \vert }
\newcommand{\dke}[1]{\vert #1 \rangle \hspace{-0.5 mm} \rangle }
\newcommand{\res}[1]{\S \ref{#1}}
\newcommand{\bv}[1]{\big \vert_{#1}}
\newcommand{\Bv}[1]{\Big \vert_{#1}}
\newcommand{\RM}[1]{{\rm{#1}}}
\newcommand{\Refe}[1]{Ref.\cite{#1}}
\newcommand{\rec}[1]{Chap.\ref{#1}}

\makeatletter
\@addtoreset{equation}{section}
\makeatother

\def\department#1{\def\@department{#1}}

\title{Theoretical studies on quantum pump and excess entropy production: Quantum master equation approach
\footnote{This paper is a doctoral thesis submitted to University of Tsukuba. 
The supervisor is Yasuhiro Tokura.}}

\author{Satoshi Nakajima\footnote{Email: subarusatosi@gmail.com} \\
$Graduate$ $School$ $of$ $Pure$ $and$ $Applied$ $Sciences$, $University$ $of$ $Tsukuba$, \\
${\it 1-1-1}$ , $Tennodai$, $Tsukuba$, ${\it 305-8571}$, $Japan$}

\date{February 27, 2017}

\begin{document}

\maketitle

\begin{abstract}
In this thesis, we considered quantum systems coupled to several baths. 
We supposed that the system state is governed by the quantum master equation (QME).
We investigated the quantum pump and the excess entropy production.
When the set of control parameters $\al=\{ \al^n \}_n$ is modulated between times $t=0$ and $t=\tau$,   
the average change of a {\it time-independent} observable $O$ of the baths is given by
\bea
\bra \Dl o \ket=\int_0^\tau dt \ i^\st_O(\al_t)+\int_C d\al^n \ A_n^O(\al)+\bra \Dl o \ket^{(\RM{na})}. \no
\eea
Here, the summation symbol for $n$ is omitted, $\al_t$ is $\al$ at time $t$, $C$ is the trajectory in the control parameter space, 
$i^\st_O(\al_t)$ is the instantaneous steady current of $O$ and $A_n^O(\al)$ is called the Berry-Sinitsyn-Nemenman (BSN) vector.  
$\bra \Dl o \ket^{(\RM{na})}$ is a non-adiabatic term and order of $\om/\Ga$ where $\om$ is the modulation frequency of the control parameters and 
$\Ga$ is the coupling strength between the system and the baths. 
If $\om/\Ga$ is sufficiently small, this pumping is called the quantum adiabatic pump. 
Similarly, the average entropy production $\sig$ under quasistatic ($\om/\Ga \to 0$) modulation is given by 
\bea
\sig=\int_0^\tau dt \ j_\sig(\al_t)+\int_C d\al^n \ A_n^\sig(\al).\no
\eea
Here, $j_\sig(\al_t)$ is the instantaneous steady entropy production rate and $A_n^\sig(\al)$ is called the BSN vector for entropy production. 
The second term of the right hand side (RHS) of the above equation is called the excess entropy production, $\sig_\RM{ex}$. 

First, we investigated the quantum pump using the full counting statistics with quantum master equation (FCS-QME) approach. 
We studied the non-adiabatic effect and showed that the general solution of the QME $\rho(t)$ is decomposed as 
$\rho(t)=\rho_0(\al_t)+\sum_{n=1}^\infty \rho^{(n)}(t)+\sum_{n=0}^\infty \tl \rho^{(n)}(t) $. 
Here, $\rho_0(\al_t)$ is the instantaneous steady state of the QME, 
$\rho^{(n)}(t)$ and $ \tl \rho^{(n)}(t) $ are calculable and order $(\om/\Ga)^n$. 
$ \tl \rho^{(n)}(t) $ exponentially damps (like $e^{-\Ga t}$) as a function of time. 
We showed that the generalized mater equation (GME) approach provides $\bm{p}(t)=\bm{p}_{(\RM{ss})}(t)+\dl \bm{p}(t)$ in the Born approximation.
Here, $\bm{p}$ corresponds to the set of the diagonal components of $\rho$ in the matrix representation by the energy eigenstates,  
$\bm{p}_{(\RM{ss})}(t)$ corresponds to $\rho_0(\al_t)+\sum_{n=1}^\infty \rho^{(n)}(t)$ and the term $\dl \bm{p}(t)$ originates from non-Markovian effects. 
We showed that the FCS-QME method provides $(n+1)$-th order pump current from $\rho^{(n)}(t)$. 
We showed that the quantum pump dose not occur in all orders of the pumping frequency when the system control parameters and 
the thermodynamic parameters (the temperatures and the chemical potentials of the baths) are fixed under the zero-bias condition.

Next, we studied the quantum adiabatic pump of the quantum dot (QD) system weakly coupled to two leads ($L$ and $R$)
using the FCS-QME. 
We confirmed the consistency between the FCS-QME approach and the GME approach for a QD of one quantum level with finite Coulomb interaction. 
We showed that the pumped charge and spin coming from the instantaneous steady current are not negligible when the thermodynamic parameters are not fixed to zero bias.
To observe the spin effects, we considered collinear magnetic fields, which affect the spins through the Zeeman effect, with different amplitudes applying to the QDs ($B_S$) 
and the leads ($B_L$ and $B_R$). 
We focused on the dynamic parameters ($B_S$, $B_{L/R}$ and the coupling strength between QDs and leads, $\Dl_{L/R}$) as control parameters. 
In one level QD with the Coulomb interaction $U$, we studied $(B_L,B_S)$ pump 
and $(\Dl_L,B_S)$ pump for the noninteracting limit ($U=0$) and the strong interaction limit ($U=\infty$) at zero-bias.
The difference depending on $U$ appeared through $n_U(s B_S)$ which is the average number of the electrons with spin $s$ in the QD. 
For $(B_L, B_S)$ pump, the energy dependences of the line-width functions are essential. 
Moreover, we studied the $(\Dl_L,B_S)$ pump for finite $U$ at zero-bias. 
The effect of $U$ appeared through $n_U(s B_S)$. 
When half-filling condition satisfies, the charge pump does not occur.

We studied quantum diabatic pump for spinless one level QD coupled to two leads. 
We calculated $\{\rho^{(n)}(t)\}_{n=1}^5$, $\{\tl \rho^{(n)}(t)\}_{n=1}^5$ and particle current up to 6th order and pumped particle numbers. 
 
In the latter part of the thesis, we investigated the excess entropy production. 
In weakly nonequilibrium regime, we analyzed the BSN vector for the entropy production and showed
\bea 
A_n^\sig(\al)=-\tr_S\big[\ln \rho_0^{(-1)}(\al)\f{\p \rho_0(\al)}{\p \al^n} \big]+\mO(\ep^2). \no
\eea 
Here, 
$\tr_S$ denotes the trace of the system, and $\ep$ is a measure of degree of nonequilibrium. 
$\rho_0^{(-1)}(\al)$ is the instantaneous steady state obtained from the QME with reversing the sign of the Lamb shift term. 
In general, the potential $\mS(\al)$ such that $A_n^\sig(\al)=\f{\p \mS(\al)}{\p \al^n} +\mO(\ep^2)$ dose not exist. 
This is the most important result of this thesis. 
The origins of the non-existence of the potential $\mS(\al)$ are a quantum effect (the Lamb shift term) and the breaking of the time-reversal symmetry. 
The non-existence of the potential means that the excess entropy essentially depends on the path of the modulation. 
In contrast, if the system Hamiltonian is non-degenerate or the Lamb shift term is negligible, we obtain 
$\sig_\RM{ex} =S_\RM{vN}(\rho_0(\al_\tau))-S_\RM{vN}(\rho_0(\al_0))+\mO(\ep^2 \dl)$. 
Here, $S_\RM{vN}(\rho) = -\tr_S[\rho \ln\rho]$ is the von Neumann entropy,  
and $\dl$ describes the amplitude of the change of the control parameters. 
For systems with time-reversal symmetry, there exists a potential $\mS(\al)$, which is the symmetrized von Neumann entropy.  
Additionally, we pointed out that the expression of the entropy production obtained in the classical Markov jump process is different from our result 
and showed that these are approximately equivalent only in the weakly nonequilibrium regime.

\end{abstract}

\newpage

\tableofcontents

\newpage

\section{Introduction} \la{Introduction}

\subsection{Background}

The properties of the isolated static quantum system in the equilibrium state have been studied deeply. 
The studies of more general systems are important, however, uncompleted and are actively being studied. 
This thesis focus on the following three points of view.
The first is (1) time-dependence. 
In the isolated quantum system with time-dependent parameters, the Berry phase \cite{Berry} is important. 
The second is (2) open quantum system. 
The quantum dot (QD) system coupled to several leads is an instance of the open quantum system. 
A theoretical method to study the open quantum system is the quantum master equation (QME). 
The third is the (3) nonequilibrium steady state (NESS). 
The entropy production under operations between NESSs of the classical system is being studied actively. 

In particular, in this thesis, we study the quantum pump and the excess entropy production. 
In a mesoscopic system, even at zero bias, a charge or spin current is induced by a modulation of the control parameters 
\cite{Thouless83, ex1, ex2, ex3, ex4, ex5, ex6, ex7, ex8, ex9, ex10}.
This phenomenon, called the quantum pump, is theoretically interesting because its origins are quantum effects and nonequilibrium effects. 
The entropy production under operations between NESSs is composed of the time integral of the instantaneous steady entropy production rate and 
the excess entropy production. 
The excess entropy production is intensively being studied as a generalization of the entropy concept. 

Recently, Ref.\cite{Sagawa} had been applied the Berry-Sinitsyn-Nemenman (BSN) phase to the excess entropy production in the 
classical system. 
The BSN phase is the ``Berry phase'' of the modified master equation including the counting fields which is a tool of the 
 full counting statistics (FCS). 
For quantum system described by the QME, Ref.\cite{Yuge12} had applied the BSN phase using the FCS-QME \cite{FCS-QME} 
to study the quantum adiabatic pump. 
The FCS-QME had also been applied the excess entropy production in the quantum system \cite{Yuge13}. 
However, we point out that this study has serious flaws \cite{Nakajima2}.

\subsection{Full counting statistics}

In this section, we consider two terminals system. 
In a mesoscopic system, we can see quantum properties through the conducting property.
By recent development of experimental techniques, the transfered charge $Q$ within a time interval $\tau$ and the variance $\bra (Q-\bra Q \ket)^2 \ket$ 
and higher cumulants can be measured ($\bra \cdots \ket$ is the statistical average). 
The notion of obtaining all cumulants is called the full counting statistics (FCS)\cite{FCS-QME, FCS, Utumi, Saito}.
The $n$-th order cumulant $\bra Q^n \ket_c$ is defined by 
\bea
\bra Q^n \ket_c \aeqd \f{\partial^n S_\tau(\chi)}{\partial (i\chi)^n} \Big \vert_{\chi=0},
\eea
where 
\bea
S_\tau(\chi) \aeq \ln \int dQ \ P_\tau(Q) e^{iQ\chi},
\eea
is the cumulant generating function of $Q$. 
$P_\tau(Q)$ is the probability distribution of $Q$. 
$\chi$ is called the counting field. 
The cumulants up to fourth order are given by
\bea
\bra Q \ket_c \aeq \bra Q \ket ,\no\\
\bra Q^2 \ket_c \aeq \bra Q^2 \ket -\bra Q \ket^2 ,\no\\
\bra Q^3 \ket_c \aeq \bra Q^3 \ket-3\bra Q^2 \ket \bra Q \ket+2\bra Q \ket^3 \no\\
\aeq \bra (Q-\bra Q \ket)^3 \ket ,\no\\
\bra Q^4 \ket_c \aeq \bra Q^4 \ket-4\bra Q^3 \ket \bra Q \ket-3\bra Q^2 \ket^2+12\bra Q^2 \ket \bra Q \ket^2-6\bra Q \ket^4 \no\\
\aeq \bra (Q-\bra Q \ket)^4 \ket-3\bra (Q-\bra Q \ket)^2 \ket^2 \no.
\eea
The third and fourth cumulants describe the skewness and sharpness, respectively.

The noise $\bra Q^2 \ket_c$ is composed of the thermal noise (the Johnson-Nyquis noise) and the shot noise.
The shot noise appears when $\abs{eV} > k_\RM{B}T$ where $V$ is the voltage and $T$ is the temperature.
The shot noise $S$ relates with the current $I=\f{\bra Q \ket}{\tau}$ as 
\bea
S \aeq 2e F  I,
\eea
where $F$ is the Fano factor.
For classical shot noise (Poisson noise), $F=1$ holds. 
Then, effective charge $e^\ast$ is defined by
\bea
S \aeq 2e^\ast I.
\eea
$e^\ast=e/3$ had been observed for the fractional quantum Hall state $\nu=1/3$ \cite{FQH1, FQH2}.

The FCS \cite{FCS-QME, FCS, Utumi, Saito} is the method to calculate the generating function.
From the FCS of entropy production, the fluctuation theorem \cite{yuragi1, yuragi2, yuragi3} is derived \cite{FCS-QME, Saito}.
The fluctuation theorem leads to 
\bea
S^{(0)} \aeq 2k_BT G^{(1)} ,\la{FT1}\\
S^{(1)} \aeq k_BT G^{(2)}.\la{FT2}
\eea
Here, the noise $S$ and the current $I$ are expanded as
\bea
S \aeq S^{(0)}+S^{(1)}V+S^{(2)}V^2+\cdots ,\\
I \aeq G^{(1)}V+G^{(2)}V^2+\cdots .
\eea
\re{FT1} is the Johnson-Nyquist relation, which can be derived from the linear response theory.
\re{FT2} is a relation of the non-linear response.
This relation had been tested by experiments \cite{Nakamura10, Nakamura11}.

\subsection{Quantum adiabatic pump} \la{sQAP}

In a mesoscopic system, even at zero bias, a charge or spin current is induced by a slow modulation of control parameters
\cite{Thouless83, ex1, ex2, ex3, ex4, ex5, ex6, ex7, ex8, ex9, ex10}. 
This phenomenon, called the quantum adiabatic pump, is theoretically interesting because its origins are quantum effects and nonequilibrium effects.
The quantum adiabatic pump is also expected to be applied to the single electron transfer devices and the current standard\cite{CS1, CS2}.

\subsubsection{Closed system}

For a closed quantum system, the Berry phase \cite{Berry, Berry_rev} appear when the parameter of the Hamiltonian is 
changed adiabatically (slowly). 
The quantization of the quantum Hall coefficient is proposed by Thouless {\it et al.} \cite{Thouless83} in 1982.
In 1983, for the system which only the $x$-direction is periodic, Thouless showed \cite{Thouless83} that the transfered charge by the cyclic adiabatic modulation of the potential 
is quantized.

\subsubsection{Brouwer formula}

The adiabatically pumped quantity is described by a geometric expression in the control parameter space, 
although the pumped quantity coming from second or higher order of the pumping frequency is not geometric. 
In noninteracting systems, the quantum adiabatic pump had extensively been studied by the Brouwer formula
\cite{Brouwer98, BrouwerGeo, BrouwerFCS1, BrouwerFCS2, Brouwer02, Buiittiker02, Shutenko, Wei, Wohlman1},
which describes the pumped charge by the scattering matrix. 
The Brouwer formula is discovered in 1998 by Brouwer \cite{Brouwer98}.

When $n$-th control parameter $\al^n$ is changed to $\al^n+\dl \al^n$, 
the change of the average number of the electrons of the bath $b$ is given by $\mathcal{E}^{N_b}_n(\al) \dl \al^n$. 
$\mathcal{E}^{N_b}_n(\al)$ is called emissivity. 
For absolute zero temperature, 
\bea
\mathcal{E}^{N_b}_n(\al) \aeq \f{1}{2\pi}\sum_B \sum_{A \in b} \mbox{Im}\Big[ \f{\partial S_{AB}(\al)}{\partial \al^n}S_{AB}^\ast (\al) \Big], \la{Brouwer1}
\eea
holds \cite{Buiittiker94}. 
Here, $A$ and $B$ are labels of electron in the baths and $S(\al)$ is the scattering matrix. 
By slow modulation of the control parameters between time $t=0$ and $t=\tau$, 
the change of the average number of the electrons of the bath $\bra\Dl N_b\ket$ is given by 
\bea
\bra\Dl N_b\ket \aeq \int_0^\tau dt \ \f{d\al^n_t}{dt}\mathcal{E}^{N_b}_n(\al_t)= \int_C d\al^n \ \mathcal{E}^{N_b}_n(\al).
\eea
The summation symbol $\sum_{n}$ is omitted.  
$C$ is the trajectory in the control parameters. 
In particular, for cyclic modulation $\al_0=\al_\tau$, using the Stokes theorem,
\bea
\bra\Dl N_b\ket \aeq \int_S d\al^m \wedge d\al^n \ \half F^{N_b}_{mn}(\al),
\eea
holds.  $S$ is the surface enclosed by $C$. $F^{N_b}_{mn}$ is given by
\bea
F^{N_b}_{mn}(\al) \aeqd \f{\p \mathcal{E}^{N_b}_n(\al)}{\p \al^m}-\f{\p \mathcal{E}^{N_b}_m(\al)}{\p \al^n} \no\\
\aeq  \f{1}{\pi}\sum_B \sum_{A \in b} \mbox{Im}\Big[\f{\partial S_{AB}(\al)}{\partial \al^n}\f{\partial S_{AB}^\ast(\al)}{\partial \al^m} \Big].
\eea
If the electrons of the bath $b$ are labeled by $b$ and energy $\ep$ and the scattering is elastic
\bea
S_{b\ep,b^\pr \ep^\pr} (\al) \aeq S_{b,b^\pr}(\ep,\al)\dl_{\ep,\ep^\pr},
\eea
$F^{N_b}_{mn}(\al)$ at zero-bias is given by
\bea
F^{N_b}_{mn}(\al) \aeq \f{1}{\pi}\sum_{b^\pr} \mbox{Im}\Big[\f{\partial S_{b,b^\pr}(\mu,\al)}{\partial \al^n}\f{\partial S_{b,b^\pr}^\ast(\mu,\al)}{\partial \al^m} \Big] .
\eea
Here, $\mu$ is the chemical potential of the baths.

On the other hand, it is difficult to calculate the scattering matrix in the interacting systems.
In the interacting system, the Brouwer formula had only been applied in mean field treatments \cite{aono03, aono04} or in the Toulouse limit \cite{Schiller}.

\subsubsection{Recent studies of the quantum pump}

Recently, the quantum pump in interacting systems have been actively researched. 
There are three theoretical approaches. The first is the Green's function approach \cite{G1, G2, G3, N16, N17}.
The second is the  generalized master equation (GME)
\cite{Splettstoesser06, Konig09-, RT09, Splettstoesser10-1, Splettstoesser10-2 ,Splettstoesser12, Splettstoesser13, Konig13-2} 
 approach 
which uses the GME that is equivalent\cite{a9, a10} to the quantum master equation (QME) derived using the Nakajima-Zwanzig projection operator technique
\cite{open}. 
Particularly, \Refe{Splettstoesser12} derived a geometric expression similar to the Brouwer formula and the Berry-Sinitsyn-Nemenman (BSN) vector explained later.
The third is the full counting statistics\cite{FCS-QME, FCS,Utumi} (FCS) with quantum master equation 
(FCS-QME, which is also called the generalized quantum master equation\cite{FCS-QME}) approach 
proposed in \Refe{Yuge12}.

The adiabatic modulation of the control parameters induces a Berry-phase-like\cite{Berry} quantity called the BSN phase in the FCS-QME with the Markov approximation.
Sinitsyn and Nemenman\cite{Sinitsyn} studied the adiabatically pumped charge using the FCS and had shown that it is 
characterized by the BSN vector, which results from the BSN phase. 
The BSN vector was applied to the spin boson system \cite{Spin-boson}.
The FCS-QME approach can treat the Coulomb interaction, which can not be treated in the Brouwer formula. 
The derived formula of the BSN vector depends on the approximations used for the QME. 
The Born-Markov approximation with or without the rotating wave approximation \cite{open}(RWA)
is frequently used. 
The QME in the Born-Markov approximation without RWA sometimes violates the non-negativity of the system reduced density operator \cite{Mar}.
The  QME of the RWA or the coarse-graining approximation\cite{CG, CG13}(CGA) is the Lindblad type which guarantees the non-negativity \cite{open}.

Some recent papers \cite{Splettstoesser10-2, Splettstoesser12, Yuge12} showed that the Coulomb interaction induces the quantum pump. 
In Refs.\cite{Splettstoesser10-2, Splettstoesser12}, it was shown that in a one level interacting quantum dot (QD) weakly coupled to two leads, 
the pumped charge (also spin in \Refe{Splettstoesser12}) induced by an adiabatic modulation of the energy level of the QD and the bias between the two leads 
vanishes in the noninteracting limit. 
In particular, Yuge {\it et al.}\cite{Yuge12} studied the pumped charge coming from the BSN curvatures by adiabatic modulation of the thermodynamic parameters 
(the chemical potentials and the temperatures) in spinless QDs weakly coupled to two spinless leads and showed that the BSN curvatures 
are zero in noninteracting QDs although they are nonzero for finite interaction.

\subsection{Thermodynamic entropy}

We review the thermodynamic entropy based on Ref.\cite{Shimizu}. 

\subsubsection{Principles of thermodynamics} \la{yousei}

A macro system $A$ is generally imposed internal constraints which describe the characters of the
internal structures.
For instance, the subsystem of $A$ is enclosed by the wall which does not transmit heat.
$A$ can be decomposed to the simple systems $\{A_i\}_i$. 
The simple system is the macro system which has not internal constraints and of which spatial non-uniformity in the equilibrium state  
due to the external fields is negligible. 
The equilibrium state is the state which all macro variables of the system do not change (as functions of time).  
As a principle, for arbitrary macro system $A$, it is requested that if $A$ is isolated (static external fields can exist) and is left sufficiently long time, 
$A$ becomes the equilibrium state. 
As principles, the followings are requested: 
(i) If $A$ is in the equilibrium state, the entropy $S$ exist uniquely. \\
(ii) The entropy $S_i$ of $A_i$ is a function of the internal energy $U_i$ of $A_i$ and the set of additive variables of $A_i$, $\{X_i^\al\}_{\al=1}^{m_i}$: 
$S_i=S_i(U_i,X_i^1,\cdots,X_i^{m_i})$. $U_i,X_i^1,\cdots,X_i^{m_i}$ are called the natural variables. \\   
(iii) $S_i(U_i,X_i^1,\cdots,X_i^{m_i})$ is continuously differentiable for the natural variables. 
In particular, $k_\RM{B}\be_i\defe \p S_i/\p U_i$ is positive and its lower limit is 0 and the upper limit does not exist. 
Here, $k_\RM{B}$ is the Boltzmann constant and $\be_i$ is the inverse temperature of $A_i$. \\
(iv) $A$ is in the equilibrium state if and only if all $A_i$ are in the equilibrium states and $\hat{S}\defe \sum_i S_i(U_i,X_i^1,\cdots,X_i^{m_i})$ is 
maximized. 
The entropy $S$ of $A$ is given by $S=\max_{\{U_i,X_i^1,\cdots,X_i^{m_i}\}}\hat{S}$ where $\max$ is the maximization under the permitted area.
The values of the natural variables which provide the $S$ are those in the equilibrium state.

\subsubsection{Heat and entropy}

The work $W$ is the transfered energy described by the macro variables. 
In general, $W$ is the summation of the mechanical work $W^M$ and the work due to particle transfer $W^C$ and etc. 
The heat $Q$ is defined by $\mU-W$ where $\mU$ is the total transfered energy from the external system. 
Because $\mU$ is the change of the internal energy $\Dl U$, $Q=\Dl U-W$ holds.

The process in which a system $B$ can be regarded as always be in the equilibrium state is called the quasistatic process for $B$. 

From the principles of \res{yousei}, the following theorem is derived. 
We consider a process that a general system $A$ receives the heat from external systems $B_1$, $B_2$, $\cdots, B_M$ 
exchanging mechanical work with the external systems $C_1$, $C_2$, $\cdots, C_N$. 
When $A$ contacts with $B_i$, $A$ does not contact with $\{B_b\}_{b \ne i}$. 
The set $\{(b,k) \in \{1,2,\cdots,M \}\times \{1,2,\cdots,N \} \vert B_b=C_k\}$ may not be an empty set. 
We suppose that this process is quasistatic process for $\{B_b\}_{b=1}^M$. 
Then, the change of the entropy $\Dl S$ of $A$ satisfies
\bea
\Dl S \le \sum_b \int_{i^{(b)}}^{f^{(b)}} k_\RM{B}\be_b d^\pr Q \la{Cl}.
\eea
Here, $\be_b$ is the inverse temperature of $B_b$, and ${i^{(b)}}$(${f^{(b)}}$) denotes the initial (final) state contacting $B_b$.  
In particular, the equality holds if the following conditions satisfy: (i) This process is also quasistatic process for $A$. 
(ii) While $A$ contacts to $B_b$, the inverse temperature of $A$ equals to $\be_b$.

In the following of this subsection, we consider a simple system $A$. 
We denote the natural variables of the entropy $S$ of $A$ by $U$ and $\{X^\al \}_{\al=1}^m$.
From the principles of \res{yousei}, $S(U,\{X^\al \}_{\al=1}^m)$ is convex upward for each natural variable.
The equation $S=S(U,\{X^\al \}_{\al=1}^m)$ can solve for $U$ uniquely: $U=U(S,\{X^\al \}_{\al=1}^m)$.
We introduce  $T\defe \p U/\p S$ and $P_\al \defe \p U/\p X^\al$. 
$T$ is the temperature of $A$ and $T=1/(k_\RM{B}\be)$ holds with $k_\RM{B}\be=\p S/\p U$. 
If $X^\al$ is the total particle number $N$, $\mu\defe \p U/\p N$ is the chemical potential. 

We denote the work by changing of $X^\al $ by $W^\al$. 
For a quasistatic process for $A$, the work is defined by
\bea
d^\pr W^\al \defe P_\al dX^\al \ (\RM{quasistatic\ process}). \la{def_dW}
\eea
Using $dU=TdS+\sum_\al P_\al dX^\al$ and the definition of the heat, 
\bea
dS=\be d^\pr Q \ (\RM{quasistatic \ process}),
\eea
hold. 
This is called the Clausius equality.
For the general system (which is not simple system), the Clausius equality holds if the temperature is uniform in the system.
In particular, if $\sum_{\al=1}^M  P_\al dX^\al=\mu dN $ holds, 
$d^\pr Q=dU-\mu dN$ and
\bea
dS=\be (dU-\mu dN) \ (\RM{quasistatic \ process}), \la{key_dS}
\eea
hold. Here and in the following this thesis, we set $k_\RM{B}=1$.
In general process, it is difficult to define the heat. 
For a quasistatic process for $B$, $Q_B$ can be defined as explained above. 
In \re{Cl}, $d^\pr Q$ is defined by $-d^\pr Q_b$ where $d^\pr Q_b$ is the heat to $B_b$.

In the equilibrium classical (quantum) system, the entropy is given by the Shannon entropy of the probability distribution 
(von Neumann entropy of the density matrix) of states. 

\subsection{Nonequilibrium steady state}

Let us consider a system $A$ coupled to the baths $\{B_b \}_{b=1}^M$ ($M>1$).   
We suppose that $\{B_b\}_{b \in \RM{\bm{\mC}}}$ are the canonical baths and $\{B_b\}_{b \in \RM{\bm{\mG}}}$ are the grand canonical baths. 
We denote the inverse temperature of $B_b$ by $\be_b$ and the chemical potential of $B_b$ ($b\in \RM{\bm{\mG}}$) by $\mu_b$. 
If all $\be_b$ and $\mu_b$ are the same ($\be_b=\be$ for all $b$ and $\mu_b=\mu$ for all $b\in \RM{\bm{\mG}}$), 
the total system is referred as zero-bias or equilibrium. 
For the nonequilibrium total system fixing (control) parameters, 
if $A$ is left sufficiently long time  and becomes a steady state, 
this state of $A$ or the total system is called the nonequilibrium steady state (NESS). 
For quantum system described by the QME, the NESS  exists uniquely. 

As the instance, we consider spinless one level QD coupled to several leads.   
$\ke{0}$ ($\ke{1}$) denotes the state that the QD is empty (occupied).  
The diagonal components $p_n(t)=\br{n}\rho(t)\ke{n}$ ($n=0,1$) of the system state $\rho$ are governed by the master equation:
\bea
\f{d}{dt}\begin{pmatrix} 
p_0(t) \\
p_1(t) \\ 
\end{pmatrix} \aeq K\begin{pmatrix} 
p_0(t) \\
p_1(t) \\ 
\end{pmatrix}. 
\eea
The Liouvillian is given by
\bea
K \aeq \sum_b \Ga_b \begin{pmatrix} 
-f_b &&1-f_b \\
f_b &&-(1-f_b) 
\end{pmatrix}.
\eea
Here, $\Ga_b$ is the line-width function of the lead $b$, 
$f_b=[e^{\be_b(\ep-\mu_b)}+1]^{-1}$ is the Fermi distribution function, 
$\ep$ is the energy level of the QD. 
In this section the parameters are fixed. 
The solution of the master equation is 
\bea
\begin{pmatrix} 
p_0(t) \\
p_1(t) \\ 
\end{pmatrix} \aeq \begin{pmatrix} 
1-F \\
F \\ 
\end{pmatrix}+e^{-\Ga t}\begin{pmatrix} 
-p_1(0)+F \\
p_1(0)-F
\end{pmatrix}, \la{ex_NESS}
\eea
where $\Ga=\sum_b \Ga_b$ and 
\bea
F \aeq \sum_b f_b \f{\Ga_b}{\Ga}.
\eea
The first term of the RHS of \re{ex_NESS} is the NESS. 

\subsection{Excess entropy} \la{Introduction_EE}

The investigation of thermodynamic structures of NESSs has been a topic of active research in nonequilibrium statistical mechanics 
\cite{18, Oono, SasaTasaki, S10, S11, S12, K17, K33, K35}. 
For instance, the extension of the relations in equilibrium thermodynamics, such as the Clausius equality, to NESSs has been one of the central subjects. 
Recently there has been a progress in the extension of the Clausius equality to NESSs \cite{Komatsu08, Komatsu11, Saitou} (see also Refs.\cite{22, 5, 6,Hatano, Sasa14, K13}).
In these studies, the excess heat $Q_{b,\rm{ex}}$ (of the bath $b$) \cite{Oono}, which describes an additional heat induced by a transition between NESSs with time-dependent external
control parameters, has been introduced instead of the total heat $Q_b$. 
The excess heat $Q_{b,\rm{ex}}$ is defined by subtracting from $Q_b$ the time integral of the instantaneous steady heat current from the bath $b$. 
In the weakly nonequilibrium regime, it is proposed that there exists a scalar potential $\mS$ in the control parameter space 
which approximately satisfies the extended Clausius equality
\bea
\sum_b \be_b Q_{b,\rm{ex}} \approx \Dl \mS. \la{2}
\eea
Here, $\be_b$ is the inverse temperature of the bath $b$, $\Dl \mS=\mS(\al_{t_f})-\mS(\al_{t_i})$, $\al_t$ is the value of the set of the control parameters at time $t$, 
and $t_i$ and $t_f$ are initial and final times of the operation. 
In classical systems, $\mS$ is the symmetrized Shannon entropy \cite{Komatsu11}. 
In quantum systems with the time-reversal symmetry, $\mS$ is the symmetrized von Neumann entropy \cite{Saitou}.
In general, the left hand side (LHS) of \re{2} is replaced by the excess entropy $\sig_\RM{ex} \defe \sig-\int_{t_i}^{t_f} dt\ J^\sig_\st(\al_t)$ 
where $\sig$ is the average entropy production and $J^\sig_\st(\al_t)$ is the instantaneous steady entropy production rate \cite{Sagawa, Yuge13, Komatsu15}. 
In the quasistatic operation, the excess entropy is given by  
\bea
\sig_\RM{ex} =\Dl \mS+\mO(\ep^2 \dl),
\eea
where $\ep$ is a measure of degree of nonequilibrium and $\dl$ describes the amplitude of the change of the control parameters.
Sagawa and Hayakawa \cite{Sagawa} studied the full counting statistics (FCS) of the entropy production for classical systems described by the Markov jump process and showed that
the excess entropy is characterized by the Berry-Sinitsyn-Nemenman (BSN) phase \cite{Sinitsyn}. 

The method of Ref.\cite{Sagawa} was generalized to quantum systems and applied to studies of the quantum pump \cite{Yuge12, Watanabe, Nakajima}. 
We explain the studies of the quantum pump.
At $t=0$ and $t=\tau$, we perform projection measurements of a {\it time-independent} observable $O$ of the baths and obtain the outcomes $o(0)$ and $o(\tau)$.
The generating function of $\Dl o=o(\tau)-o(0)$ is $Z_\tau(\chi)=\int d\Dl o \ P_\tau(\Dl o)e^{i\chi \Dl o}$ 
where $P_\tau(\Dl o)$ is the probability density distribution of $\Dl o$ and $\chi$ is called the counting field.
To calculate the generating function, the method using the quantum master equation (QME) with the counting field (FCS-QME) \cite{FCS-QME} had been proposed.
The solution of the FCS-QME $\rho^\chi(t)$ provides the generating function as $Z_\tau(\chi)=\tr_S[\rho^\chi(\tau)]$. 
$\tr_S$ denotes the trace of the system. 
The Berry phase \cite{Berry} of the FCS-QME is the BSN phase. 
The average of the difference of the outcomes is given by $\bra \Dl o \ket=\int_0^\tau dt \ i^O(t)$ where $i^O(t)$ is the current of an operator $O$.
If the state of the system at $t=0$ is the instantaneous steady state and the modulation of the control parameters is slow, the relation
\bea
\bra \Dl o \ket=\int_0^\tau dt \ i_\st^O(\al_t)+\int_C d\al^n \ A_n^O(\al), \la{I4}
\eea
holds. 
Here, the summation symbol for $n$ is omitted. 
$i_\st^O(\al_t)$ is the instantaneous steady current of $O$ and $A_n^O(\al)$ is the BSN vector derived from the BSN phase. 
$\al^n$ is $n$-th component of the control parameters, and $C$ is the trajectory from $\al_0$ to $\al_\tau$. 
The derived formula of the BSN vector depends on the approximations used for the QME.

Because of \re{key_dS}, the entropy production rate of the bath $b$ is $\dot{\sig}_b(t)= \be_b(t)[i^{H_b}(t)-\mu_b(t)i^{N_b}(t)]$ 
where $\mu_b$ is the chemical potential of the bath $b$, and $i^{H_b}(t)$ and $i^{N_b}(t)$ are energy and particle currents from the system to the bath $b$, respectively. 
$H_b$ and $N_b$ are the Hamiltonian and the total particle number of the bath $b$, respectively.
Then, it is natural to identify $\dot{\sig}(t)\defe -\sum_b \dot{\sig}_b(t)= \sum_b \be_b(t)[-i^{H_b}(t)-\mu_b(t)\{ -i^{N_b}(t)\}]$ with the average entropy production rate 
of the system. 
$\sig \defe \int_0^\tau dt \ \dot{\sig}(t)$ is the average entropy production.
Because of \re{I4}, $\sig=\int_0^\tau dt \ J^\sig_\st(\al_t)+\int_C \ d\al^n \ A_n^\sig(\al)$ holds with
$J^\sig_\st(\al)\defe \sum_b \be_b[-i_\st^{H_b}(\al)-\mu_b\{-i_\st^{N_b}(\al)\}]$ and 
\bea
A_n^\sig(\al)\defe \sum_b \be_b[-A_n^{H_b}(\al)-\mu_b\{-A_n^{N_b}(\al)\}] .
\eea
Here, $i_\st^{H_b}(\al)$ and $i_\st^{N_b}(\al)$ are the instantaneous steady currents of the energy and particle from the system to the bath $b$. 
$A_n^{H_b}(\al)$ and $A_n^{N_b}(\al)$ are the BSN vectors of $H_b$ and $N_b$. 
The excess entropy production is given by
\bea
\sig_\RM{ex}=\int_C \ d\al^n \ A_n^\sig(\al) .
\eea

Yuge {\it et al.} \cite{Yuge13} applied the FCS-QME approach to the excess entropy production of the quantum system.
They identified $\sig^\pr \defe \bra a(\tau)-a(0)\ket$ with the average entropy production. 
Here, $a(0)$ and $a(\tau)$ are the outcomes of $A(t)=-\sum_b \be_b(t)[H_b-\mu_b(t)N_b]$ at $t=0$ and $t=\tau$.
However, $\sig^\pr$ is not the average entropy production $\sig$. 
$\sig^\pr \approx \tr_\tot[A(\tau)\rho_\tot(\tau)]-\tr_\tot[A(0)\rho_\tot(0)]$ can be rewritten as 
\bea
\sig^\pr \aeqap -\int_0^\tau dt \ \sum_b \Big[\f{d\be_b(t)}{dt}\bra H_b \ket_t-\f{d[\be_b(t)\mu_b(t)]}{dt}\bra N_b \ket_t  \Big] \no\\
&&+\int_0^\tau dt \ \sum_b \Big[\be_b(t)\{ - \f{d}{dt} \bra H_b \ket_t\}-\be_b(t)\mu_b(t)\{ - \f{d}{dt} \bra N_b \ket_t\} \Big]. \la{sig^pr}
\eea
Here, $\bra \bu \ket_t \defe \tr_\tot[\bu \rho_\tot(t)]$, $\rho_\tot(t)$ is the total system state and $\tr_\tot$ denotes the trace of the total system. 
The integrand of the second term of the RHS of \re{sig^pr} roughly equals to $\dot{\sig}$ \cite{FN}. 
However, the physical meaning of the first term is not clear.  
Then, because of the presence of the first term, $\sig^\pr \ne \sig$ is concluded. 
Moreover, they improperly used the FCS-QME applicable only for {\it time-independent} observable to calculate $\sig^\pr$ although $A(t)$ is time-dependent.
These two issues are the problems of Ref.\cite{Yuge13}.

\subsection{Aim of this thesis}

There are several theoretical approaches to analyze the quantum pump. 
However, the relations among these are not clear. 
Then, the first aim of this thesis is to clarify these relations (in particular, 
the relation between the FCS-QME approach and the GME approach). 
Moreover, in the previous works, the charge pump had been studied mainly. 
However, for applications to the spintronics and quantum information processing, 
the spin degree of freedom is important.  
Then, we consider the spin degree of freedom and study the spin pump. 

Recently, the excess entropy of the classical system is established. 
However, one for the quantum system is not sufficient as we explained in \res{Introduction_EE}. 
The second aim of this thesis is to develop the excess entropy of the quantum system. 
Moreover, we compare between our results and previous results of both classical and quantum systems.

\subsection{Outline of the thesis}

The outline of the thesis is as follows. 
First, we review the FCS and the FCS-QME (\rec{s2}). 
In \res{sFCS}, we derive the modified von Neumann equation including the counting fields. 
In \res{sQME}, we derive and the FCS-QME with the CGA. 
In \res{RWA}, we explain the RWA. 
In \res{sKMS}, we derive the detailed balance condition. 

Next, we move to the original results (\res{BSNP},\res{Cyclic} and before \re{BSN_Nakajima} are review parts). 
\rec{sFCS-QME} and \rec{QAP} are based on Ref.\cite{Nakajima}. 
\rec{sGQME}, \rec{GEEE} and \rec{Other_def} are based on Ref.\cite{Nakajima2}.
We apply the FCS-QME to the quantum pump (\rec{sFCS-QME}). 
In \res{Currents}, we derive the expression for current without any approximation and introduce the BSN vector.
The BSN vector is also derived from the BSN phase (\res{BSNP}). 
In \res{Cyclic}, we introduce the BSN curvature used to cyclic adiabatic pump. 
In \res{Expansion}, we expand the general solution of the QME $\rho(t)$ by the modulation frequency $\om$ as
\bea 
\rho(t)=\rho_0(\al_t)+\sum_{n=1}^\infty \rho^{(n)}(t)+\sum_{n=0}^\infty \tl \rho^{(n)}(t) . \la{E}
\eea
Here, $\rho_0(\al_t)$ is the instantaneous steady state of the QME, 
$\rho^{(n)}(t)$ and $ \tl \rho^{(n)}(t) $ are calculable and order $(\om/\Ga)^n$. 
$\Ga$ is the coupling strength between the system and the baths. 
$ \tl \rho^{(n)}(t) $ exponentially damps as a function of time. 
In the expansion \re{E}, a pseudo-inverse of the Liouvillian is used. 
In \res{PI}, we proof the expansion \re{E} is independent of the choice of the pseudo-inverse. 
In \res{GMEa}, we show that the GME provides an expansion corresponding to $\rho_0(\al_t)+\sum_{n=1}^\infty \rho^{(n)}(t)$.

In \rec{QAP}, we apply the FCS-QME with the RWA to the quantum adiabatic pump of the quantum dots (QDs) coupled to two leads ($L$ and $R$).  
 In \res{model}, we explain the model.  
We show that the pumped charge and spin coming from the instantaneous steady current are not negligible when the thermodynamic parameters are not fixed to zero bias 
in noninteracting QDs (\res{SC}) and an interacting QD (\res{current,inf}). 
To observe the spin effects, we consider collinear magnetic fields, which relate to spins through the Zeeman effect, with different amplitudes applying to the QDs ($B_S$) 
and the leads ($B_L$ and $B_R$). 
We focus on the dynamic parameters ($B_S$, $B_{L/R}$ and the coupling strength between QDs and leads, $\Dl_{L/R}$) as control parameters. 
In one level QD with the Coulomb interaction $U$, we analytically calculate the BSN curvatures of spin and charge of $(B_L,B_S)$ pump 
and $(\Dl_L,B_S)$ pump for the noninteracting limit ($U=0$, \res{pump,U=0}) and the strong interaction limit ($U=\infty$, \res{pump}) at zero-bias.
Moreover, we study the $(\Dl_L,B_S)$ pump for finite $U$ at zero-bias (\res{pump_finite}).

We study the  quantum diabatic pump for spinless one level QD coupled to two leads (\rec{diabatic}). 
We calculate $\{\rho^{(n)}(t)\}_{n=1}^5$, $\{\tl \rho^{(n)}(t)\}_{n=1}^5$ and particle current up to 6th order and 
pumped particle numbers.  

Next, we introduce the generalized QME (\rec{sGQME}) used to analyze the BSN vector of the entropy production. 
In \rec{GEEE} and \rec{Other_def}, we focus on the RWA. 
In \res{eq}, the BSN vector $A_n^\sig$ in the equilibrium is discussed. 
In \res{noneq}, one of the main result of this thesis
\bea
A_n^\sig(\al)=-\tr_S\big[\ln \rho_0^{(-1)}(\al)\f{\p \rho_0(\al)}{\p \al^n} \big]+\mO(\ep^2), \la{Main}
\eea
 is derived  without any assumption on the time-reversal symmetry \cite{Nakajima2}. 
$\rho_0^{(-1)}(\al)$ is the instantaneous steady state obtained from the QME with reversing the sign of the Lamb shift term. 
In \res{Discussion}, we consider the time-reversal operation.  
We show that if the time-reversal symmetry is broken and the system Hamiltonian is degenerated,  $\mS(\al)$ such that $A_n^\sig(\al)=\p \mS(\al)/\p \al^n+\mO(\ep^2)$ 
dose not exist. 
This is the most important result of this thesis.
Next we mention the results in the Born-Markov approximation (\res{Born_Markov}). 
In \rec{Other_def}, we compare preceding study on of the entropy production in the classical Markov jump process \cite{Komatsu15, Jarzynski} with ours.

At last (\rec{Conclusion}), we summarize this thesis. 
In Appendix \ref{BMA}, the Liouvillian for the  Born-Markov approximation is discussed. 
In Appendix \ref{Liouville space}, the Liouville space\cite{Fano, FCS-QME} and the matrix representation of the Liouvillian are explained. 
In Appendix \ref{co}, we derive \re{exp}. 
In Appendix \ref{val}, we discuss the validity of the adiabatic expansion in \rec{sFCS-QME}.
In Appendix \ref{proof}, we discuss the derivation of \re{W_R}. 
In the Appendix \ref{A_GME}, we discuss the solutions of the GME expanded by the modulation frequency and the coupling strength between the system and the baths. 
In the Appendix \ref{Current_op}, we calculate the energy current operator. 
In the Appendix \ref{Appendix A}, we derive the formula of the derivative of the von Neumann entropy. 
In the Appendix \ref{proof_ln A+B}, we proof \re{ln A+B}. 
In the Appendix \ref{Appendix B}, we explain the definition of entropy production of the Markov jump process and a result of Ref.\cite{Komatsu15}.

\newpage

\section{Full counting statistics and quantum master equation} \la{s2}

\subsection{Full counting statistics} \la{sFCS}

We consider the system $S$ coupled with the bath system $B$:
\bea
H_\tot(t) \aeq H_S(t) +H_B(t)+ H_{\RM{int}}(t).
\eea
The bath system may contain several baths. 
The simultaneous eigenstate of a set of the bath's observables $\{ O_\mu \}$ is given by
\bea
O_\mu \ke{ \{ o_\nu \} ,r} \aeq o_\mu \ke{ \{ o_\nu \} ,r} ,\\
\br{ \{ o_\nu \} ,r} \{ o_\nu^\pr \} ,s \ket \aeq \dl_{r,s} \dl_{\{o_\nu \},\{o_\nu^\pr \} } .
\eea
Here, $r$ and $s$ denote the label of degeneracy, and $\dl_{\{o_\nu \},\{o_\nu^\pr \} } = \prod_{\nu=1}^n \dl_{o_\nu,o_\nu^\pr}  $ is the kronecker delta.
The projection operator to $\{ o_\mu \}$ is given by
\bea
P_{\{ o_\mu \}} \aeq \sum_r \ke{ \{ o_\mu \} ,r} \br{ \{ o_\mu \} ,r} .
\eea
This has the following properties 
\bea
P_{\{ o_\nu \}} P_{\{ o_\nu^\pr \}} \aeq \dl_{\{o_\nu \},\{o_\nu^\pr \} }  P_{\{ o_\nu \}} \la{PP1},\\
\sum_{\{ o_\nu \}} P_{\{ o_\nu \}} \aeq 1 \la{PP2}.
\eea

The total system state $\rho_\tot(t)$ is governed by the von Neumann equation:
\bea
\f{d}{dt} \rho_\tot(t) \aeq  -i[H_\tot(t),\rho_\tot(t)]. \la{vNeq}
\eea
In this thesis, we set $\hbar=1$. The formal solution is given by 
\bea
\rho_\tot(t) \aeq V(t) \rho_\tot(0) V\dg(t) ,
\eea
where $V(t)$ is the solution of 
\bea
\f{d}{dt}  V(t) \aeq -iH_\tot(t)V(t) ,
\eea
with $V(0)=1$. 
At $t=0$, we perform projection measurements of $\{ O_\mu \}$. 
The probability getting $\{ o_\mu^{(0)} \} $ is given by
\bea
P[\{ o_\mu^{(0)} \}] \aeq \tr_\tot [P_{\{ o_\mu^{(0)} \}}\rho_\tot(0)P_{\{ o_\mu^{(0)} \}}] .
\eea
$\tr_\tot$ denotes the trace over the total system. 
The state just after measuring $\{ o_\mu^{(0)} \} $ is
\bea
\rho_\tot^{\{ o_\mu^{(0)} \}}(0) \aeq \f{P_{\{ o_\mu^{(0)} \}}\rho_\tot(0)P_{\{ o_\mu^{(0)} \}}}{P[\{ o_\mu^{(0)} \}]} .
\eea
After the time evolution by \re{vNeq}, the state at time $t$ is 
\bea
\rho_\tot^{\{ o_\mu^{(0)} \}}(t) \aeq V(t)\rho_\tot^{\{ o_\mu^{(0)} \}}(0) V\dg(t) \no\\
\aeq \f{V(t)P_{\{ o_\mu^{(0)} \}}\rho_\tot(0)P_{\{ o_\mu^{(0)} \}}V\dg(t)}{P[\{ o_\mu^{(0)} \}]} .
\eea
Under this condition, we perform projection measurements of $\{ O_\mu \}$ at $t=\tau$. 
The probability getting $\{ o_\mu^{(\tau)} \} $ is given by 
\bea
P[\{ o_\mu^{(\tau)} \}  \vert \{ o_\mu^{(0)} \} ] \aeq  \tr_\tot [P_{\{ o_\mu^{(\tau)} \}}\rho_\tot^{\{ o_\mu^{(0)} \}}(\tau) P_{\{ o_\mu^{(\tau)} \}}]  \no\\
\aeq \f{1}{P[\{ o_\mu^{(0)} \}]} \tr_\tot[P_{\{ o_\mu^{(\tau)} \}}V(\tau)P_{\{ o_\mu^{(0)} \}}\rho_\tot(0)P_{\{ o_\mu^{(0)} \}}V\dg(\tau) P_{\{ o_\mu^{(\tau)} \}} ].
\eea
The probability getting $\{ o_\mu^{(0)} \}$ at $t=0$ and $\{ o_\mu^{(\tau)} \}$ at $t=\tau$ is 
\bea
P[\{ o_\mu^{(\tau)} \}  , \{ o_\mu^{(0)} \} ] \aeq P[\{ o_\mu^{(0)} \}] \cdot P[\{ o_\mu^{(\tau)} \}  \vert \{ o_\mu^{(0)} \} ] \no\\
\aeq \tr_\tot[P_{\{ o_\mu^{(\tau)} \}}V(\tau)P_{\{ o_\mu^{(0)} \}}\rho_\tot(0)P_{\{ o_\mu^{(0)} \}}V\dg(\tau) P_{\{ o_\mu^{(\tau)} \}} ].  \la{p_mn}
\eea

The probability density distribution of $\{ o_\mu^{(\tau)}-o_\mu^{(0)} \}$ is given by 
\bea
P_\tau[\{\Dl  o_\mu \}] \aeqd \mbox{Prob.}[\{ o_\mu^{(\tau)}-o_\mu^{(0)}=\Dl o_\mu \}] \\
\aeq \sum_{\{ o_\mu^{(0)} \},\{ o_\mu^{(\tau)} \}}  P[\{ o_\mu^{(\tau)} \}  , \{ o_\mu^{(0)} \} ] \prod_\mu \dl(o_\mu^{(\tau)}-o_\mu^{(0)}-\Dl o_\mu) . \la{p_da}
\eea
The generating function is defined by 
\bea
Z_\tau(\chi ) \aeqd \int( \prod_{\nu=1}^n d\Dl o_\nu ) \ P_\tau(\{\Dl o_\mu \} ) e^{i \sum_{\mu=1}^n \chi_\mu \Dl o_\mu } . \la{Z}
\eea
Here, $\chi_\mu$ is a real number called the counting field for $O_\mu$. 
$\chi$ denotes the set of the counting fields. 
The cumulant generating function is defined by
\bea
S_\tau(\chi) \aeqd \ln Z_\tau(\chi ).
\eea
The $n$-th order cumulant $\bra \Dl o_{\mu_1} \Dl o_{\mu_2} \dots \Dl o_{\mu_n} \ket_c$ is given by
\bea
\bra \Dl o_{\mu_1} \Dl o_{\mu_2} \dots \Dl o_{\mu_n} \ket_c 
\aeq \f{\partial^n S_\tau(\chi)}{\partial (i\chi_{\mu_1} )(i\chi_{\mu_2} ) \cdots (i\chi_{\mu_n} )}  \Big \vert_{\chi=0}.
\eea
In particular, 
\bea
\bra \Dl o_{\nu} \ket_c = \f{\partial S_\tau(\chi)}{\partial (i\chi_\nu)}   \Big \vert_{\chi=0} = \bra \Dl o_{\nu} \ket ,
\eea
is the average of $\Dl o_{\nu}$. 

Substituting \re{p_da} to \re{Z}, we obtain
\bea
Z_\tau(\chi ) \aeq \sum_{\{ o_\mu^{(0)} \},\{ o_\mu^{(\tau)} \}}P[\{ o_\mu^{(\tau)} \}  , \{ o_\mu^{(0)} \} ]e^{i\sum_\mu \chi_\mu [o_\mu^{(\tau)}-o_\mu^{(0)}]} .
\eea
Substituting \re{p_mn} to the above equation, we obtain
\bea
&&\hs{-10mm}Z_\tau(\chi ) \no\\
\aeq \sum_{\{ o_\mu^{(0)} \},\{ o_\mu^{(\tau)} \}}\tr_\tot[P_{\{ o_\mu^{(\tau)} \}}V(\tau)P_{\{ o_\mu^{(0)} \}}\rho_\tot(0)
P_{\{ o_\mu^{(0)} \}}V\dg(\tau) P_{\{ o_\mu^{(\tau)} \}} ]e^{i\sum_\mu \chi_\mu [o_\mu^{(\tau)}-o_\mu^{(0)}]}. \no\\ \la{Z_2}
\eea
Now, we introduce
\bea
\Bar{\rho}_\tot(0) \aeqd \sum_{\{ o_\mu^{(0)} \}} P_{\{ o_\mu^{(0)} \}}\rho_\tot(0)P_{\{ o_\mu^{(0)} \}}.
\eea
Properties 
\bea
e^{-i\chi_\nu O_\nu/2}P_{\{ o_\mu^{(0)} \}} \aeq e^{-i\chi_\nu o_\nu^{(0)}/2}P_{\{ o_\mu^{(0)} \}} ,\ 
 P_{\{ o_\mu^{(0)} \}}e^{-i\chi_\nu O_\nu/2} = e^{-i\chi_\nu o_\nu^{(0)}/2}P_{\{ o_\mu^{(0)} \}}, \la{d1}
\eea
lead
\bea
e^{-i\sum_\mu \chi_\mu O_\mu/2}\Bar{\rho}_\tot(0)e^{-i\sum_\mu \chi_\mu O_\mu/2} \aeq 
\sum_{\{ o_\mu^{(0)} \}}  e^{-i\chi_\mu o_\mu^{(0)}}P_{\{ o_\mu^{(0)} \}}\rho_\tot(0)P_{\{ o_\mu^{(0)} \}}. \la{d2}
\eea
Then, \re{Z_2} becomes 
\bea
&&\hs{-10mm}Z_\tau(\chi ) \no\\
\aeq \sum_{\{ o_\mu^{(\tau)} \}} \tr_\tot[P_{\{ o_\mu^{(\tau)} \}}V(\tau) e^{-i\sum_\mu \chi_\mu O_\mu/2}\Bar{\rho}_\tot(0)
e^{-i\sum_\mu \chi_\mu O_\mu/2}  V\dg(\tau) P_{\{ o_\mu^{(\tau)} \}} ]e^{i\sum_\mu \chi_\mu o_\mu^{(\tau)}} .\no\\
\eea
Moreover, from
\bea
P_{\{ o_\mu^{(\tau)} \}}e^{i\chi_\nu O_\nu/2} = e^{-i\chi_\nu o_\nu^{(\tau)}/2}P_{\{ o_\mu^{(\tau)} \}} ,\  
e^{i\chi_\nu O_\nu/2}P_{\{ o_\mu^{(\tau)} \}} \aeq e^{i\chi_\nu o_\nu^{(\tau)}/2}P_{\{ o_\mu^{(\tau)} \}},   \la{d3}
\eea
we obtain
\bea
Z_\tau(\chi) \aeq \sum_{\{ o_\mu^{(\tau)} \}} \tr_\tot[P_{\{ o_\mu^{(\tau)} \}}e^{i\chi_\mu O_\mu/2} V(\tau) e^{-i \chi_\mu O_\mu/2}\Bar{\rho}_\tot(0)
e^{-i \chi_\mu O_\mu/2} V\dg(\tau)e^{i\chi_\mu O_\mu/2} P_{\{ o_\mu^{(\tau)} \}} ] \no\\
\aeq  \sum_{\{ o_\mu^{(\tau)} \}} \tr_\tot[P_{\{ o_\mu^{(\tau)} \}} P_{\{ o_\mu^{(\tau)} \}}e^{i\chi_\mu O_\mu/2} V(\tau) e^{-i \chi_\mu O_\mu/2}\Bar{\rho}_\tot(0)
e^{-i \chi_\mu O_\mu/2} V\dg(\tau)e^{i\chi_\mu O_\mu/2} ] \no\\
\aeq \tr_\tot[ \sum_{\{ o_\mu^{(\tau)} \}} P_{\{ o_\mu^{(\tau)} \}} e^{i\chi_\mu O_\mu/2} V(\tau) 
e^{-i \chi_\mu O_\mu/2}\Bar{\rho}_\tot(0)e^{-i \chi_\mu O_\mu/2} V\dg(\tau)e^{i\chi_\mu O_\mu/2} ] \no\\
\aeq \tr_\tot[ e^{i\chi_\mu O_\mu/2} V(\tau) e^{-i \chi_\mu O_\mu/2}\Bar{\rho}_\tot(0)e^{-i \chi_\mu O_\mu/2} V\dg(\tau)e^{i\chi_\mu O_\mu/2} ]\no\\
\aeq \tr_\tot[  V_\chi(\tau) \Bar{\rho}_\tot(0) V_{-\chi}\dg(\tau)] \no\\
\aeq \tr_\tot[ \rho_\tot^\chi(\tau)].  \la{d4}
\eea
Here, we used \re{PP1} and \re{PP2}. 
Here and in the following of this section, $\chi_\mu O_\mu\e \sum_\mu \chi_\mu O_\mu$.
$V_\chi(t)$ and $\rho_\tot^\chi(t)$ are defined by
\bea
V_\chi(t) \aeqd e^{i\chi_\mu O_\mu/2} V(t) e^{-i\chi_\mu O_\mu/2} ,\\
\rho_\tot^\chi(t) \aeqd V_\chi(t) \Bar{\rho}_\tot(0) V_{-\chi}\dg(t). \la{chi,t}
\eea

$V_\chi(0)$ and $\rho_\tot^\chi(0)$ are given by
\bea
V_\chi(0) \aeq  1, \no\\
\rho_\tot^\chi(0) \aeq \Bar{\rho}_\tot(0) =\sum_{\{ o_\mu^{(0)} \}} P_{\{ o_\mu^{(0)} \}}\rho(0)P_{\{ o_\mu^{(0)} \}}.
\eea
$V_\chi(t)$ is governed by
\bea
\f{d}{dt} V_\chi(t) \aeq e^{i\chi_\mu O_\mu/2} \big[ \f{d}{dt} V(t) \big] e^{-i\chi_\mu O_\mu/2} \no\\
\aeq e^{i\chi_\mu O_\mu/2} \big[ -iH_\tot(t)V(t) \big] e^{-i\chi_\mu O_\mu/2} \no\\
\aeq -ie^{i\chi_\mu O_\mu/2} H_\tot(t)e^{-i\chi_\mu O_\mu/2} e^{i\chi_\mu O_\mu/2} V(t)  e^{-i\chi_\mu O_\mu/2} \no\\
\aeq -iH_{\tot,\chi}(t) V_\chi(t) , \la{eq.Vc}
\eea
with 
\bea
H_{\tot,\chi}(t) \aeqd e^{i\chi_\mu O_\mu/2} H_\tot(t)e^{-i\chi_\mu O_\mu/2} .
\eea
$H_{\tot,\chi}(t)$ is a Hermitian operator:
\bea
H_{\tot,\chi} \dg(t) \aeq H_{\tot,\chi}(t) .
\eea
From the Hermitian conjugate of \re{eq.Vc}, we obtain
\bea
\f{d}{dt} V_{-\chi} \dg (t) \aeq i V_{-\chi} \dg (t)H_{\tot,-\chi}(t).
\eea
From \re{chi,t}, \re{eq.Vc} and the above equation, $\rho_\tot^\chi(t)$ is governed by
\bea 
\f{d}{dt} \rho_\tot^\chi(t) \aeq  \f{d}{dt} [V_\chi(t) \Bar{\rho}_\tot(0) V_{-\chi}\dg(t)] \no\\
\aeq -iH_{\tot,\chi}(t) V_\chi(t)\Bar{\rho}_\tot(0) V_{-\chi}\dg(t)+iV_\chi(t) \Bar{\rho}_\tot(0) V_{-\chi}\dg(t)H_{\tot,-\chi}(t) \no\\
\aeq -i[H_{\tot,\chi}(t)\rho_\tot^\chi(t)-\rho_\tot^\chi(t)H_{\tot,-\chi}(t)] .\la{G_von}
\eea

\subsection{Quantum master equation with counting fields} \la{sQME}

\subsubsection{Derivation of FCS-QME}

We consider system $S$ weakly coupled to several baths. 
The total Hamiltonian is given by 
\bea
H_\tot(\al^\pr(t)) = H_S(\al_S(t))+\sum_b [H_b(\al_b^\pr(t))+H_{Sb}(\al_{Sb}(t))].
\eea
$H_S(\al_S)$ is the system Hamiltonian and $\al_S$ denotes a set of control parameters of the system. 
$H_b(\al_b^\pr)$ is the Hamiltonian of the bath $b$ and $\al_b^\pr$ is a set of control parameters. 
$H_{Sb}(\al_{Sb})$ is the coupling Hamiltonian between $S$ and the bath $b$, and $\al_{Sb}$ is a set of control parameters.
We suppose that the states of the baths for $b=1,2,\cdots,n_\RM{C}$ are the canonical distributions and these for $b=n_\RM{C}+1,\cdots,n_\RM{C}+n_\RM{GC}$
are the grand canonical distributions. 
We denote $\{1,\cdots,n_\RM{C} \}$ and $\{n_\RM{C}+1,\cdots,n_\RM{C}+n_\RM{GC}\}$ by $\RM{\bm{\mC}}$ and $\RM{\bm{\mG}}$. 
We denote the inverse temperature of the bath $b$ by $\be_b$ and the chemical potential of the bath $b\in \RM{\bm{\mG}}$ by $\mu_b$. 
$\al_b^{\pr\pr}$ denotes  $\be_b$ for $b \in \RM{\bm{\mC}}$ and the set of $\be_b$ and $\be_b \mu_b$ for $b \in \RM{\bm{\mG}}$.
We symbolize the set of all control parameters ($\al_S$, $\{\al_{Sb}\}_b$, $\{\al_b^\pr \}_b$, $\{\al_b^{\pr\pr} \}_b$) by $\al$, 
($\al_S$, $\{\al_{Sb}\}_b$, $\{\al_b^\pr \}_b$) by $\al^\pr$, $\{\al_b^{\pr\pr} \}_b$ by $\al^{\pr\pr}$, 
 $(\al_b^\pr$, $\al_b^{\pr\pr})$ by $\al_b$, and $\{\al_b \}_b$ by $\al_B$.
While $\al^\pr$ are dynamical parameters, $\al^{\pr\pr}$ are thermodynamical parameters. 
We denote the set of all the linear operators of $S$ by $\RM{\bm{B}}$. 

The modified von Neumann equation \re{G_von} \cite{FCS-QME} is
\bea 
\f{d}{dt} \rho_\tot^\chi(t) =  -i[H_\tot(t), \rho_\tot^\chi(t)]_\chi . \la{GLN}
\eea
Here, $[A,B]_\chi \defe A_{\chi}B-BA_{-\chi}$ and $A_\chi \defe e^{i \sum_\mu \chi_{O_\mu} O_\mu/2}Ae^{-i\sum_\mu \chi_{O_\mu} O_\mu/2} $. 
$\chi_{O_\mu}$ is $\chi_\mu$ of \res{sFCS}. 
We suppose 
\bea
\rho_\tot(0)=\rho(0) \otimes \rho_B(\al_B(0)),
\eea
 where 
$\rho_B (\al_B(0)) \defe \bigotimes_{b} \rho_b(\al_b(0))$ and $\rho_b(\al_b(0)) \defe e^{-\beta_b(0) H_b(\al_b^\pr(0)) }/Z_b(\al_b(0)) $
with $Z_b(\al_b)\defe \tr_b[e^{-\beta_b H_b(\al_b^\pr) }]$ for $b \in \RM{\bm{\mC}}$ and 
$\rho_b(\al_b(0)) \defe e^{-\beta_b(0) [H_b(\al_b^\pr(0)) -\mu_b(0) N_b] }/\Xi_b(\al_b(0)) $ 
with $\Xi_b(\al_b)\defe \tr_b[e^{-\beta_b[ H_b(\al_b^\pr) -\mu_b N_b] }]$ for $b \in \RM{\bm{\mG}}$. 
$\tr_b$ denotes the trace of the bath $b$ and $N_b$ ($b \in \RM{\bm{\mG}}$) is the total number operator of the bath $b$. 
Then, 
\bea
\rho_\tot^\chi(0)=\rho(0) \otimes \sum_{\{ o_\nu \}} P_{\{ o_\nu \}} \rho_B(\al_B(0)) P_{\{ o_\nu \}},
\eea
 obeys. 
We suppose $[H_b,N_b]=0$. 
We suppose that $O_\mu$ commute with $H_b$ and $N_b$:
\bea
[O_\mu,H_b]=0,\ [O_\mu,N_b]=0.
\eea 
Then, $P_{\{ o_\nu \}}$ commutes with $\rho_B(\al_B(0))$ and 
\bea
\rho_\tot^\chi(0)=\rho(0) \otimes \rho_B(\al_B(0)),
\eea
holds because \re{PP1} and \re{PP2}.

We defined 
\bea
\rho^\chi(t)\defe \tr_B[\rho_\tot^\chi(t)],
\eea
 which provides the generating function 
\bea
Z_\tau(\chi)=\tr_S[\rho^\chi(t=\tau)].
\eea  
$\tr_B$ denotes the trace over all baths' degrees of freedom. 
We assume $\rho_\tot(t)\approx \rho(t) \otimes \rho_B(\al_B(t))$ $(0<t\le \tau)$, where 
\bea
 \rho_B (\al_B(t)) \aeqd \bigotimes_{b} \rho_b(\al_b(t)), \\
\rho_b(\al_b(t)) \aeqd  \left \{ \begin{array}{ll}
e^{-\beta_b(t) H_b(\al_b^\pr(t)) }/Z_b(\al_b(t)) \hs{19.3mm} b \in \RM{\bm{\mC}} \\
e^{-\beta_b(t) [H_b(\al_b^\pr(t)) -\mu_b(t) N_b] }/\Xi_b(\al_b(t)) \hs{5mm} b \in \RM{\bm{\mG}}
\end{array} \right. .
 \eea
and 
\bea
\rho(t)\defe \tr_B[\rho_\tot(t)].  
\eea
 
First, we introduce the CGA. 
An operator in the interaction picture corresponding to $A(t)$ is defined by 
\bea
A^I(t)=U_0\dg(t)A(t)U_0(t),
\eea
with 
\bea
\f{dU_0(t)}{dt}=-i[H_S(\al_S(t))+\sum_{b}H_b(\al_b^\pr(t))]U_0(t),
\eea
and $U_0(0)=1$. 
The system reduced density operator in the interaction picture is given by 
\bea
\rho^{I,\chi}(t)=\tr_B[\rho_\tot^{I,\chi}(t)],
\eea
where 
\bea
\rho_\tot^{I,\chi}(t)=U_0\dg(t)\rho_\tot^\chi(t)U_0(t).
\eea  
$\rho_\tot^{I,\chi}(t)$ is governed by
\bea
\f{d\rho_\tot^{I,\chi}(t)}{dt} = -i[H_{\rm{int}}^I(t),\rho_\tot^{I,\chi}(t)]_\chi, \la{vN,I}
\eea
with 
\bea
H_{\rm{int}}\defe \sum_{b}H_{Sb}.
\eea 
Up to the second order perturbation in $H_{\rm{int}}$, we obtain
\bea
\rho^{I,\chi}(t+\tau_{\rm{CG}}) 
\aeq \rho^{I,\chi}(t) \no\\
&&-\int_t^{t+\tau_{\rm{CG}}} du \int_t^{u} ds \ \tr_B \big\{ [ H_{\rm{int}}^I(u), [H_{\rm{int}}^I(s), \rho^{I,\chi}(t) \rho_B (\al_B(t)) ]_\chi ]_\chi \big \} \no\\
\aeqe \rho^{I,\chi}(t) +\tau_{\rm{CG}}\hat{L}_{\tau_{\rm{CG}}}^\chi(t)\rho^{I,\chi}(t) , \la{FCS-QME_I_CG}
\eea
using the large-reservoir approximation 
\bea
 \rho_\tot^{I,\chi}(t) \approx \rho^{I,\chi}(t) \otimes \rho_B (\al_B(t)),
\eea
 and supposing 
\bea
\tr_B[H_{\rm{int}}^I(u)\rho_B (\al_B(t))]=0. \la{H_int_sup}
\eea
The arbitrary parameter $\tau_{\rm{CG}}$ $(>0)$ is called the coarse-graining time. 
The CGA \cite{CG, CG13} is defined by 
\bea
\f{d}{dt}\rho^{I,\chi}(t)=\hat{L}_{\tau_{\rm{CG}}}^\chi(t)\rho^{I,\chi}(t). \la{def,CGA}
\eea
In the Schr\"{o}dinger picture, \re{def,CGA} is described as 
\bea
\f{d\rho^\chi(t)}{dt}=-i[H_S(\al_S(t)),\rho^\chi(t)]+\sum_b \mL_{b,\tau_{\rm{CG}}}^\chi(\al_t) \rho^\chi(t) .\la{B7}
\eea
At $\chi=0$, this is the Lindblad type.
If $\tau_{\rm{CG}} \ll \tau$, the super-operator $\mL_{b,\tau_{\rm{CG}}}^\chi$ is described as a function of the set of control parameters at time $t$.  
$\al_t=\al(t)$ is the value of $\al$ at time $t$.  
In this thesis, we suppose 
\bea
\tau_{\rm{CG}} \ll \tau.
\eea
Moreover, $\tau_{\rm{CG}}$ should be much shorter than the relaxation time of the system, $\tau_S$:
\bea
\tau_{\rm{CG}} \ll \tau_S.
\eea
For the adiabatic modulation, $\tau_S \ll \tau $ should hold, then $\tau_{\rm{CG}} \ll \tau_S \ll \tau$ holds. 

In general, the FCS-QME is given by 
\bea
\f{d\rho^\chi(t)}{dt}=-i[H_S(\al_S(t)),\rho^\chi(t)]+\sum_b \mL_b^\chi(t) \rho^\chi(t),
\eea
with the initial condition 
\bea
\rho^\chi(0)=\rho(0).
\eea
$\mL_b^\chi(t) $ describes the coupling effects between $S$ and the bath $b$ and depends on used approximations. 
In this thesis, we suppose 
\bea
\mL_b^\chi(t)=\mL_b^\chi(\al_t).
\eea
The Born-Markov approximation without or within the RWA and the CGA satisfy this equation. 
Then, the FCS-QME is given by 
\bea
\f{d \rho^\chi(t)}{dt}=\hat{K}^\chi(\al_t)\rho^\chi(t). \la{FCS-QME}
\eea
Here,
\bea
\hat{K}^\chi(\al)\bu=-i[H_S(\al_S),\bu]+\sum_b \mL_b^\chi(\al) \bu, 
\eea
is the Liouvillian. 
Here and in the following, $\bu$ denotes an arbitrary liner operator of the system. 

The Born-Markov approximation is given by 
\bea
\f{d\rho^{I,\chi}(t)}{dt} = -\int_0^\infty ds \ \tr_B\Big\{[H_{\rm{int}}^I(t),[H_{\rm{int}}^I(t-s),\rho^{I,\chi}(t) \rho_B (\al_B(t))]_\chi]_\chi \Big\}.\la{BM}
\eea

\subsubsection{Coarse-graining approximation}

In general, the interaction Hamiltonian is given by
\bea
H_{Sb}(\al_{Sb}) \aeq \sum_\mu s_{b\mu} R_{b,\mu}(\al_{Sb}) = \sum_\mu R_{b,\mu} \dg (\al_{Sb}) s_{b\mu} \dg . \la{H_Sb_G}
\eea
Here, $s_{b\mu}$ is an operator of the system and $R_{b,\mu}(\al_{Sb})$ is an operator of the bath $b$. 
We suppose
\bea
\tr_b[ \rho_b(\al_b(t)) R_{b,\mu}(\al_{Sb}(s))] =0 ,
\eea
corresponding to \re{H_int_sup}.
Then, 
\bea
&&\hs{-10mm}\tr_B\Big\{[H_{\rm{int}}^I(u),[H_{\rm{int}}^I(s),\rho^{I,\chi}(t) \rho_B (\al_B(t))]_\chi]_\chi \Big\} \no\\
\aeq \sum_b \sum_{\mu,\nu}\Big( s_{b\nu}^{I\dagger}(u)s_{b\mu}^I(s)\rho^{I,\chi}(t)\tr_b[ R_{b,\nu,\chi}^{I\dagger}(u) R_{b,\mu,\chi}^I(s)\rho_b(\al_b(t))] \no\\
&&-s_{b\mu}^I(s)\rho^{I,\chi}(t)s_{b\nu}^{I\dagger}(u)\tr_b[  R_{b,\mu,\chi}^I(s)\rho_b(\al_b(t))R_{b,\nu,-\chi}^{I\dagger}(u)] \no\\
&&- s_{b\nu}^{I}(u)\rho^{I,\chi}(t)s_{b\mu}^{I\dagger}(s) \tr_b[ R_{b,\nu,\chi}^{I}(u)\rho_b(\al_b(t)) R_{b,\mu,-\chi}^{I\dagger}(s)] \no\\
&&+ \rho^{I,\chi}(t)s_{b\mu}^{I\dagger}(s)s_{b\nu}^{I}(u) \tr_b[ \rho_b(\al_b(t)) R_{b,\mu,-\chi}^{I\dagger}(s)R_{b,\nu,-\chi}^{I}(u) ] \Big), \la{H_H}
\eea
holds. 
In the calculation of $\tr_b[ R_{b,\nu,\chi}^{I\dagger}(u) R_{b,\mu,\chi}^I(s)\rho_b(\al_b(t))]$, 
the values of the control parameters can be approximated by $\al_t$.
Then, we obtain
\bea
\tr_b[ R_{b,\nu,\chi}^{I\dagger}(u) R_{b,\mu,\chi}^I(s)\rho_b] \aeqap \tr_b[\rho_b R_{b,\nu}^{\dagger}(u-s)R_{b,\mu} ]
\e C_{b,\nu\mu}(u-s) ,\\
\tr_b[  R_{b,\mu,\chi}^I(s)\rho_bR_{b,\nu,-\chi}^{I\dagger}(u)] \aeqap \tr_b[\rho_b R_{b,\nu,-2\chi}^{\dagger}(u-s)R_{b,\mu}  ] 
\e C_{b,\nu\mu}^\chi(u-s),\\
\tr_b[ R_{b,\nu,\chi}^{I}(u)\rho_b R_{b,\mu,-\chi}^{I\dagger}(s)] \aeqap \tr_b[\rho_b R_{b,\mu,-2\chi}^{\dagger}(s-u)R_{b,\nu}  ] 
= C_{b,\mu\nu}^\chi(s-u) ,\\
\tr_b[ \rho_b R_{b,\mu,-\chi}^{I\dagger}(s)R_{b,\nu,-\chi}^{I}(u) ] \aeqap \tr_b[\rho_b R_{b,\mu}^{\dagger}(s-u)R_{b,\nu}  ]
=C_{b,\mu\nu}(s-u) ,
\eea
with 
\bea
R_{b,\nu}^{\dagger}(v)\aeq e^{iH_b(\al_b(t))v}R_{b,\nu}^{\dagger}(\al_{Sb}(t))e^{-iH_b(\al_b(t))v}.
\eea
Here, $\rho_b=\rho_b(\al_b(t))$ and $R_{b,\mu}=R_{b,\mu}(\al_b(t))$. 
Then, \re{H_H} becomes 
\bea
&&\hs{-10mm}\tr_B\Big\{[H_{\rm{int}}^I(u),[H_{\rm{int}}^I(s),\rho^{I,\chi}(t) \rho_B (\al_B(t))]_\chi]_\chi \Big\} \no\\
\aeq \sum_b \sum_{\mu,\nu}\Big( s_{b\nu}^{I\dagger}(u)s_{b\mu}^I(s)\rho^{I,\chi}(t)C_{b,\nu\mu}(u-s) 
-s_{b\mu}^I(s)\rho^{I,\chi}(t)s_{b\nu}^{I\dagger}(u)C_{b,\nu\mu}^\chi(u-s) \no\\
&&- s_{b\nu}^{I}(u)\rho^{I,\chi}(t)s_{b\mu}^{I\dagger}(s) C_{b,\mu\nu}^\chi(s-u) 
+ \rho^{I,\chi}(t)s_{b\mu}^{I\dagger}(s)s_{b\nu}^{I}(u) C_{b,\mu\nu}(s-u) \Big), \la{tr_u_s}
\eea
and 
\bea
&&\hs{-5mm}\mL_{b,\tau_{\rm{CG}}}^\chi(\al_t)\bu \no\\
\aeq -\f{1}{\tau_\RM{CG}}\int_t^{t+\tau_{\rm{CG}}} du \int_t^{u} ds \ \sum_{\mu,\nu}\Big( s_{b\nu}^{I\dagger}(u,t)s_{b\mu}^I(s,t)\bu C_{b,\nu\mu}(u-s) \no\\
&&-s_{b\mu}^I(s,t)\bu s_{b\nu}^{I\dagger}(u,t)C_{b,\nu\mu}^\chi(u-s) \no\\
&&- s_{b\nu}^{I}(u,t)\bu s_{b\mu}^{I\dagger}(s,t) C_{b,\mu\nu}^\chi(s-u) 
+ \bu s_{b\mu}^{I\dagger}(s,t)s_{b\nu}^{I}(u,t) C_{b,\mu\nu}(s-u) \Big),
\eea
holds. 
Here, 
\bea
s_{b\mu}^I(s,t) \aeq U_S(t)U_S\dg(s) s_{b\mu} U_S(s)U_S\dg(t) .
\eea
and $U_S(t)$ is the solution of $\f{dU_S(t)}{dt}=-iH_S(\al_S(t))U_S(t)$ for $U_S(0)=1$.
In the calculation of $s_{b\mu}^I(s,t)$, the values of the control parameters can be approximated by $\al_t$.
Then, we obtain
\bea
s_{b\mu}^I(s,t) \aeq \sum_\om  e^{-i\om(s-t)}s_{b\mu}(\om), \\
s_{b\nu}^{I\dagger}(u,t) \aeq \sum_\om  e^{i\om(u-t)}[s_{b\nu}(\om)]\dg.
\eea
Here, the eigenoperator $s_{b\mu}(\om)$ is defined by
\bea
s_{b\mu}(\om) \aeq \sum_{n,m,r,s} \dl_{\om_{m n},\om} \ke{E_n,r}\br{E_n,r} s_{b\mu} \ke{E_m,s} \br{E_m,s}, 
\eea
with $ \om_{mn}= E_m-E_n$ and 
\bea
H_S\ke{E_n,r}=E_n\ke{E_n,r}.
\eea
 $r$ denotes the label of the degeneracy. 
$\om$ is one of the elements of \\
$\{ \om_{mn} \vert \ {  \br{E_n,r} s_{b\mu} \ke{E_m,s} \ne 0 \hs{2mm} }^\exists \mu \} $.
$s_{b\mu}(\om)$ and $\om$ depend on $\al_S$. 
The eigenoperators satisfy 
\bea
\sum_\om s_{b\mu}(\om)=s_{b\mu},
\eea
and 
\bea
[H_S,s_{b\mu}(\om)]=-\om s_{b\mu}(\om) \la{H_S_a(om)} .
\eea
Then, we obtain
\bea
\mL_{b,\tau_{\rm{CG}}}^\chi(\al)\bu \aeq -\f{1}{\tau_\RM{CG}}\int_t^{t+\tau_{\rm{CG}}} du \int_t^{t+\tau_{\rm{CG}}} ds \ \sum_{\mu,\nu}\sum_{\om,\om^\pr}\theta(u-s) \no\\
&&\times\Big( \Big\{ [s_{b\nu}(\om^\pr)]\dg s_{b\mu}(\om)\bu C_{b,\nu\mu}(u-s) \no\\
&&-s_{b\mu}(\om)\bu [s_{b\nu}(\om^\pr)]\dg C_{b,\nu\mu}^\chi(u-s) \Big\} e^{-i\om(s-t)} e^{i\om^\pr(u-t)}\no\\
&&+\Big\{- s_{b\mu}(\om) \bu [s_{b\nu}(\om^\pr)]\dg C_{b,\nu\mu}^\chi(s-u) \no\\
&&+ \bu [s_{b\nu}(\om^\pr)]\dg s_{b\mu}(\om) C_{b,\nu\mu}(s-u)\Big\}e^{i\om^\pr(s-t)}e^{-i\om(u-t)} \Big).
\eea
In last two terms, we swapped $\mu$ and $\nu$. 
$\theta(u-s)$ is the step function. 

Now, we introduce
\bea
\Phi_{b,\nu\mu}^\chi(\Om) \aeqd \int_{-\infty}^\infty du \ C_{b,\nu\mu}^\chi(u)e^{i\Om u}.
\eea
Then, 
\bea
 \int_{-\infty}^\infty du \ C_{b,\nu\mu}^\chi(u)\theta(u) e^{i\om u} 
\aeq \f{1}{2\pi}\int_0^\infty du \int_{-\infty}^\infty d\Om \ \Phi_{b,\nu\mu}^\chi(\Om) e^{-i\Om u}e^{i\om u} \no\\
\aeq \f{1}{2\pi}\int_{-\infty}^\infty d\Om \ \Big[\pi \dl(\Om-\om)-i\f{\RM{P}}{\Om-\om} \Big]\Phi_{b,\nu\mu}^\chi(\Om) \no\\
\aeq \half \Phi_{b,\nu\mu}^\chi(\om)-\f{i}{2}\Psi_{b,\nu\mu}^\chi(\om)= \Phi_{b,\nu\mu}^{(+)\chi}(\om), \la{FT_C}
\eea
holds. 
Here, $\RM{P}$ denotes the Cauchy principal value and 
\bea
\Psi_{b,\nu\mu}^\chi(\om) \aeqd \f{\RM{P}}{\pi}\int_{-\infty}^\infty d\Om \ \f{\Phi_{b,\nu\mu}^\chi(\Om)}{\Om-\om} \la{def_psi} ,\\
 \Phi_{b,\nu\mu}^{(\pm)\chi}(\Om)\aeqd \half \Phi_{b,\nu\mu}^\chi(\om)\mp\f{i}{2}\Psi_{b,\nu\mu}^\chi(\om).
\eea
\re{FT_C} leads 
\bea
C_{b,\nu\mu}^\chi(u-s) \theta(u-s) 
\aeq \int_{-\infty}^\infty d\Om \ \f{\Phi_{b,\nu\mu}^{(+)\chi}(\Om)}{2\pi}e^{-i\Om (u-s)} .
\eea
Similarly,
\bea
C_{b,\nu\mu}^\chi(s-u) \theta(u-s) 
\aeq \int_{-\infty}^\infty d\Om \ \f{\Phi_{b,\nu\mu}^{(-)\chi}(\Om)}{2\pi} e^{i\Om (u-s)}, 
\eea
holds.
Then, we obtain
\bea
\mL_{b,\tau_{\rm{CG}}}^\chi(\al)\bu\aeq -\f{1}{\tau_\RM{CG}}\int_t^{t+\tau_{\rm{CG}}} du \int_t^{t+\tau_{\rm{CG}}} ds \int_{-\infty}^\infty \f{d\Om}{2\pi} \ 
 \sum_{\mu,\nu}\sum_{\om,\om^\pr} \no\\
&&\times\Big( \Big\{ [s_{b\nu}(\om^\pr)]\dg s_{b\mu}(\om)\bu \Phi_{b,\nu\mu}^{(+)}(\Om)\no\\
&&-s_{b\mu}(\om)\bu [s_{b\nu}(\om^\pr)]\dg \Phi_{b,\nu\mu}^{(+)\chi}(\Om) \Big\} e^{-i\Om (u-s)}e^{-i\om(s-t)} e^{i\om^\pr(u-t)}\no\\
&&+\Big\{- s_{b\mu}(\om) \bu [s_{b\nu}(\om^\pr)]\dg \Phi_{b,\nu\mu}^{(-)\chi}(\Om)\no\\ 
&&+ \bu [s_{b\nu}(\om^\pr)]\dg s_{b\mu}(\om) \Phi_{b,\nu\mu}^{(-)}(\Om)\Big\} e^{i\Om (u-s)}e^{i\om^\pr(s-t)}e^{-i\om(u-t)}\Big),
\eea
with $\Phi_{b,\nu\mu}^{(\pm)}=\Phi_{b,\nu\mu}^{(\pm)\chi}\bv{\chi=0}$.
The integrals for $u$ and $s$ are performed as
\bea
\int_t^{t+\tau_{\rm{CG}}} du \ e^{-i\Om u}e^{i\om^\pr(u-t)} 
\aeq \tau_{\rm{CG}}e^{-i\Om t-i[\Om-\om^\pr]\tau_{\rm{CG}}/2 } \sinc ([\Om-\om^\pr]\tau_{\rm{CG}}/2) ,\\
\int_t^{t+\tau_{\rm{CG}}} ds \ e^{i\Om s}e^{-i\om(s-t)} \aeq \tau_{\rm{CG}}e^{i\Om t+i[\Om-\om]\tau_{\rm{CG}}/2 } \sinc ([\Om-\om]\tau_{\rm{CG}}/2),
\eea
then 
\bea
\mL_{b,\tau_{\rm{CG}}}^\chi(\al)\bu\aeq - \sum_{\mu,\nu}\sum_{\om,\om^\pr}\f{e^{-i(\om-\om^\pr)/\tau_{\rm{CG}}}}{2\pi}\int_{-\infty}^\infty d\Om \
\Big(  [s_{b\nu}(\om^\pr)]\dg s_{b\mu}(\om)\bu \Phi_{b,\nu\mu}^{(+)}(\Om) \no\\
&&-s_{b\mu}(\om)\bu [s_{b\nu}(\om^\pr)]\dg \Phi_{b,\nu\mu}^{(+)\chi}(\Om)  \no\\
&&\hs{-10mm}- s_{b\mu}(\om) \bu [s_{b\nu}(\om^\pr)]\dg \Phi_{b,\nu\mu}^{(-)\chi}(\Om) 
+ \bu [s_{b\nu}(\om^\pr)]\dg s_{b\mu}(\om) \Phi_{b,\nu\mu}^{(-)}(\Om) \Big) \no\\
&&\times \tau_{\rm{CG}} \sinc \f{[\Om-\om^\pr]\tau_{\rm{CG}}}{2}\sinc \f{[\Om-\om]\tau_{\rm{CG}}}{2}, \no\\
\eea
holds. Here, $\sinc (x)=\sin x/x$. 
The above equation can be rewritten as  
\bea
\mL_{b,{\tau_{\rm{CG}}}}^\chi(\al) \bu 
\aeq    -i [ h_{b,\tau_{\rm{CG}}}(\al) ,\bu ]  +\Pi_{b,{\tau_{\rm{CG}}}}^\chi(\al) \bu ,\\
\Pi_{b,{\tau_{\rm{CG}}}}^\chi(\al) \bu \aeq  \sum_{\om,\om^\pr }  \sum_{\mu,\nu} \Big[
 \Phi^{\chi}_{b,\mu \nu}(\tau_{\rm{CG}},\om,\om^\pr) s_{b\nu}(\om^\pr) \bu [s_{b\mu}  (\om)]\dg \no\\
&&-\half \Phi_{b,\mu \nu}(\tau_{\rm{CG}},\om,\om^\pr) \bu [s_{b\mu}(\om)] \dg s_{b\nu}(\om^\pr) \no\\
&&-\half \Phi_{b,\mu \nu}(\tau_{\rm{CG}},\om,\om^\pr) [s_{b\mu}(\om)] \dg s_{b\nu}(\om^\pr) \bu \Big],
 \eea
with 
 \bea
 h_{b,{\tau_{\rm{CG}}}}(\al) \aeq -\half \sum_{\om,\om^\pr } \sum_{\mu,\nu} \Psi_{b,\mu \nu}(\tau_{\rm{CG}},\om,\om^\pr)  [s_{b\mu}  (\om)]\dg s_{b\nu}(\om^\pr).  
\eea
Here, 
\bea
&&\hs{-10mm}X^{\chi}(\tau_{\rm{CG}},\om,\om^\pr) \no\\ 
\aeq \f{e^{i( \om - \om^\pr)\tau_{\rm{CG}}/2} }{2\pi} \int_{-\infty}^\infty d\Om \hs{0.7mm}  X^{\chi}(\Om) 
\tau_{\rm{CG}}\sinc \big(\f{\tau_{\rm{CG}}(\Om - \om)}{2}\big) \sinc \big(\f{\tau_{\rm{CG}}(\Om - \om^\pr)}{2}\big) ,
\eea
with $X=\Phi_{b,\mu\nu},\Psi_{b,\mu\nu}$. 
$\Pi_{b,{\tau_{\rm{CG}}}}=\Pi_{b,{\tau_{\rm{CG}}}}^\chi \bv{\chi=0}$ is the Lindblad type.
By the way, from
\bea
[C_{b,\mu\nu}(t)]^\ast \aeq C_{b,\nu\mu}(-t),
\eea
relations 
\bea
[\Phi_{b,\mu\nu}(\Om)]^\ast \aeq \Phi_{b,\nu\mu}(\Om) ,
\eea
and $[\Psi_{\mu\nu}(\Om)]^\ast = \Psi_{\nu\mu}(\Om)$ hold. 
Then, 
\bea
[\Phi_{b,\mu\nu}(\tau_{\rm{CG}},\om,\om^\pr)]^\ast \aeq \Phi_{b,\nu\mu}(\tau_{\rm{CG}},\om^\pr,\om), \la{Phi^ast}
\eea
holds. 

For super-operator $\mJ$, $\mJ \dg$ is defined by
\bea
\tr_S(Y\dg \mJ X) =\tr_S([\mJ \dg Y]\dg X),
\eea
where $X, Y \in \RM{\bm{B}}$. If 
$\mJ \bu = \sum_a A_a \bu B_a$ holds, 
\bea
\mJ \dg \bu = \sum_a A_a \dg \bu B_a \dg,
\eea
is obtained. Here, $A_a,B_a \in \RM{\bm{B}}$.
\re{Phi^ast} leads 
\bea
\Pi_{b,{\tau_{\rm{CG}}}}\dg (\al)\bu \aeq \sum_{\om,\om^\pr }  \sum_{\mu,\nu} \Big[
 \Phi_{b,\mu \nu}(\tau_{\rm{CG}},\om,\om^\pr) [s_{b\mu}(\om)]\dg \bu s_{b\nu}  (\om^\pr) \no\\
&&-\half \Phi_{b,\mu \nu}(\tau_{\rm{CG}},\om,\om^\pr) \bu [s_{b\mu}(\om)] \dg s_{b\nu}(\om^\pr)\no\\
&&-\half \Phi_{b,\mu \nu}(\tau_{\rm{CG}},\om,\om^\pr) [s_{b\mu}(\om)] \dg s_{b\nu}(\om^\pr) \bu \Big]. \la{Pidg}
\eea
This leads 
\bea
\Pi_{b,{\tau_{\rm{CG}}}}\dg(\al)1=0 , \la{Pidg1=0}
\eea
which means the conservation of the probability.

\subsubsection{Concrete model}

In this subsection, we consider $b=n_\RM{C}+1,\cdots,n_\RM{C}+n_\RM{GC}$. 
Now we suppose
\bea
H_{Sb}(\al_{Sb})= \sum_{\al}a_{\al}\dg B_{b\al}+\hc ,\ B_{b\al}=\sum_{k,\sig}V_{bk\sig,\al }(\al_{Sb})c_{bk\sig} \ (b \in \RM{\bm{\mG}}), \la{H_Sb}
\eea
where $a_{\al}$ and $c_{bk\sig}$ are single-particle annihilation operators of the system and of the bath $b$. 
Using 
\bea
\tr_b [\rho_b B_{b\al}^I(t^\pr)B_{b\be}^I(t^{\pr\pr})]=0=\tr_b [\rho_b B_{b\al}^{I\dagger}(t^\pr)B_{b\be}^{I\dagger}(t^{\pr\pr})], \la{katei}
\eea 
we obtain 
\bea
\mL_{b,{\tau_{\rm{CG}}}}^\chi(\al) \bu  \aeq  -i [ h_{b,\tau_{\rm{CG}}}(\al) ,\bu ]+\Pi_{b,{\tau_{\rm{CG}}}}^\chi(\al)\bu,  \no\\
\Pi_{b,{\tau_{\rm{CG}}}}^\chi(\al)\bu \aeq \sum_{\om,\om^\pr }  \sum_{\al,\be} \Big[
 \Phi^{-,\chi}_{b,\al \be}(\tau_{\rm{CG}},\om,\om^\pr) a_{\be}(\om^\pr) \bu [a_{\al}  (\om)]\dg \no\\
&&-\half \Phi^-_{b,\al \be}(\tau_{\rm{CG}},\om,\om^\pr) \bu [a_{\al}(\om)] \dg a_{\be}(\om^\pr)\no\\
&&-\half \Phi^-_{b,\al \be}(\tau_{\rm{CG}},\om,\om^\pr) 
[ a_{\al} (\om)]\dg a_{\be}(\om^\pr) \bu \no\\
&&+\Phi^{+,\chi}_{b,\al \be}(\tau_{\rm{CG}},\om,\om^\pr) [a_{\be} (\om^\pr)]\dg \bu a_{\al} (\om) \no\\
&&-\half \Phi^+_{b,\al \be}(\tau_{\rm{CG}},\om,\om^\pr) \bu a_{\al}(\om) [a_{\be}  (\om^\pr)]\dg \no\\
&&-\half \Phi^+_{b,\al \be}(\tau_{\rm{CG}},\om,\om^\pr)  a_{\al}  (\om)[a_{\be}  (\om^\pr)]\dg \bu  \Big] , \la{CG_go}
 \eea
 and 
 \bea
 h_{b,{\tau_{\rm{CG}}}}(\al) \aeq \sum_{\om,\om^\pr }  \sum_{\al,\be} \Big[-\half \Psi^-_{b,\al \be}(\tau_{\rm{CG}},\om,\om^\pr) 
[ a_{\al}  (\om)]\dg a_{\be}(\om^\pr) \no\\
&&+\half \Psi^+_{b,\al \be}(\tau_{\rm{CG}},\om,\om^\pr)  a_{\al}  (\om)[a_{\be}  (\om^\pr)]\dg \Big]. \la{CG_go_L}
\eea
The eigenoperators $a_{\al}(\om)$ are given by
\bea
a_{\al}(\om) \aeq \sum_{n,m,r,s} \dl_{\om_{m n},\om} \ke{E_n,r}\br{E_n,r} a_{\al} \ke{E_m,s} \br{E_m,s}. \la{Def_a_om}
\eea
$\om$ is one of the elements of $\{ \om_{mn} \vert \ {  \br{E_n,r} a_{\al} \ke{E_m,s} \ne 0 \hs{2mm} }^\exists \al \} $.
$a_{\al}(\om)$ satisfy 
\bea
\sum_\om a_{\al}(\om)=a_{\al},
\eea
and 
\bea
[H_S,a_{\al}(\om)]=-\om a_{\al}(\om),\ [N_S,a_{\al}(\om)]=-a_{\al}(\om) .\la{N_S_a(om)}
\eea 
$N_S$ is total number operator of the system. 
Here and in the following, we suppose 
\bea
[N_S,H_S]=0.
\eea
If $n_\RM{GC}=0$, existence of $N_S$ and the above equation are not required. 
In \re{CG_go} and \re{CG_go_L}, 
\bea
&&\hs{-10mm}X^{\pm,\chi}(\tau_{\rm{CG}},\om,\om^\pr) \no\\
\aeq \f{e^{\pm i( \om - \om^\pr)\tau_{\rm{CG}}/2} }{2\pi} \int_{-\infty}^\infty d\Om \hs{0.7mm}  X^{\pm,\chi}(\Om) 
\tau_{\rm{CG}}\sinc \big(\f{\tau_{\rm{CG}}(\Om - \om)}{2}\big) \sinc \big(\f{\tau_{\rm{CG}}(\Om - \om^\pr)}{2}\big) ,\no\\
\eea
and $X^\pm(\tau_{\rm{CG}},\om,\om^\pr)=X^{\pm,\chi}(\tau_{\rm{CG}},\om,\om^\pr)\bv{\chi=0}$. 
Here, $X^{\pm,\chi}(\Om)$ denotes one of $\Phi_{b,\al,\be}^{\pm,\chi}(\Om)$, $\Psi_{b,\al\be}^{\pm,\chi}(\Om) $, where 
\bea
\Phi_{b,\al \be}^{-,\chi}(\Om)\aeq \int_{-\infty}^\infty du \ \tr_b[\rho_b B_{b\al,-2\chi}^I(u)B_{b\be} \dg ]e^{i\Om u} ,\\
\Phi_{b,\al \be}^{+,\chi}(\Om)\aeq \int_{-\infty}^\infty du \ \tr_b[\rho_b B_{b\al,-2\chi}^{\dagger I}(u)B_{b\be} ] e^{-i\Om u},\\
\Psi_{b,\al\be}^{\pm,\chi}(\Om) \aeqd \f{\RM{P}}{\pi}\int_{-\infty}^\infty d\Om^\pr \ \f{\Phi_{b,\al\be}^{\pm,\chi}(\Om^\pr)}{\Om^\pr-\Om}.
\eea
We set $\{O_\mu\}=\{N_b\}_{b \in \RM{\bm{\mG}}}+\{H_b\}_b$, where
\bea
N_{b} \aeq \sum_{k,\sig}c_{bk\sig} \dg c_{bk\sig} .
\eea
Whenever $H_b$ is an element of $\{O_\mu\}$, we suppose $\al_b^\pr$ are fixed.  
We introduce the eigenoperator
\bea
B_{b\al}(\Om_b) \aeq \sum_{n,m,r,s} \dl_{\Om_{b,m n},\Om_b} \ke{E_{b,n},r}\br{E_{b,n},r} B_{b\al} \ke{E_{b,m},s} \br{E_{b,m},s}, \la{def_B(om)}
\eea
with $ \Om_{b,mn}= E_{b,m}-E_{b,n}$ and $H_b\ke{E_{b,n},r}=E_{b,n}\ke{E_{b,n},r}$. $r$ denotes the label of the degeneracy. 
$\Om_b$ is one of the elements of $\{ \Om_{b,mn} \vert \ {  \br{E_{b,n},r} B_{b\al} \ke{E_{b,m},s} \ne 0 \hs{2mm} }^\exists \al \} $. 
The relations 
\bea
\sum_{\Om_b} B_{b\al}(\Om_b)=B_{b\al}, \la{B_sum}
\eea
and 
\bea
[H_b,B_{b\al}(\Om_b)]=-\Om_b B_{b\al}(\Om_b),\ [N_b,B_{b\al}(\Om_b)]=-B_{b\al}(\Om_b) \la{B_com}
\eea
hold. 
Then, we obtain
\bea
B_{b\al,-2\chi}^I(u) \aeq \sum_{\Om_b} B_{b\al}(\Om_b)e^{-i\Om_bu+i\chi_{H_b}\Om_b+i\chi_{N_b}} ,\\
B_{b\al,-2\chi}^{\dagger I}(u) \aeq \sum_{\Om_b} [B_{b\al}(\Om_b)]\dg e^{i\Om_bu-i\chi_{H_b}\Om_b-i\chi_{N_b}},
\eea
and 
\bea
\Phi_{b,\al \be}^{-,\chi}(\Om)\aeq 2\pi \sum_{\Om_b} \dl(\Om-\Om_b)e^{i\chi_{H_b}\Om_b+i\chi_{N_b}}\tr_b(\rho_b B_{b\al}(\Om_b)B_{b\be} \dg ) \no\\
\aeq e^{i\chi_{H_b}\Om+i\chi_{N_b}} 2\pi \sum_{\Om_b} \dl(\Om-\Om_b)\tr_b(\rho_b B_{b\al}(\Om_b)[B_{b\be}(\Om_b)] \dg ), \la{Phi^+}\\
\Phi_{b,\al \be}^{+,\chi}(\Om)\aeq  2\pi\sum_{\Om_b} \dl(\Om-\Om_b)e^{-i\chi_{H_b}\Om_b-i\chi_{N_b}} \tr_b(\rho_b [B_{b\al}(\Om_b)]\dg B_{b\be} )  \no\\
\aeq  e^{-i\chi_{H_b}\Om-i\chi_{N_b}}2\pi\sum_{\Om_b} \dl(\Om-\Om_b) \tr_b(\rho_b [B_{b\al}(\Om_b)]\dg B_{b\be}(\Om_b) ) \la{Phi^-}.
\eea
Here, we used \re{B_sum} and $\tr_b(\rho_b B_{b\al}(\Om_b)[B_{b\be}(\Om_b^\pr)] \dg )=0$ and \\
$\tr_b(\rho_b [B_{b\al}(\Om_b)]\dg B_{b\be}(\Om_b^\pr)  )=0$ for $\Om_b \ne \Om_b^\pr$. 
Then, we obtain 
\bea
\Phi_{b,\al \be}^{\pm,\chi}(\Om) =e^{\mp(i\chi_{H_b}\Om+i\chi_{N_b})}\Phi_{b,\al \be}^{\pm}(\Om), \la{Phi_chi}
\eea
with $\Phi_{b,\al \be}^\pm(\Om)=\Phi_{b,\al \be}^{\pm,\chi}(\Om)\bv{\chi=0}$ and 
\bea
\Psi_{b,\al \be}^{-}(\Om)\aeq 2 \sum_{\Om_b} \RM{P}\f{1}{\Om_b-\Om}\tr_b(\rho_b B_{b\al}(\Om_b)[B_{b\be}(\Om_b)] \dg ),\\
\Psi_{b,\al \be}^{+}(\Om)\aeq  2\sum_{\Om_b} \RM{P}\f{1}{\Om_b-\Om}\tr_b(\rho_b [B_{b\al}(\Om_b)]\dg B_{b\be}(\Om_b) ) .
\eea
$\Phi_{b,\al \be }^{\pm}(\Om)$ satisfy 
\bea
[\Phi_{b,\al \be }^{\pm}(\Om)]^\ast \aeq \Phi_{b, \be \al}^{\pm}(\Om) ,\\
\Phi_{b,\al \be }^+(\Om) \aeq e^{-\be_b(\Om-\mu_b)}\Phi_{b, \be \al}^-(\Om) .\la{KMS}
\eea
The latter is the Kubo-Martin-Schwinger (KMS) condition. 
\re{KMS} is derived from \\ 
$\rho_b B_{b\al}(\Om_b)=e^{\be_b(\Om_b-\mu)}B_{b\al}(\Om_b)\rho_b$ (derived from \re{B_com}) and 
\re{Phi^+} and \re{Phi^-}.

Here, we suppose the free Hamiltonian of the bath $b$:
\bea
H_b(\al_b^\pr) = \sum_{k,\sig} \ep_{bk\sig}(\al_b^\pr)c_{bk\sig} \dg c_{bk\sig},
\eea
and $\{O_\mu\}=\{N_{b\sig} \}_{b\sig}$ with 
\bea
N_{b\sig} \aeq \sum_{k}c_{bk\sig} \dg c_{bk\sig}.
\eea 
In this case, $\al_b^\pr$ can depend on time and
\bea
\Phi_{b,\al \be}^{-,\chi}(\Om) \aeq 2\pi\sum_{k,\sig}  V_{bk\sig,\al} V_{bk\sig,\be}^\ast
F_b^-(\ep_{bk\sig}) e^{i\chi_{b\sig} } \dl(\ep_{bk\sig}-\Om) , \la{phi-}\\
\Phi_{b,\al \be}^{+,\chi}(\Om) \aeq 2\pi\sum_{k,\sig} V_{bk\sig,\al}^\ast V_{bk\sig, \be}
F_b^+(\ep_{bk\sig}) e^{-i\chi_{b\sig} }  \dl(\ep_{bk\sig}-\Om), \la{phi+} \\
\Psi_{b,\al \be}^{-,\chi}(\Om) \aeq 2\sum_{k,\sig}  V_{bk\sig,\al} V_{bk\sig,\be}^\ast  
F_b^-(\ep_{bk\sig}) e^{i\chi_{b\sig} }  \RM{P}\f{1 }{\ep_{bk\sig}-\Om} , \\
\Psi_{b,\al \be}^{+,\chi}(\Om) \aeq 2\sum_{k,\sig} V_{bk\sig,\al}^\ast V_{bk\sig, \be}
F_b^+(\ep_{bk\sig}) e^{-i\chi_{b\sig} }  \RM{P}\f{1 }{\ep_{bk\sig}-\Om} ,
\eea
hold. 
$\chi_{b\sig}$  denotes the counting fields for $N_{b\sig}$. 
If the baths are fermions, 
$F_b^+(\ep)=f_b(\ep)\defe[\exp(\be_b(\ep-\mu_b))+1]^{-1}$ and $F_b^-(\ep)=1-f_b(\ep)$. 
If  the baths are bosons,
$F_b^+(\ep)=n_b(\ep)\defe[\exp(\be_b(\ep-\mu_b))-1]^{-1}$ and $F_b^-(\ep)=1+n_b(\ep)$. 

\re{H_Sb} can be generalized as
\bea
H_{Sb}(\al_{Sb})= \sum_{n,\xi}s_{(n)\xi}\dg B_{b,(n)\xi}+\hc \ (b \in \RM{\bm{\mG}}), \la{H_SbG}
\eea
with
\bea
[s_{(n)\xi}(\om),N_S]=-ns_{(n)\xi}(\om),\ [B_{b,(n)\xi}(\Om_b),N_b]=-nB_{b,(n)\xi}(\Om_b).
\eea
Here, $n=1,2,\cdots$, and $s_{(n)\xi}(\om)$ and $B_{b,(n)\xi}(\Om_b)$ are the eigenoperators.

\subsection{Rotating wave approximation} \la{RWA}

In the CGA or Born-Markov approximation, the FCS-QME is described by $a_{\al}(\om)$ and $[a_{\al}(\om^\pr)]\dg$ $(\om,\om^\pr \in \mathcal{W})$. 
If $H_S$ is time dependent, the generalization of usual RWA \cite{open} with static $H_S$ is unclear. 
In this thesis, the RWA is defined as the limit $\tau_\RM{CG} \to \infty$ ($\tau_\RM{CG} \cdot \min_{\om \ne \om^\pr}\abs{\om-\om^\pr} \gg 1$) of the CGA. 
In this limit, 
\bea
\Phi_{b,\mu\nu}^{\chi}(\tau_{\rm{CG}},\om,\om^\pr) \approx \Phi_{b,\mu\nu}^{\chi}(\om)\dl_{\om,\om^\pr},\
\Psi_{b,\mu\nu}^{\chi}(\tau_{\rm{CG}},\om,\om^\pr) \approx \Psi_{b,\mu\nu}^{\chi}(\om)\dl_{\om,\om^\pr},
\eea
 hold  because of the fact that 
\bea
\lim_{\tau_{\rm{CG}} \to \infty}\tau_{\rm{CG}}\sinc \f{\tau_{\rm{CG}}(\Om-\om)}{2} \sinc \f{\tau_{\rm{CG}}(\Om-\om^\pr)}{2}=2\pi \dl_{\om,\om^\pr}\dl(\Om-\om).
\eea
If $H_S$ is time independent, this RWA is equivalent to usual RWA. 
$\mL_b^\chi(\al)$ is given by 
\bea
\mL_b^\chi(\al) \bu =\Pi_b^\chi(\al) \bu -i[h_b(\al),\bu], \la{def_Pi_b}
\eea
where $h_b(\al)$ is a Hermitian operator describing the Lamb shift.
$H_\RM{L}(\al)\defe \sum_b h_b(\al)$ is called the Lamb shift Hamiltonian. 
$\Pi_{b}^\chi(\al)$ and $ h_b(\al)$ are given by
\bea
\Pi_{b}^\chi(\al) \bu \aeq  \sum_{\om }  \sum_{\mu,\nu} \Big[
 \Phi^{\chi}_{b,\mu \nu}(\om) s_{b\nu}(\om) \bu [s_{b\mu}  (\om)]\dg \no\\
&&-\half \Phi_{b,\mu \nu}(\om) \bu [s_{b\mu}(\om)] \dg s_{b\nu}(\om)
-\half \Phi_{b,\mu \nu}(\om) [s_{b\mu}(\om)] \dg s_{b\nu}(\om) \bu \Big] ,\\
 h_b(\al) \aeq -\half \sum_{\om } \sum_{\mu,\nu} \Psi_{b,\mu \nu}(\om)  [s_{b\mu}  (\om)]\dg s_{b\nu}(\om) .\la{RWA_G}
\eea
Because of \re{H_S_a(om)}, $h_b(\al)$ commutes with $H_S(\al_S)$:
\bea
[ h_b(\al),H_S(\al_S)]=0. \la{H_L_H_S=0}
\eea

We introduce projection super-operators $\mP(\al_S)$ and $\mQ(\al_S)$ by 
\bea
\mP(\al_S) \ke{E_n,r}\br{E_m,s}=\dl_{E_n,E_m}\ke{E_n,r}\br{E_m,s},
\eea
and $\mQ(\al_S)=1-\mP(\al_S)$. We define ${\rm\bm{B}_P}\defe\{X\in \RM{\bm{B}} \vert \mP X=X  \}$
 and ${\rm\bm{B}_Q}\defe\{X\in \RM{\bm{B}} \vert \mQ X=X  \}$.
$\hat{K}^\chi\mP \bu \in {\rm\bm{B}_P}$ holds. 
Then, $\hat{K}^\chi\mQ \bu \in {\rm\bm{B}_Q}$ and
\bea
\mQ\hat{K}^\chi \mP=0=\mP\hat{K}^\chi \mQ ,\la{bunri}
\eea
hold. 
This implies that the right eigenvalue equations \re{rig} are decomposed into two closed systems of equations 
for $\mP\rho_n^\chi$ and for $\mQ\rho_n^\chi$.
Thus, $\rho_n^\chi$ is an element of ${\rm\bm{B}_P}$ or ${\rm\bm{B}_Q}$.
In particular, $\rho_0^\chi \in {\rm\bm{B}_P}$.
Then, the matrix representation of $\rho_0(\al)$ by $\ke{E_n,r}$ is block diagonalized.
This implies
\bea
[H_S(\al_S),\rho_0(\al)]= 0 . \la{H_S,rho_0}
\eea

For \re{H_Sb}, $\Pi_b^\chi(\al)$ in \re{def_Pi_b} is given by
\bea
\Pi_b^\chi(\al) \bu 
\aeq    \sum_{\om }  \sum_{\al,\be} \Big[
 \Phi^{-,\chi}_{b,\al  \be }(\om) a_{\be }(\om) \bu [a_{\al} (\om)]\dg 
-\half \Phi^-_{b,\al \be}(\om) \bu [a_{\al}(\om)] \dg a_{\be}(\om)\no\\
 \no\\
&&-\half \Phi^-_{b,\al  \be }(\om) [ a_{\al }  (\om)]\dg a_{\be }(\om) \bu+\Phi^{+,\chi}_{b,\al  \be }(\om) [a_{\be }  (\om)]\dg \bu a_{\al } (\om) \no\\
&&-\half \Phi^+_{b,\al  \be }(\om) \bu a_{\al }(\om) [a_{\be }  (\om)]\dg 
-\half \Phi^+_{b,\al  \be }(\om)  a_{\al }  (\om)[a_{\be }  (\om)]\dg \bu  \Big] . \la{Pi_RWA}
\eea
The Lamb shift is given by
\bea
h_b(\al) \aeq \sum_\om \sum_{\al,\be} \Big( -\half \Psi^-_{b,\al  \be }(\om) [ a_{\al }  (\om)]\dg a_{\be }(\om) 
+\half \Psi^+_{b,\al \be}(\om)  a_{\al }  (\om)[a_{\be }  (\om)]\dg \Big).
\eea
The second equation of \re{N_S_a(om)} leads 
\bea
[h_b(\al),N_S]=0.\la{H_L_N_S=0-}
\eea

\subsection{Detailed balance condition} \la{sKMS}

In this section, we consider the RWA. 
If we suppose \re{H_Sb}, 
\bea
\Pi_b(\bu e^{-\be_b(H_S-\mu_bN_S)})=(\Pi_b\dg \bu)e^{-\be_b(H_S-\mu_bN_S)} \ (b \in \RM{\bm{\mG}}) \la{KMS2-},
\eea
holds using \re{KMS}. 
This is the detailed balance condition. 
If we suppose \re{H_SbG}, the above relation also holds. 
From $\mL_b \dg 1=\Pi_b \dg 1=0$ (see \re{Pidg1=0}) and \re{KMS2-} for $\bu=1$ lead 
\bea
\Pi_b e^{-\be_b(H_S-\mu_bN_S)}=\mL_b e^{-\be_b(H_S-\mu_bN_S)}=0, \la{KMS3}
\eea
using \re{H_L_H_S=0} and \re{H_L_N_S=0-}. 
If the bath $b$ is fermion, \re{H_Sb} or \re{H_SbG} are general.

In the following of this section, we consider canonical baths ($b \in \RM{\bm{\mC}}$). 
\re{RWA_G} leads
\bea
\Pi_{b}\dg(\al) \bu \aeq  \sum_{\om }  \sum_{\mu,\nu} \Big[
 \Phi_{b,\mu \nu}(\om) [s_{b\mu}(\om)]\dg \bu s_{b\nu} (\om) \no\\
&&-\half \Phi_{b,\mu \nu}(\om) \bu [s_{b\mu}(\om)] \dg s_{b\nu}(\om)
-\half \Phi_{b,\mu \nu}(\om) [s_{b\mu}(\om)] \dg s_{b\nu}(\om) \bu \Big] .\la{Pi_dg}
\eea
Then, we obtain
\bea
(\Pi_{b}\dg(\al) \bu)e^{-\be_b H_S} \aeq  \sum_{\om }  \sum_{\mu,\nu} \Big[
 \Phi_{b,\mu \nu}(\om)e^{-\be_b \om} [s_{b\mu}(\om)]\dg \bu e^{-\be_b H_S} s_{b\nu} (\om) \no\\
&&-\half \Phi_{b,\mu \nu}(\om) \bu e^{-\be_b H_S} [s_{b\mu}(\om)] \dg s_{b\nu}(\om) \no\\
&&-\half \Phi_{b,\mu \nu}(\om) [s_{b\mu}(\om)] \dg s_{b\nu}(\om) \bu e^{-\be_b H_S} \Big] ,
\eea
using \re{H_S_a(om)}. Then,
\bea
&&\hs{-10mm}\Pi_b(\bu e^{-\be_bH_S})-(\Pi_b\dg \bu)e^{-\be_bH_S} \no\\
\aeq
\sum_{\om }  \sum_{\mu,\nu} \Big[\Phi_{b,\mu \nu}(\om) s_{b\nu}(\om) \bu e^{-\be_bH_S} [s_{b\mu}(\om)]\dg \no\\
&&- \Phi_{b,\mu \nu}(\om)e^{-\be_b \om} [s_{b\mu}(\om)]\dg \bu e^{-\be_b H_S} s_{b\nu} (\om) \Big] \no\\
 \aeq \sum_{\om } \sum_{\mu,\nu} \Big[\phi_{b,\mu \nu}(\om) [s_{b\nu}(\om)]\dg \bu e^{-\be_b H_S} s_{b\mu}(\om) \no\\
&&-\Phi_{b,\mu \nu}(\om)e^{-\be_b \om} [s_{b\mu}(\om)]\dg \bu e^{-\be_b H_S} s_{b\nu} (\om) \Big], \la{R-L}
\eea
holds. Here, we used 
\bea
\sum_{\om }  \sum_{\mu,\nu} \Phi_{b,\mu \nu}(\om) s_{b\nu}(\om) \bu [s_{b\mu}(\om)]\dg=
\sum_{\om}  \sum_{\mu,\nu} \phi_{b,\mu \nu}(\om) [s_{b\nu}(\om)]\dg \bu s_{b\mu}(\om),
\eea
with 
\bea
 \phi_{b,\mu \nu}(\om) \aeq \int_{-\infty}^\infty du \ D_{b,\mu\nu}(u)e^{-i\om u} ,\\
D_{b,\mu\nu}(u) \aeq \tr_b[\rho_b R_{b,\mu}^I(u)R_{b,\nu} \dg],\ \rho_b=e^{-\be_b H_b}/\tr_b(e^{-\be_b H_b}).
\eea
Using 
\bea
\tr_b[\rho_b R_\mu^I(u)R_\nu \dg]\aeq \tr_b[ R_{b,\mu}^I(u+i\be_b) \rho_b R_{b,\nu} \dg] \no\\
\aeq \tr_b[  \rho_b R_{b,\nu}^{I\dagger}(-u-i\be_b) R_{b,\mu}]=C_{b,\nu\mu}(-u-i\be_b),
\eea
$\phi_{b,\mu \nu}(\om)$ is given by
\bea
 \phi_{b,\mu \nu}(\om) \aeq \int_{-\infty}^\infty du \ C_{b,\nu\mu}(-u-i\be_b)e^{-i\om u} \no\\
 \aeq \int_{-\infty}^\infty du \int_{-\infty}^\infty d\Om \  \Phi_{b,\nu\mu}(\Om)e^{i\Om u-\be_b \Om}e^{-i\om u} \no\\
 \aeq \Phi_{b,\nu\mu}(\om)e^{-\be_b \om}.
\eea
Substituting this into \re{R-L}, we obtain 
\bea
\Pi_b(\bu e^{-\be_bH_S})=(\Pi_b\dg \bu)e^{-\be_bH_S} \ (b \in \RM{\bm{\mC}}).
\eea
Substituting $\bu=1$ to this equation, we get
\bea
\Pi_b e^{-\be_bH_S}=0.
\eea
If $n_\RM{GC}>0$, we suppose
\bea
[s_{b\mu}(\om),N_S]=0 \ (b \in \RM{\bm{\mC}}). \la{s_N_S}
\eea
Then, 
\bea
\Pi_b(\bu e^{-\be_b(H_S-\mu_b^\pr N_S)})=(\Pi_b\dg \bu)e^{-\be_b(H_S-\mu_b^\pr N_S)} \ (b \in \RM{\bm{\mC}}), \la{KMS4-}
\eea
and 
\bea
\Pi_b e^{-\be_b(H_S-\mu_b^\pr N_S)}=\mL_b e^{-\be_b(H_S-\mu_b^\pr N_S)}=0, \la{KMS4}
\eea
hold. \re{KMS4-} is the detailed balance condition. 
Here, $\mu_b^\pr$ is an arbitrary real number, and we used 
\bea
[h_b(\al),N_S]=0, \la{h_bN_S}
\eea
derived from \re{s_N_S}. 
\re{s_N_S} and \re{Pidg} lead
\bea
\Pi_b \dg N_S=0 \ (b \in \RM{\bm{\mC}}). \la{PidgN_S}
\eea

\newpage

\section{FCS-QME and quantum pump} \la{sFCS-QME}

\subsection{Currents} \la{Currents}

Generally, $\mL_b^\chi(\al) $ has the form: 
\bea
\mL_b^\chi(\al)  \bu=\sum_a c_{ba}^\chi(\al) A_a \bu B_a , \la{mL_b}
\eea
where $A_a$ and $B_a$ belong to $\RM{\bm{B}}$ and depend on $\al_S$, and $c_{ba}^\chi(\al)$ is a complex number which depends on $\al_S$, $\al_{Sb}$ and $\al_b$.  
If and only if $A_a, B_a \ne 1$, $c_{ba}^\chi(\al)$ depends on $\chi$. 
In this chapter, we assume only Markov property (i.e., $\hat{K}^\chi$ just depends on $\al_t$).
At $\chi=0$, the FCS-QME becomes the quantum master equation (QME)
\bea
 \f{d\rho(t)}{dt} = \hat{K}(\al_t)\rho(t). \la{QME}
\eea
$\hat{K}(\al_t)$ equals $\hat{K}^\chi(\al_t)$ at $\chi=0$. In the following, a symbol $X$ without $\chi$ denotes $X^\chi \vert_{\chi=0}$.

In the Liouville space \cite{Nakajima, FCS-QME}, the left and right eigenvalue equations of the Liouvillian are
\bea
\hat{K}^\chi(\al)\dke{\rho_n^\chi(\al)}\aeq \lm_n^\chi(\al)\dke{\rho_n^\chi(\al)} \la{rig} ,\\
\dbr{l_n^\chi(\al)}\hat{K}^\chi(\al)\aeq \lm_n^\chi(\al)\dbr{l_n^\chi(\al)} \la{left}.
\eea
In the Liouville space, $A \in \RM{\bm{B}}$ is described by $\dke{A}$. 
The inner produce is defined by $\dbr{A}B\dket= \tr_S(A\dg B)$ ($A,B \in \RM{\bm{B}}$). 
In particular, $\dbr{1}A\dket= \tr_S A$ holds. 
A super-operator which operates to a liner operator of the system becomes an operator of the Liouville space. 
The left eigenvectors $l_n^\chi(\al)$ and the right eigenvectors $\rho_m^\chi(\al)$ satisfy 
\bea
\dbra l_n^\chi(\al) \dke{\rho_m^\chi(\al)}=\dl_{nm}.
\eea 
The mode which has the eigenvalue with the maximum real part is assigned by the label $n=0$.  
Because the conservation of the probability $\f{d}{dt}\dbra 1 \dke{\rho(t)}=\dbr{1}\hat{K}(\al_t)\dke{\rho(t)}=0$ leads
\bea
\dbr{1}\hat{K}(\al)=0,
\eea
in the limit $\chi \to 0$, $\lm_0^\chi(\al)$ becomes $0$ and $\dbr{l_0^\chi(\al)}$ becomes $\dbr{1}$
(i.e., $l_0(\al)$ is identity operator). 
In addition, $\dke{\rho_0(\al)}$ determined by 
\bea
\hat{K}(\al)\dke{\rho_0(\al)}=0 ,
\eea
represents the {\it instantaneous steady state}.

The formal solution of the FCS-QME \re{FCS-QME} is
\bea
\dke{\rho^\chi(t)} = {\rm{T}} \exp \Big[\int_0^t ds \ \hat{K}^\chi(\al_s) \Big]\dke{\rho(0)} ,
\eea
where ${\rm{T}}$ denotes the time-ordering operation. Using this, we obtain the averages
\bea
\bra \Dl o_\mu \ket_t \aeq \f{\partial }{\partial(i\chi_{O_\mu}) } \dbra 1 \dke{\rho^\chi(t)}   \Big \vert _{\chi=0} \no\\
\aeq  \int_0^t du \ \dbr{1}\hat{K}^{O_\mu}(\al_u)\dke{\rho(u)} \e  \int_0^t du \  i_{O_\mu} (u), \la{wat}
\eea
where $X^{O_\mu}(\al)\defe \f{\partial X^\chi(\al) }{\partial (i\chi_{O_\mu})} \big \vert_{\chi=0}$ when $X$ is an (super)operator or c-number.
Here, we used $\dbr{1}\hat{K}(\al)=0$.
Moreover, using $\dbr{l_0(\al)}=\dbr{1}$, $\lm_0(\al)=0$ and \re{left}, we obtain
\bea
\dbr{1}\hat{K}^{O_\mu}(\al)
= \lm_0^{O_\mu} (\al) \dbr{1} - \dbr{l_0^{O_\mu} (\al)}\hat{K}(\al). \la{wata}
\eea
Here, $\dbr{l_0^{O_\mu} (\al)}$ is defined by $\f{\partial \dbr{l_0^\chi(\al)} }{\partial (i\chi_{O_\mu})} \big \vert_{\chi=0}$, 
then 
\bea
l_0^{O_\mu}=-\f{\partial l_0^\chi(\al) }{\partial (i\chi_{O_\mu})} \Big \vert_{\chi=0},
\eea 
holds.
The current $i_{O_\mu} (t)$ is given by \cite{Watanabe}
\bea
i_{O_\mu} (t) \aeq \dbr{1}\hat{K}^{O_\mu}(\al_t)\dke{\rho(t)} \no\\
\aeq \lm_0^{O_\mu} (\al_t) - \dbr{l_0^{O_\mu} (\al_t)}\hat{K}(\al_t) \dke{\rho(t)}  \no\\
\aeq \lm_0^{O_\mu} (\al_t)- \dbr{l_0^{O_\mu} (\al_t)}\f{d}{dt}\dke{\rho(t)}. \la{watan}
\eea
The current can also be written as
\bea
i_{O_\mu}(t) 
= \dbr{1} W^{O_\mu} (\al_t) \dke{\rho(t)} , \la{I_mu}
\eea
where $W^{O_\mu}(\al)$ is the current operator defined by 
\bea
\dbr{1}W^{O_\mu}(\al) = \dbr{1}\hat{K}^{O_\mu} (\al) \la{defW},
\eea
i.e., $\tr_S[W^{O_\mu}(\al) \bu] =\tr_S[\hat{K}^{O_\mu}(\al)\bu]$ for any $\bu \in \RM{\bm{B}}$. 
Therefore, using \re{mL_b}, the current operator is given by 
\bea
 W^{O_\mu}(\al)=\sum_{b,a} c_{ba}^{O_\mu}(\al) B_a A_a . \la{W^O}
\eea
Using \re{wata}, the {\it instantaneous steady current} is given by 
\bea
\dbr{1}W^{O_\mu} (\al) \dke{\rho_0(\al)} = \lm_0^{O_\mu}(\al)\e i_{O_\mu}^\st(\al) . \la{touka_sc}
\eea
In the following, we suppose $\rho(0)=\rho_0(\al_0)$. 
In this case, as we will show, $\rho(t)=\rho_0(\al_t)+\mO(\om/\Ga)$ holds  where 
$\om=2\pi/\tau$ and  
\bea
\Ga=\min_{n \ne 0, \al \in C}\{-\RM{Re}[\lm_n(\al)]\}. \la{Ga_G}
\eea 
In $\om \ll \Ga$ limit, we obtain  
\bea
i_{O_\mu}(t)  =  i_{O_\mu}^\st(\al_t)-\dbr{l_0^{O_\mu} (\al_t)}\f{d}{dt}\dke{\rho_0(\al_t)}+\mO\big(\f{\om^2}{\Ga}\big), \la{i^O}
\eea
which leads to
\bea
\bra \Dl o_\mu \ket_\tau = \int_0^\tau dt \ i_{O_\mu}^\st(\al_t)+\int_C d\al^n \ A_n^{O_\mu}(\al) +\mO\big(\f{\om}{\Ga}\big). \la{Dl o}
\eea
Here, $\al^n$ is the $n$-th component of the control parameters,  $C$ is the trajectory from $\al_0$ to $\al_\tau$, 
\bea
A_n^{O_\mu}(\al) \defe -\dbr{l_0^{O_\mu} (\al)}\f{\p }{\p \al^n}\dke{\rho_0(\al)}, \la{BSN}
\eea
is the BSN vector, and the summation symbol $\sum_{n}$ is omitted. 
As we will show, the BSN vector is also given by \cite{Nakajima}
\bea
A_n^{O_\mu}(\al) =\dbr{1}W^{O_\mu}(\al)\mR(\al)\f{\p }{\p \al^n}\dke{\rho_0(\al)}, \la{BSN_Nakajima}
\eea 
where  $\mR(\al)$ is the pseudo-inverse of the Liouvillian defined by
\bea
\mR(\al)\hat{K}(\al)=1-\dke{\rho_0(\al)}\dbr{1}. \la{defR}
\eea

In the research of adiabatic pumping, the expression of \re{Dl o} is essential. 
In Refs.\cite{Yuge12,Watanabe,Nakajima}, \re{Dl o} with \re{BSN} was used to study the quantum pump.
On the other hand, in Ref.\cite{RT12}, \re{Dl o} was derived using the generalized master equation \cite{RT09} and without using the FCS. 
In Ref.\cite{RT12}, $A_n^{O_\mu}(\al)$ was described by the quantity corresponding to the current operator and the pseudo-inverse of the Liouvillian, 
as shown in \re{BSN_Nakajima}.    
In this chapter, we show the equivalence between the FCS-QME approach and the generalized master equation approach (with the Born-approximation)
for all orders of the pumping frequency \cite{Nakajima} (see also Ref.\cite{RT16}).

\subsection{Berry-Sinitsyn-Nemenman phase} \la{BSNP}

The expression of \re{Dl o} was originally derived like the following.
The formal solution of the FCS-QME is expanded as
\bea
\dke{\rho^\chi(t)} = \sum_n c_n^\chi(t)e^{\int_0^t ds \ \lm_n^\chi(\al_s)}\dke{\rho_n^\chi(\al_t)} \la{exp}.
\eea
Because $e^{\int_0^t ds \ \lm_n^\chi(\al_s)}$ $(n \ne 0)$ exponentially damps as a function of time, 
only $n=0$ term remains if $\Ga \tau \gg 1$. 
Solving the time evolution equation of $c_0^\chi(t)$ in $\om \ll \Ga$ limit, we obtain
\bea
c_0^\chi(\tau) = c_0^\chi(0)\exp \Big[-\int_0^\tau dt \  \dbr{l_0^\chi(\al_t)} \f{d }{dt}\dke{\rho_0^\chi(\al_t)} \Big], \la{c_0^chi}
\eea
using \re{comment2} and the fact that the second term of RHS of \re{comment2} for $m=0$ exponentially damps as a function of time.  
Here, the argument of the exponential function is called the BSN phase. 
Substituting this  expression and $c_0^\chi(0)=\dbra l_0^\chi(\al_0)\dke{\rho_0(\al_0)}$ into \re{exp}, 
we obtain  the expression of $\rho^\chi(\tau)$ which provides \re{Dl o}. 
However, when we consider only the average of $\Dl o_\mu$, the BSN phase is not essential. 
All informations of the counting fields up to the first order are included in $W^{O_\mu}$.

Substituting \re{c_0^chi} and $c_0^\chi(0)=\dbra l_0^\chi(\al_0)\dke{\rho(0)}$ into \re{exp}, we obtain
\bea
\dke{\rho^\chi(\tau)} 
\aeqap \dbra l_0^\chi(\al_0) \dke{\rho(0)}  e^{ -\int_0^\tau dt \ \dbr{l_0^\chi(\al_t)}\f{d}{dt} \dke{\rho_0^\chi(\al_t)}} e^{\int_0^\tau dt \ \lm_0^\chi(\al_t) }\dke{\rho_0^\chi(\al_\tau)} , \la{sol_ap}
\eea
and the cumulant generating function $S_\tau(\chi)=\ln Z_\tau(\chi)=\ln \dbra 1 \dke{\rho^\chi(\tau)}$ :
\bea
S_\tau(\chi) \aeq \int_0^\tau dt \ \lm_0^\chi(\al_t)-\int_C d\al^n \  \dbr{l_0^\chi(\al)} \f{\partial \dke{\rho_0^\chi(\al)}}{\partial \al^n} \no\\
&&+\ln \dbra l_0^\chi(\al_0)\dke{\rho(0)} +\ln \dbra 1 \dke{\rho_0^\chi(\al_\tau)}. \la{S_tau}
\eea
(\ref{S_tau}) is the same with Yuge {\it et al.}\cite{Yuge12} except for that $\chi$ denotes a multi-counting field. 
The averages $\bra \Dl o_\mu \ket_\tau= \f{\partial S_\tau(\chi) }{\partial (i\chi_\mu)} \big \vert_{\chi=0}$ are 
\bea
\bra \Dl o_\mu \ket_\tau 
\aeq \int_0^\tau dt \ \lm_0^{O_\mu}(\al_t)+\int_C d\al^n \  A_n^{O_\mu}(\al) +\dbra l_0^{O_\mu}(\al_0)\dke{\rho(0)}+\dbra 1 \dke{\rho_0^\mu(\al_0)} .\la{ave}
\eea
Here, we used $-\int_C d\al^n \ \dbr{l_0(\al)} \f{\partial \dke{\rho_0^\mu(\al)}}{\partial \al^n}=-\dbra 1 \dke{\rho_0^\mu(\al_\tau)}+\dbra 1 \dke{\rho_0^\mu(\al_0)} $ 
because \\
$\dbr{l_0(\al)} \f{\partial \dke{\rho_0^\mu(\al)}}{\partial \al^n}= \f{\partial }{\partial \al^n}\dbra 1 \dke{\rho_0^\mu(\al)}$. 
The integrand of the first time integral, $\lm_0^{O_\mu}(\al_t)$,  are the instantaneous steady currents of $O_\mu$ at time $t$;
if the control parameters are fixed to $\al$ and the state is $\rho_0(\al)$, the current of $O_\mu$ is $\lm_0^{O_\mu}(\al)$.
The third and fourth terms of the right side of \re{ave} cancel if the initial condition is the instantaneous steady state $\rho_0(\al_0)$.

\subsection{Cyclic pump} \la{Cyclic}

For $\al_\tau=\al_0$, the second term of the right side of \re{ave} can be described as a surface integral over the surface $S$ enclosed by $C$ using the Stokes theorem :
\bea
\bra \Dl o_\mu \ket_\tau \aeq \bra \Dl o_\mu \ket^\st_\tau+\bra \Dl o_\mu \ket^{\rm{Berry}}_S ,\la{yuge} \\
\bra \Dl o_\mu \ket^\st_\tau \aeq \int_0^\tau dt \ \lm_0^{O_\mu}(\al_t) \la{yuge_S},\\
\bra \Dl o_\mu \ket^\RM{Berry}_S \aeq \int_S d\al^m \wedge d\al^n \ \half F^{O_\mu}_{mn}(\al) \la{yuge_B}  .
\eea
Here, $\wedge$ is the wedge product and the summation symbol $\sum_{n,m}$ is omitted. BSN curvature $F^{O_\mu}_{mn}(\al)$ is given by 
\bea
F^{O_\mu}_{mn}(\al)\aeq \f{\partial A_n^{O_\mu}(\al)}{\partial \al^m}-\f{\partial A_m^{O_\mu}(\al)}{\partial \al^n}. 
\eea
Yuge {\it et al.}\cite{Yuge12} focus on only the second term of \re{yuge} subtracting the first term, and they did not evaluate $\bra \Dl o_\mu \ket^\st_\tau$.
In \res{SC}, we show that this contribution is usually dominant if the thermodynamic parameters are modulated 
although the steady currents $\lm_0^{O_\mu}(\al_t)$ are zero if the thermodynamic parameters are fixed to zero bias.

\subsection{Expansion by frequency} \la{Expansion}

Applying the pseudo-inverse $\mR(\al)$ to the QME \re{QME}, we obtain
\bea
\Big[1-\mR(\al_t)\f{d}{dt} \Big]\dke{\dl \rho(t)} \aeq  \mR(\al_t) \f{d}{dt}\dke{\rho_0(\al_t)}, \la{QME_R}
\eea
with $\dl \rho(t)\defe \rho(t)-\rho_0(\al_t)$. 
One of the solution of \re{QME_R} is 
\bea
\dke{\dl \rho_{(\RM{ss})}(t)}\aeq \sum_{n=1}^\infty \Big[ \mR(\al_t) \f{d}{dt} \Big]^n \dke{\rho_0(\al_t)} \e \sum_{n=1}^\infty \dke{\rho^{(n)}(t)}. \la{key6}
\eea
$\dbra 1 \dke{\rho^{(n)}(t)}=0$ holds (we show this at \res{PI}). 
The general solution of \re{QME_R} is 
\bea
\dke{\dl \rho(t)} \aeq \dke{\dl \rho_{(\RM{ss})}(t)}+\dke{\tl \rho(t)} ,
\eea
where $\tl \rho(t)$ is the solution of
\bea
\Big[1- \mR(\al_t)\f{d}{dt} \Big]\dke{\tl \rho(t)} = 0 ,\la{QME3}
\eea
with $\tl \rho(0)=\dl \rho(0)-\dl \rho_{(\st)}(0)$. 
By the way, applying $\Hat{K}(\al)$ to \re{defR} from the left, we obtain
\bea
\Hat{K}(\al)\mR(\al)\Hat{K}(\al)=\Hat{K}(\al).
\eea
This leads 
\bea
\Hat{K}(\al)\mR(\al)=1-\dke{\sig(\al)}\dbr{1},\ \dbra 1 \dke{\sig(\al)}=1.
\eea
Applying $\Hat{K}(\al)$ to \re{QME_R} from the left and using the above relation and $\dbra 1\dke{\tl \rho(t)}=0$, 
we obtain
\bea
\f{d}{dt} \dke{\tl \rho(t)} \aeq \Hat{K}(\al)\dke{\tl \rho(t)},
\eea
which is the same form with the original QME. 
The solution is $\dke{\tl \rho(t)} = \Hat{U}(t) \dke{\tl \rho(0)} $
with 
\bea
\Hat{U}(t) \defe {\rm{T}}\exp \Big[\int_0^t ds \ \Hat{K}(\al_s) \Big].
\eea
Because $\dbra 1\dke{\tl \rho(0)}=0$, $\dke{\tl \rho(t)}$ is described as
$\dke{\tl \rho(t)} = \sum_{n \ne 0} c_n^\pr(t) e^{\int_0^t ds \ \lm_n(\al_s)}\dke{\rho_n(\al_s)} $.
This damps exponentially as a function of time. 
Then, the state reaches to a ``steady state"
\bea
\rho_{(\st)}\defe \rho_0(\al_t)+\dl \rho_{(\st)}(t).
\eea
$\tl \rho(0)=\dl \rho(0)-\dl \rho_{(\st)}(0)$ is the difference of the initial state from the ``steady state". 
We introduce 
\bea
\dke{\tl \rho^{(n)}(t)} \aeqd -\Hat{U}(t)\dke{\rho^{(n)}(0)}, \la{def_tl_rho^n} \\
\dke{ \tl \rho^{(0)}(t)} \aeqd \Hat{U}(t)\dke{\dl \rho(0)}.
\eea
The general solution of the QME is given by
\bea
\dke{\dl \rho(t)} \aeq \sum_{n=1}^\infty \Big[\dke{\rho^{(n)}(t)} +\dke{\tl \rho^{(n)}(t)} \Big]
+\dke{ \tl \rho^{(0)}(t)}.
\eea
$\dbra 1 \dke{\tl \rho^{(n)}(t)}=-\dbra 1 \dke{\rho^{(n)}(0)}=0$ and $\dbra 1 \dke{\tl \rho^{(0)}(t)}=\dbra 1 \dke{\dl \rho(0)}=0$ holds. 
The current $i_{O_\mu}(t)$ is given by 
\bea
i_{O_\mu}(t) \aeq i_{O_\mu}^\st(\al_t)+\dl i_{O_\mu}^{(\st)}(t)+\tl i_{O_\mu}(t),\la{key7} \\
\dl i_{O_\mu}^{(\st)}(t) \aeqd \dbr{1}W^{O_\mu}(\al_t)\dke{\dl \rho_{(\st)}(t)},\\
\tl i_{O_\mu}(t)  \aeqd \dbr{1}W^{O_\mu}(\al_t)\dke{\tl \rho(t)}.
\eea
\bea
\dl i_{O_\mu}^{(\st)}(t) \aeq \sum_{n=1}^\infty  i_{O_\mu}^{(n)}(t),\ i_{O_\mu}^{(n)}(t)\defe \dbr{1}W^{O_\mu}(\al_t)\dke{\rho^{(n)}(t)},\\
\tl i_{O_\mu}(t)  \aeq  \sum_{n=0}^\infty  \tl i_{O_\mu}^{(n)}(t) ,\ \tl i_{O_\mu}^{(n)}(t) \defe \dbr{1}W^{O_\mu}(\al_t)\dke{\tl \rho^{(n)}(t)}.
\eea

Let's consider the relation between \re{watan} and \re{key7}. 
In \res{BSNP}, we used $\chi$-adiabatic approximation \re{sol_ap}, which becomes $\dke{\rho(t)} \approx  \dke{\rho_0(\al_t)}$ at $\chi=0$. 
Substituting it to \re{I_mu}, we obtain 
$ i_{O_\mu} (t) \approx  i_{O_\mu}^{\rm{ss}}(t) $. So, we cannot obtain $\dl i_{O_\mu}^{(\st)}(t)+\tl i_{O_\mu}(t)$. 
However, from the $\chi_{O_\mu}$ derivative of \re{sol_ap}, we obtain
\bea
 i_{O_\mu} (t) \aeqap \lm_0^{O_\mu}(\al_t)-\dbr{l_0^{O_\mu}(\al_t)}\f{d}{dt} \dke{\rho_0(\al_t)} \la{key8}.
\eea
This is equivalent to \re{ave} for $\rho(0)=\rho_0(\al_0)$. 
(\ref{key8}) suggests
\bea
i_{O_\mu}^{(1)}(t)= -\dbr{l_0^{O_\mu}(\al_t)}\f{d}{dt} \dke{\rho_0(\al_t)}  \la{I^a} .
\eea
In fact, this is equivalent to $i_{O_\mu}^{(1)}(t)=\dbr{1} W^{O_\mu}(\al_t)\dke{\rho^{(1)}(t)}$, namely 
\bea
i_{O_\mu}^{(1)}(t)=\dbr{1} W^{O_\mu}(\al_t)\mR(\al_t)\f{d}{dt}\dke{\rho_0(\al_t)}  \la{key9},
\eea
because of
\bea
\dbr{1} W^{O_\mu}(\al) \mR (\al)=-\dbr{l_0^{O_\mu}(\al)}+c^{O_\mu}(\al) \dbr{1}. \la{W_R}
\eea
 Here, $c^{O_\mu}(\al)$ are constants shown in \re{Y2}. 
We prove \re {W_R} in Appendix \ref{proof}. 
\re{W_R} leads \re{BSN_Nakajima} and 
\bea
i_{O_\mu}^{(n+1)}(t)\aeq \dbr{1}W^{O_\mu}(\al_t)\mR(\al_t)\f{d}{dt}\dke{ \rho^{(n)}(t)} \no\\
\aeq -\dbr{l_0^{O_\mu}(\al_t)} \f{d}{dt}\dke{\rho^{(n)}(t)} .
\eea
By the way, \re{watan} is 
\bea
i_{O_\mu}(t) \aeq i_{O_\mu}^\st(\al_t)-\dbr{l_0^{O_\mu}(\al_t)} \f{d}{dt}\dke{\rho(t)}. \la{wa}
\eea
Substituting 
\bea
\rho(t)=\rho_0(\al_t)+\sum_{n=1}\rho^{(n)}(t)+\sum_{n=0}\tl \rho^{(n)}(t)=\rho_{(\st)}(t)+\sum_{n=0}\tl \rho^{(n)}(t), 
\eea
to the RHS of \re{wa}, $\rho_0$ provides $i_{O_\mu}^{(1)}$ and $\rho^{(n)}$ provides $i_{O_\mu}^{(n+1)}$.
$\rho_{(\st)}$ provides $\dl i_{O_\mu}^{(\st)}$. 
$\tl \rho^{(n)}$ provides $\tl i_{O_\mu}^{(n)}$:
\bea
&&\hs{-10mm}-\dbr{l_0^{O_\mu}(\al_t)} \f{d}{dt}\dke{\tl \rho^{(n)}(t)} \no\\
 \aeq -\dbr{l_0^{O_\mu}(\al_t)} \hat{K}(\al_t)\dke{\tl \rho^{(n)}(t)} \no\\
\aeq \dbr{1} W^{O_\mu}(\al_t) \mR (\al_t) \hat{K}(\al_t)\dke{\tl \rho^{(n)}(t)}-c^{O_\mu}(\al_t) \dbr{1}\hat{K}(\al_t)\dke{\tl \rho^{(n)}(t)} \no\\
\aeq \dbr{1} W^{O_\mu}(\al_t) (1-\dke{\rho_0(\al_t)}\dbr{1})\dke{\tl \rho^{(n)}(t)} \no\\
\aeq \tl i_{O_\mu}^{(n)}.
\eea

The third and fourth terms of \re{ave}, $\bra \Dl o_\mu \ket_\tau^{3+4}=\dbra l_0^{O_\mu}(\al_0)\dke{\rho(0)}+\dbra 1 \dke{\rho_0^{O_\mu}(\al_0)}$, 
result from this relaxation.  
The contribution of $\bra \Dl o_\mu \ket_\tau$ from $\dl \rho(0)$ is 
\bea
\bra \Dl o_\mu \ket_\tau^{\rm{ini}} \aeqd -\int_0^\tau dt \ \dbr{l_0^{O_\mu} (\al_t)}\f{d}{dt}\dke{\tl \rho^{(0)}(t)} \no\\
\aeq \dbra l_0^{O_\mu} (\al_0)\dke{\tl \rho^{(0)}(0)}-\dbra l_0^{O_\mu} (\al_\tau)\dke{\tl \rho^{(0)}(\tau)} \no\\
&&+\int_0^\tau dt \ \f{d\dbr{l_0^{O_\mu} (\al_t)}}{dt}\dke{\tl \rho^{(0)}(t)}. \la{con_ini}
\eea
The first term of the right side of \re{con_ini} is $\bra \Dl o_\mu \ket_\tau^{3+4}$. 
Because we can obtain \\
$\dbr{l_0^{O_\mu}(\al)} \rho_0(\al) \dket+\dbr{1} \rho_0^{O_\mu}(\al) \dket=0$ from the normalization $\dbra l_0^\chi(\al) \dke{\rho_0^\chi(\al)}=1$, \\
$\bra \Dl o_\mu \ket_\tau^{3+4}$ is given by 
\bea
\bra \Dl o_\mu \ket_\tau^{3+4}=\dbra l_0^{O_\mu} (\al_0)\big[\dke{\rho(0)}-\dke{\rho_0(\al_0)} \big]=\dbra l_0^{O_\mu} (\al_0)\dke{\tl \rho^{(0)}(0)}.
\eea
The second term of the right side of \re{con_ini} is exponentially small since $\tl \rho^{(0)}(\tau) \sim e^{-\Ga \tau}$. 
The order of the third term is $\mO(\f{\om}{\Ga})$ with $\om=2\pi/\tau$ because $\f{d\dbr{l_0^{O_\mu} (\al_t)}}{dt}=\mO(\om)$ and the integral range is restricted up to $1/\Ga$ 
since $\tl \rho^{(0)}(t) \sim e^{-\Ga t}$. Hence 
\bea
\bra \Dl o_\mu \ket_\tau^{\rm{ini}}=\bra \Dl o_\mu \ket_\tau^{3+4}+\mO(\f{\om}{\Ga}).
\eea

Since $\f{d}{dt}\al_t=\mO(\om)$ and $\mR(\al_t)=\mO(\f{1}{\Ga})$,
\bea
\rho^{(n)}(t)=\mO \big(\f{\om}{\Ga} \big)^n. \la{order}
\eea
In Appendix \ref{val}, we discuss the reasonable range of $n$ of $\rho^{(n)}(t)$ and show that with the larger non-adiabaticity ($\f{\om}{\Ga}$), the reasonable range becomes wider.
We have 
\bea
\ \tl \rho^{(n)}= \mO(\f{\om^n}{\Ga^n}e^{-\Ga t} ),\ \tl \rho^{(0)}= \mO(e^{-\Ga t} ).
\eea
The above equations and $W^{O_\mu}=\mO(\Ga)$ lead
\bea
i_{O_\mu}^{(n)}= \mO(\f{\om^{n}}{\Ga^{n-1}}) ,\ \tl i_{O_\mu}^{(n)}= \mO(\f{\om^n}{\Ga^{n-1}}e^{-\Ga t}),\ \tl i_{O_\mu}^{(0)}= \mO(\Ga e^{-\Ga t}).
\eea
This leads 
\bea
\bra \Dl o_\mu \ket_\tau^{(n)} \aeqd \int_0^\tau dt \ i_{O_\mu}^{(n)}(t)=\mO(\f{\om^{n-1}}{\Ga^{n-1}}) ,\\
\widetilde{\bra \Dl o_\mu \ket}_\tau^{(n)} \aeqd \int_0^\tau dt \ \tl i_{O_\mu}^{(n)}(t)=\mO(\f{\om^{n}}{\Ga^n}) ,\\
\bra \Dl o_\mu \ket_\tau^{\rm{ini}} \aeq \int_0^\tau dt \ \tl i_{O_\mu}^{(0)}(t)=\mO(1).
\eea
In particular, the contribution from the BSN vector is 
\bea
\bra \Dl o_\mu \ket_\tau^{\RM{BSN}} \defe \bra \Dl o_\mu \ket_\tau^{(1)} =-\int_C d\al^n \  \dbr{l_0^{O_\mu} (\al)} \f{\partial \dke{\rho_0(\al)}}{\partial \al^n}=\mO(1).
\eea

Moreover, although the BSN phase is derived under 
the $\chi$-adiabatic condition which makes \re{c_0^chi} and $c_n^\chi(\tau)e^{\Lm_n^\chi(\tau)} \approx 0$ $(n \ne 0)$ appropriate, its origin is probably a 
non-adiabatic effect that comes from $\f{\om}{\Ga}$, 
because \re{I^a} shows that the BSN phase has the information of the non-adiabatic part of the QME $(\dl \rho(t)=\rho(t)-\rho_0(\al_t))$. 

For the RWA, at equilibrium (zero-bias) case,  
$\rho_0(\al)$ is the grand canonical distribution
\bea
\rho_\RM{gc}(\al_S;\be,\be \mu) \defe \f{e^{-\be(H_S(\al_S)- \mu N_S)}}{\Xi(\al_S;\be,\be \mu)} . \la{rho_gc}
\eea
Cf.\re{BM_rho_ss}. Here, $\Xi(\al_S;\be,\be \mu)\defe\tr_S[e^{-\be(H_S(\al_S)-\mu N_S)}]$, 
$\be$ is the inverse temperature of all baths and $\mu$ is the chemical potential for $b \in \RM{\bm{\mG}}$.
\re{rho_gc} is derived from \re{KMS3} and \re{KMS4}. 
At zero-bias, for pumping by only $\al^\pr$ ($\al^{\pr\pr}$ are fixed), \re{BSN}, \re{key6} and \re{def_tl_rho^n}
 lead that the 
pumping dose not occur in all orders of $\om$ when $\al_S$ are fixed. 

\subsection{Arbitrariness of pseudo-inverse} \la{PI}

General solution of $\mR \Hat{K}(\al_t)=1- \dke{\rho_0(\al_t)}\dbr{1}$ is given by
\bea
\mR(t) \aeq \dke{\rho_i(t)}\dbr{1} +\mR_0(\al_t),
\eea
where $\mR_0(\al)$ is one of the solution of
\bea
\mR_0(\al)\Hat{K}(\al) \aeq 1- \dke{\rho_0(\al)}\dbr{1}. \la{R_0}
\eea
$\rho_i(t)$ can depend on the initial values of the QME. 
In the following of this section, we show that $\dke{\rho^{(n)}(t)} = \Big[ \mR(t) \f{d}{dt} \Big]^n \dke{\rho_0(\al_t)}$ is 
independent of $\rho_i(t)$. 
Then, $\rho^{(n)}$ and $\tl \rho^{(n)}$ are independent of the choice of the pseudo-inverse.

$\rho^{(1)}(t)$ is given by
\bea
\dke{\rho^{(1)}(t)} \aeq \dke{\rho_i(t)}\dbr{1} \f{d}{dt}\dke{\rho_0(\al_t)}+\mR_0(\al_t) \f{d}{dt}\dke{\rho_0(\al_t)} \no\\
\aeq \mR_0(\al_t) \f{d}{dt}\dke{\rho_0(\al_t)} .\la{a1}
\eea
Then, $\rho^{(1)}(t) $ is independent of the choice of the pseudo-inverse. 
Next, $\rho^{(2)}(t)$ is given by
\bea
\dke{\rho^{(2)}(t)} 
\aeq \dke{\rho_i(t)} \f{d}{dt}\big[\dbr{ 1}\mR_0(\al_t) \f{d}{dt}\dke{\rho_0(\al_t)}\big] +\mR_0(\al_t) \f{d}{dt}\big[\mR_0(\al_t) \f{d}{dt}\dke{\rho_0(\al_t)}\big].
\eea
By the way, applying $\dbr{1}$ to \re{R_0}, we obtain
\bea
\dbr{1}\mR_0(\al)\Hat{K}(\al) \aeq \dbr{1}- \dbra 1\dke{\rho_0(\al)}\dbr{1}=0 .
\eea
This leads 
\bea
\dbr{1}\mR_0(\al)\aeq \mC(\al)\dbr{1}. \la{z}
\eea
Then, $\dke{\rho^{(2)}(t)} $ and $\dke{\rho^{(n)}(t)} $ do not depend on the choice of the pseudo-inverse. 
In fact, 
\bea
\dke{\rho^{(n+1)}(t)} 
\aeq \dke{\rho_i(t)} \f{d}{dt}\dbra 1\dke{\rho^{(n)}(t)}+\mR_0(\al_t) \f{d}{dt}\dke{\rho^{(n)}(t)}, \la{ze}
\eea
leads 
\bea
\dbra 1\dke{\rho^{(n+1)}(t)} \aeq \dbra 1\dke{\rho_i(t)} \f{d}{dt}\dbra 1\dke{\rho^{(n)}(t)} +\mC(\al_t) \f{d}{dt}\dbra 1\dke{\rho^{(n)}(t)} .
\eea
Then, $\dbra 1\dke{\rho^{(n)}(t)}=0$ leads $\dbra 1\dke{\rho^{(n+1)}(t)}=0$. 
Because of this and $\dbra 1 \dke{\rho^{(1)}(t)}=0$ derived from \re{a1} and \re{z}, we obtain
\bea
\dbra 1\dke{\rho^{(n)}(t)}=0 \hs{3mm}(n=1,2,3,\cdots). \la{a_n}
\eea
This and \re{ze} lead
\bea
\dke{\rho^{(n+1)}(t)} \aeq \mR_0(\al_t) \f{d}{dt}\dke{\rho^{(n)}(t)} .
\eea

\subsection{Generalized mater equation approach} \la{GMEa}

It is important to recognize the relations between the FCS-QME approach and the GME approach 
\cite{Konig09-, RT09, Splettstoesser10-1, Splettstoesser10-2, Splettstoesser12, Splettstoesser13, Konig13-2}.
In the GME approach, $p_i(t)=\br{i}\rho(t)\ke{i}$ are governed by the generalized master equation (GME)
\bea
\f{d}{dt} p_i(t) \aeq \sum_j \int_{-\infty}^t dt^\pr \ W_{i j}(t,t^\pr)p_j(t^\pr), \la{GME}
\eea
where $\ke{i}$ are the energy eigenstates of the system Hamiltonian. 
The kernel $W_{i j}(t,t^\pr)$ can include the higher order contribution of the tunneling interaction between baths 
and the system. 
In the GME, $p_j(t^\pr)$ is given by $p_j(t)+\sum_{k=1}^\infty\f{(t^\pr-t)^k}{k!}\f{d^kp_j(t)}{dt^k}$ \cite{Konig09-, RT09}. 
Moreover, $W_{i j}(t,t^\pr)$ and $p_j(t)$ are expanded as $W_{i j}(t,t^\pr)=\sum_{n=0}^\infty \sum_{m=1}^\infty W_{i j(m)}^{(n)}(t;t-t^\pr)$ 
and $p_j(t)=\sum_{n=0}^\infty \sum_{m=-n}^\infty p_{j(m)}^{(n)}(t)$, where 
$W_{i j(m)}^{(n)}(t;t-t^\pr)$ and $p_{j(m)}^{(n)}(t)$ are of the order of $\om^n\Ga^m$. 
In particular, $W_{i j(m)}^{(0)}(t;t-t^\pr)=W_{i j(m)}^{(0)}(\al_t;t-t^\pr)$ is the kernel 
where the control parameters are fixed to $\al_t$.
Up to the second order of the tunneling interaction (in the following we consider this level of approximation), we obtain (see Appendix \ref{A_GME}) 
\cite{RT09, Splettstoesser13}
\bea
0 \aeq \sum_j K_{i j}^{(0)}(\al_t) p_{j}^{(0)}(\al_t) ,\la{ss_RT}\\
 \f{dp_{i(-n)}^{(n)}(t)}{dt} \aeq \sum_j K_{i j}^{(0)}(\al_t)p_{j(-n-1)}^{(n+1)}(t) \la{n_RT},
\eea
for $n=0,1,\cdots$, with
\bea
K_{i j}^{(0)}(\al_t)=\int_{-\infty}^t dt^\pr \ W_{i j(1)}^{(0)}(\al_t,t-t^\pr),
\eea
which is the instantaneous Liouvillian corresponding to our $\Hat{K}(\al_t)$. 
(\ref{ss_RT}) is just the definition of the instantaneous steady state 
$p_{j}^{(0)}(\al_t)\e p_{j(0)}^{(0)}(t)$, which satisfies $\sum_i  p_{i}^{(0)}(\al_t)=1$. 
Additionally, $p_{i(m)}^{(n)}(t)$ for $n\ge 1$ satisfies $\sum_i  p_{i(m)}^{(n)}(t)=0$.
The conservation of the probability leads to $\sum_i K_{i j}^{(0)}(\al_t)=0$, which corresponds to our $\dbr{1}\Hat{K}(\al_t)=0$. 
The charge or spin current $i_{O_\mu}(t)$ is given by \cite{Splettstoesser12, Splettstoesser13}
\bea
i_{O_\mu}(t)\aeq \sum_{i,j} w^{O_\mu}_{i j}(\al_t)p_j(t) \la{I_mu_rtd} ,
\eea
corresponding to our \re{I_mu}. $w^{O_\mu}_{i j}(\al_t)$ is the instantaneous current matrix of $O_\mu$ in the present approximation, 
which corresponds to our $W^{O_\mu}(\al_t)$ and is linear in $\Ga$ (see \re{i_X_b2}). 
$i_{O_\mu}(t)$ can be rewritten as
\bea
i_{O_\mu}(t)\aeq \sum_{j} w^{O_\mu}_{j}(\al_t)p_j(t),\ w^{O_\mu}_{j}(\al)\defe \sum_i w^{O_\mu}_{i j}(\al).
\eea
$w^{O_\mu}_{j}(\al)$ corresponds to $\br{j}W^{O_\mu}(\al_t)\ke{j}$. 
Substituting $p_j(t)\approx \sum_{n=0}^\infty p_{j(-n)}^{(n)}(t)$ into \re{I_mu_rtd}, we obtain 
\bea
i_{O_\mu}(t)\aeq \sum_{n=0}^\infty i_{O_\mu}^{(n)}(t) ,\ i_{O_\mu}^{(n)}(t)\defe \sum_{i,j} w^{O_\mu}_{i j}(\al_t)p_{j(-n)}^{(n)}(t). \la{I_mu^k}
\eea
(\ref{n_RT}) for $n=0$ leads to \cite{Splettstoesser12}
 \bea
 p_{j(-1)}^{(1)}(t) \aeq \sum_{i} R_{j i}(\al_t)  \f{dp_{i}^{(0)}(\al_t)}{dt} \la{p^1} .
 \eea
Here, $R_{j i}(\al_t) $ is the pseudo-inverse of $K_{i j}^{(0)}(\al_t)$ corresponding to our $\mR(\al_t)$ and it is given by \cite{Splettstoesser12}
\bea
R_{j i}(\al_t)=(\tl K^{-1})_{j i}, \ \tl K_{j i}= K_{j i}^{(0)}- K_{j j}^{(0)} \la{RR} .
\eea
Substituting \re{p^1} into \re{I_mu^k}, we obtain \cite{Splettstoesser12}
\bea
i_{O_\mu}^{(1)}(t)\aeq \sum_{i} \varphi_i^{O_\mu}(\al_t) \f{dp_{i}^{(0)}(\al_t)}{dt} , \la{I_mu_rtd2} \\
\varphi_i^{O_\mu}(\al_t) \aeq \sum_{k, j} w^{O_\mu}_{k j}(\al_t)R_{j i}(\al_t)=\sum_{ j} w^{O_\mu}_{j}(\al_t)R_{j i}(\al_t) \la{varphi}.
\eea
A similar method has been used in \Refe{Konig09-}. 
$\varphi_i^{O_\mu}(\al_t)$ and \re{I_mu_rtd2} respectively correspond to our $\dbr{1} W^{O_\mu}(\al) \mR (\al)$ and \re{key9}. 
Moreover, \re{n_RT} for arbitrary $n$ leads to
\bea
 p_{j(-n-1)}^{(n+1)}(t) \aeq \sum_{i} R_{j i}(\al_t)  \f{dp_{i(-n)}^{(n)}(t)}{dt} \la{p^n+1} ,
\eea
which corresponds to our \re{key6}.
Because of these relations, the GME approach is equivalent to the FCS-QME approach in the calculation up to the second order of the tunneling interaction. 
Additionally, we discuss corrections due to the non-adiabatic effect of the FCS-QME in Appendix \ref{val}. 
The first equation of \re{D1} is consistent with $p_{j(0)}^{(1)}(t)=\mO(\om \tau_B)$ derived in Appendix \ref{A_GME}. 
Here, $\tau_B$ is the relaxation time of the baths.

In this chapter, we proved the equivalence between \re{watan} and \re{key7} using a key relation \re{W_R} and showed the origin of the BSN phase is a non-adiabatic effect, 
and connected the FCS-QME approach and the GME approach \cite{Splettstoesser12}. 
These are among the most important results of the first half of this thesis.

\newpage

\section{Quantum adiabatic pump} \la{QAP}

\subsection{Model} \la{model}

In this chapter, we consider quantum dots (QDs) (denoted by a symbol $S$) weakly coupled to several leads. 
The total Hamiltonian is $H_\tot(\al^\pr(t))=H_S(\al_S(t))+\sum_{b} [H_b(\al_b^\pr(t))+H_{Sb}(\al_{Sb}(t))] $. 
Here, $H_S(\al_S(t))$ is the system (QDs) Hamiltonian, $H_b(\al_b^\pr(t))$ is the Hamiltonian of 
the lead $b$, and $H_{Sb}(\al_{Sb}(t))$ is the tunneling interaction Hamiltonian between $S$ and the lead $b$. 
To observe the spin effects, we suppose that the leads and the system are applied to collinear magnetic fields with different amplitudes, 
which relate to spins through the Zeeman effect. 
The leads are noninteracting: 
\bea
H_b(\al_b^\pr(t))=\sum_{k,\sig} (\ep_{bk}+\sig g_b B_b(t))c_{bk\sig} \dg c_{bk\sig} . \la{H_b}
\eea
Here, $\sig=\up,\dw=\pm 1$ is spin label, 
\bea
g_b=\half \mu_{\rm{B}}g_b^\ast,
\eea
where $g_b^\ast$ is the $g$-factor of the lead $b$, 
$\mu_{\rm{B}}$ is the Bohr magneton and $B_b(t)$ is the strength of the magnetic field of the lead $b$. 
$c_{bk\sig} \dg (c_{bk\sig})$ is the creation (annihilation) operator of an electron with spin $\sig$ and momentum $k$ in the lead $b$. The system Hamiltonian is
\bea
H_S(\al_S(t))=\sum_{n,m,s,s^\pr} \ep_{n s, m s^\pr}(B_S(t)) a_{n s} \dg a_{m s^\pr}+H_{\rm{Coulomb}} , \la{H_S}
\eea
where $a_{n s} \dg$ is the creation operator of an electron with orbital $n$ and spin $s$. 
$\ep_{n s, m s^\pr}(B_S(t))$ means the energy of the electron for $n=m,s=s^\pr$ and the tunneling amplitude between orbitals for $(n,s)\ne(m,s^\pr)$ 
which depends on the magnetic field of the system. $H_{\rm{Coulomb}}$ denotes Coulomb interaction.
The tunneling interaction Hamiltonian is
\bea
H_{Sb}(\al_{Sb}(t))=\sum_{k,\sig,n,s}\sqrt{\Dl_b(t)}v_{bk\sig,n s}a_{n s}\dg c_{bk\sig}+\hc , \la{H_1}
\eea
where $\Dl_b(t)$ is a dimensionless parameter, and $v_{bk\sig,n s}$ is the tunneling amplitude.

We assume $B_S$, $\{B_b\}_b$ and $\{\Dl_b\}_b$ are control parameters (denoted \\
$\al^\pr=(B_S,\{B_b\}_b,\{\Dl_b\}_b)$ and are called the dynamic parameters). 
The thermodynamic parameters (the  chemical potentials and inverse temperatures of the leads, 
$\{\mu_b\}_b$ and $\{\be_b \}_b$) are also considered as control parameters in \res{SC} and \res{current,inf}. 
We denote $\al^{\pr\pr}=\{\be_b, \mu_b\}_b$ and $\al=\al^\pr+\al^{\pr\pr}$.
Yuge {\it et al.}\cite{Yuge12} chose the set of control parameters as only $\al^{\pr\pr}$. However we are interested in $\al^\pr$ for the reason explained in \res{SC}.  

We choose the measured observables $\{ O_\mu \} =\{N_{b\sig} \}_{b,\sig=\up,\dw}$ with $N_{b\sig}=\sum_{k}c_{bk\sig} \dg c_{bk\sig}$. 
The pumped charge (spin) of the lead $b$ is given by $\bra \Dl N_{b\up} \ket\pm\bra \Dl N_{b\dw} \ket$. $\bra \Dl N_{b\sig} \ket$ are calculated by \re{yuge}. 
In fact, what we call the pumped charge, $\bra \Dl N_{b\up} \ket+ \bra \Dl N_{b\dw} \ket$, is the pumped electron number (actual pumped charge is given by 
$-e[\bra \Dl N_{b\up} \ket+ \bra \Dl N_{b\dw} \ket]$, where $e\ (>0)$ is the elementary charge).

In
\res{pump,U=0} and \res{Interacting} we consider a one level system
\bea
H_S(\al_S(t))=\sum_{s=\up,\dw} \om_s(B_S(t)) a_s \dg a_s+Ua_\up \dg a_\up a_\dw \dg a_\dw , \la{H_S,1}
\eea
as a special model of \re{H_S}. Here, $s=\up,\dw=\pm 1$, 
\bea
\om_s(B_S)=\om_0+sg_SB_S ,
\eea
with $\om_0$ the electron energy at $B_S=0$, 
and 
\bea
g_S=\half \mu_{\rm{B}}g_S^\ast ,
\eea
where $g_S^\ast$ is the $g$ factor of the QD.

In the following of this chapter, we apply the FCS-QME with RWA.

\subsection{Non-interacting system} \la{Non-interacting}

In this section, we consider a noninteracting system ($H_{\rm{Coulomb}}=0$). The system Hamiltonian \re{H_S} can be diagonalized
\bea
H_S \aeq \sum_{i=1}^{2N} \tl \om_{i} b_{i} \dg b_{i},
\eea
by a unitary transform $a_{n s} =\sum_{i=1}^{2N} U_{n s,i} b_i $. The tunneling interaction Hamiltonian \re{H_1} is 
\bea
H_{Sb} \aeq \sum_{k ,\sig,i} W_{bk\sig, i} b_{i} \dg c_{bk\sig} +\hc ,
\eea
with 
\bea
W_{bk\sig,i} =\sum_{n,s} \sqrt{\Dl_b} v_{bk\sig,n s} U_{n s,i} ^\ast .
\eea

In \res{Liouvillian,0}, the Liouvillian and its instantaneous steady state are explained. 
In \res{SC}, we consider the contribution of \re{yuge_S} and show that this cannot be neglected in general if 
the chemical potentials and the temperatures are not fixed.
In \res{pump,U=0}, we calculate the BSN curvatures for two combinations of modulated control parameters $(B_L,B_S)$ and $(\Dl_L,B_S)$.

\subsubsection{Liouvillian} \la{Liouvillian,0}

The Liouvillian in the RWA is given by
\bea
\Hat{K}^{\chi}(\al ) \aeq \sum_{i=1}^{2N}  \Hat{K}_i^{\chi}(\al) ,\la{L_non} \\
\Hat{K}_i^{\chi}(\al )\bu \aeq -i[\tl \om_i b_i \dg b_i,\bu]+ \Hat{\Pi}_i^\chi(\al )\bu-i[H_{\RM{L},i},\bu] ,
\eea
if $\{\tl \om_i \}$ are not degenerated. Here, super-operator $\Hat{\Pi}_i^\chi(\al )$ operates to an arbitrary operator $\bu$ as   
\bea
\Hat{\Pi}_i^\chi(\al )\bu 
\aeq   \Big\{ 
\Phi_{b,i}^{+,\chi} b_i \dg \bu b_i -\half  \Phi_{i}^+ \bu b_ib_i \dg 
-\half\Phi_{i}^+ b_i   b_i   \dg \bu  \no\\
&&+\Phi_{b,i}^{-,\chi} b_i \bu  b_i \dg 
-\half\Phi_{i}^- \bu b_i   \dg b_i -\half \Phi_{i}^- b_i \dg b_i \bu   \Big\} ,
\eea
with 
\bea
\Phi_{i}^{\pm,\chi} \aeq 2\pi \sum_{b,k,\sig} \abs{W_{bk\sig,i}}^2 f_b^\pm(\tl \om_i) e^{\mp i\chi_{b\sig}}  \dl(\ep_{bk}+\sig g_bB_b-\tl \om_i).
\eea
Here, $f_b^+(\om)=[e^{\be_b(\om-\mu_b)}+1]^{-1}$ is the Fermi distribution function, $f_b^-(\om)=1-f_b^+(\om)$, $\chi_{b\sig}$ is the counting field for $N_{b\sig}$. 
The Lamb shift Hamiltonian is given by
\bea
H_{\RM{L},i} \aeq \Om_i(\al^\pr)b_i\dg b_i ,
\eea
with
\bea
\Om_i(\al^\pr) \aeq -\half  \Big(\Psi_{i}^-+\Psi_{i}^+ \Big),\\
\Psi_{i}^{\pm}\aeq 2\sum_{b,k,\sig} \abs{W_{bk\sig,i}}^2 f_b^\pm(\tl \om_i) \RM{P}\f{1}{\ep_{bk}+\sig g_bB_b-\tl \om_i} .
\eea
Here, $\RM{P}$ denotes the Cauchy principal value. 
$\Phi_{i}^{\pm,\chi}$ satisfies 
\bea 
\Phi_{i}^{\pm}\aeq \Phi_{i}^{\pm,\chi}\bv{\chi=0}=\sum_{b,\sig}\Phi_{b\sig,i}^{\pm}, \la{Phi_i} 
\eea
and
\bea
\f{\Phi_{i}^{\pm,\chi}}{\p(i\chi_{b\sig})}\Bv{\chi=0}=\mp \Phi_{b\sig,i}^{\pm},
\eea
with 
\bea
\Phi_{b\sig,i}^{\pm} \aeq 2\pi \sum_{k} \abs{W_{bk\sig,i}}^2 f_b^\pm(\tl \om_i) \dl(\ep_{bk}+\sig g_bB_b-\tl \om_i).
\eea
We set 
\bea
\Ga_{i}=\sum_{b,\sig}\Ga_{b\sig,i}=\sum_b \Ga_{b,i},
\eea
with 
\bea
\Ga_{b\sig,i}=2\pi \sum_{k} \abs{W_{bk\sig,i}}^2 \dl(\ep_{bk}+\sig g_bB_b-\tl \om_i).
\eea
Then, 
\bea
\Ga_i = \Phi_{i}^{+}+\Phi_{i}^{-}, 
\eea
and 
\bea
\Phi_{b\sig,i}^{\pm} = \Ga_{b\sig,i} f_b^\pm(\tl \om_i), \la{Phi_Ga}
\eea
hold. 
The matrix representation of $\Hat{K}_i^\chi(\al)$ (see Appendix \ref{Liouville space}) by the number states of $b_i \dg b_i$ 
($\ke{0}_i$ and $\ke{1}_i$) is a $4\times 4$ matrix which is block diagonalized to $\{ \ke{0}_i {}_i\br{0}, \ke{1}_i {}_i\br{1} \}$ space and 
$\{ \ke{0}_i {}_i\br{1}, \ke{1}_i {}_i\br{0} \}$ space. The $\{ \ke{0}_i {}_i\br{0}, \ke{1}_i {}_i\br{1} \}$ part is given by
\bea
K_i^\chi(\al) \aeq
\begin{pmatrix} 
-\Phi_i^+ &\Phi_{i}^{-,\chi} \\
\Phi_{i}^{+,\chi}&-\Phi_i^- \\ 
\end{pmatrix}\begin{matrix} 
\dke{00}_i \\
\dke{11}_i \\
\end{matrix} . \la{K_U=0}
\eea
$\{\ke{0}_i {}_i\br{1},\ke{1}_i {}_i\br{0} \}$ part does not relate to the instantaneous steady state of $\Hat{K}_i^\chi(\al)$. 
The eigenvalue of the instantaneous steady state of $\Hat{K}_i^\chi(\al)$ is given by
\bea
\lm_{0,i}^\chi(\al)=-\f{\Phi_i^+(\al) +\Phi_i^-(\al)}{2} +\sqrt{D_i^\chi(\al)} , \la{lm_i}
\eea
with 
\bea
D_i^\chi(\al) = [\Phi_i^+ +\Phi_i^-]^2/4-[ \Phi_i^+ \Phi_i^--\Phi_{i}^{-,\chi}\Phi_{i}^{+,\chi}] .
\eea 
The corresponding left and right eigenvectors are $\dke{\rho_{0,i}^\chi(\al)}=C_i^\chi(\al)\dke{00}_i+E_i^\chi(\al)\dke{11}_i$ 
and $\dbr{l_{0,i}^\chi(\al)}={}_i\dbr{00}+v_i^\chi(\al){}_i\dbr{11} $ with 
$C_i^\chi(\al)=\f{\Phi_i^{-,\chi}\Phi_i^{+,\chi}}{ [\lm_{0,i}^\chi+\Phi_i^+]^2+\Phi_i^{-,\chi}\Phi_i^{+,\chi}}$, \\
$E_i^\chi(\al)=\f{\Phi_i^{+,\chi}(\lm_{0,i}^\chi+\Phi_i^+)}{ [\lm_{0,i}^\chi+\Phi_i^+]^2+\Phi_i^{-,\chi}\Phi_i^{+,\chi}}$, and
\bea
v_i^\chi(\al)=\f{\Phi_i^+-\Phi_i^-+2\sqrt{D_i^\chi(\al)} }{2\Phi_i^{+,\chi}} .
\eea
At $\chi_{b\sig}=0$, $E_i^\chi(\al)$ becomes 
\bea
E_i(\al)=\f{\Phi_i^+}{\Phi_i^++\Phi_i^-}
\eea
and 
$C_i^\chi(\al)$ becomes $C_i(\al)=1-E_i(\al)$. 
We have
\bea
\lm_0^\chi(\al)\aeq \sum_i \lm_{0,i}^\chi(\al), \la{lm_0_sum} \\
\rho_0(\al) \aeq \bigotimes_i \rho_{0,i}(\al), \\
l_0^\chi(\al) \aeq \bigotimes_i l_{0,i}^\chi(\al).
\eea 

\subsubsection{Instantaneous steady currents}  \la{SC}

The instantaneous steady current is given by $i_{b\sig}^\st(\al)= \f{\partial \lm_0^\chi(\al) }{\partial (i\chi_{b\sig})} \big \vert_{\chi=0}$. 
\re{lm_0_sum} leads to 
\bea
i_{b\sig}^\st(\al)=\sum_i i_{b\sig,i}^\st(\al). 
\eea
Here, $i_{b\sig,i}^\st(\al)= \f{\partial \lm_{0,i}^\chi(\al) }{\partial (i\chi_{b\sig})} \big \vert_{\chi=0}$ are calculated from \re{lm_i} as
\bea
i_{b\sig,i}^\st(\al)=\f{\Phi_{b\sig,i}^{-}\Phi_{i}^{+}-\Phi_{i}^{-}\Phi_{b\sig,i}^{+}}{\Ga_i} .
\eea
From \re{Phi_i}, we obtain
\bea
\sum_{b,\sig}i_{b\sig,i}^\st(\al)=0 .\la{N_hozon_non}
\eea
From \re{Phi_Ga}, we obtain
\bea
i_{b\sig,i}^\st(\al)=\f{\Ga_{b\sig,i}\sum_{b^\pr(\ne b)}\Ga_{b^\pr,i}[f_{b^\pr}(\tl \om_i)-f_{b}(\tl \om_i)]}{\Ga_i}.
\eea
$i_{b\sig,i}^\st(\al)$ vanishes at zero bias ($\be_b=\be$, $\mu_b=\mu$).
Let us consider the modulation of only the thermodynamic parameters ($\al^{\pr\pr}$) similar to Refs.\cite{Yuge12,Utiyama, Watanabe, yoshii2}. 
The factor depending on $\al^{\pr\pr}$ of $i_{b\sig,i}^\st(\al_t)$ is
$f_{\be_{b^\pr}(t),\mu_{b^\pr}(t)}(\tl \om_i)-f_{\be_b(t),\mu_b(t)}(\tl \om_i)$ with $f_{\be,\mu}(\om)=[e^{\be(\om-\mu)}+1]^{-1}$. 
Hence
\bea
\bra \Dl N_{b\sig} \ket^\st_\tau \aeq \sum_{i}\f{\Ga_{b\sig,i}}{\Ga_i} \sum_{b^\pr(\ne b)}\Ga_{b^\pr,i}
 \int_0^\tau dt \ [f_{\be_{b^\pr}(t),\mu_{b^\pr}(t)}(\tl \om_i)-f_{\be_b(t),\mu_b(t)}(\tl \om_i)],
\eea
is generally nonzero and is much lager than $\bra \Dl N_{b\sig} \ket^{\rm{Berry}}_S$ because the period $\tau$ is large for adiabatic pumps. 
Similarly, we can show that $\bra \Dl N_{b\sig} \ket^\st_\tau$ is generally nonzero for interacting system (\res{current,inf}). 
Reference\cite{Watanabe} considered special modulations of only thermodynamic parameters which satisfy 
$\bra \Dl N_{b\sig} \ket^\st_\tau=0$. 
In fact, the instantaneous steady currents are always zero for arbitrary modulations of only the dynamics parameters at zero bias. 

The pumped charge and spin due to the instantaneous steady currents (backgrounds) are generally nonzero even if the time averages of the bias are zero. \\
References\cite{Splettstoesser10-2, Splettstoesser12} (two leads case) chose $V=\mu_L-\mu_R$ as one of the modulating parameters and considered a pumping 
such that $\f{1}{\tau}\int_0^\tau dt \ V(t)=0$ and $\bra \Dl N_{b\sig} \ket^\st_\tau \ne 0$. 
In such pumping, the (thermal or voltage) bias is effectively nonzero.

Even if the backgrounds do not vanish, 
one can detect the BSN curvatures by subtracting the backgrounds by using zero-frequency measurements or by lock-in measurements. 
However, if one wants to apply the adiabatic pump to the current standard\cite{CS1, CS2}, the instantaneous steady currents should be zero at all times
because the backgrounds are sensitive to the velocity of the modulation of the control parameters and its trajectory. 
In contrast, the pumped charge and spin due to the BSN curvatures are robust against the modulation of the velocity and the trajectory. 
Hence, if one wants to directly apply the BSN curvatures to, for instance, the current standard, one should fix the thermodynamic parameters at zero bias.

\subsubsection{BSN curvatures} \la{pump,U=0}

In the following of this subsection, we consider one level system of which the Hamiltonian is \re{H_S,1} at $U=0$.  
The instantaneous steady state is given by $\dke{\rho_0(\al)}=\otimes_{s=\up,\dw} \dke{\rho_{0,s}(\al)}$ 
because the Liouvillian is described by a summation ($\Hat{K}^{\chi}= \sum_{s=\up,\dw} \Hat{K}_s^{\chi}$).
Similarly, the corresponding left eigenvalue is given by $\dbr{l_0^\chi(\al)}=\otimes_{s=\up,\dw} \dbr{l_{0,s}^\chi(\al)}$. 
The BSN vectors are given by 
\bea
A_n^{b\sig}(\al) 
\aeq -\sum_{s=\up,\dw} v_s^{b\sig}(\al^\pr) \f{\partial E_s(\al) }{\partial \al^n} \la{A_n},
\eea
where  
\bea
v_s^{b\sig}(\al^\pr) \aeq \f{\partial v_s^\chi(\al)}{\partial (i\chi_{b\sig})}  \Big \vert_{\chi=0} 
= \f{\Ga_{b\sig,s}}{\Ga_s} , \la{v_s^bsig}
\eea
with 
\bea
\Ga_{b\sig,s}(\al^\pr) \aeq 2\pi \Dl_b  \sum_{k}\abs{v_{bk\sig,s}}^2  \dl(\ep_{bk}+\sig g_bB_b-\om_0-sg_SB_S). 
\eea
$v_s^{b\sig}(\al^\pr)$ dose not depend on $\al^{\pr\pr}$. 
$\sum_{b,\sig}v_s^{b\sig}(\al^\pr)=1$ leads 
\bea
\sum_{b,\sig} A_n^{b\sig}(\al)= -\sum_{s=\up,\dw} \f{\partial E_s(\al) }{\partial \al^n} .
\eea
This equation and \re{N_hozon_non} lead
\bea
\sum_{b,\sig} \bra \Dl N_{b\sig} \ket \aeq -\sum_{s=\up,\dw}[E_s(\al_\tau)-E_s(\al_0)] .
\eea
The RHS is $(-1)$ times the change of the  total electron average number of the QD. 
The above equation describes the conservation of the total electron number. 
(\ref{A_n}) leads to an expression of the BSN curvatures
\bea
F_{mn}^{b\sig}(\al) 
\aeq -\sum_{s=\up,\dw} \Big[ \f{\partial v_s^{b\sig}(\al^\pr)  }{\partial \al^m}  \f{\partial E_s(\al) }{\partial \al^n}
-(m \leftrightarrow n) \Big] . \la{F_mn}
\eea

We emphasize that \re{F_mn} is consistent with the results of Refs.\cite{Splettstoesser10-2, Splettstoesser12, Yuge12}, which 
showed that the pumped charge (and also spin in \Refe{Splettstoesser12}) vanishes at the noninteracting limit in these settings. 
The set of control parameters $\al$ was $\al^{\pr\pr}$ (for \Refe{Yuge12}) and $\{\om_0, V=\mu_L-\mu_R\}$ (for Refs.\cite{Splettstoesser10-2, Splettstoesser12}). 
If $\al^m$ or $\al^n$ is an element of $ \al^{\pr\pr}$, $F_{mn}^{b\sig}(\al)$ is consistently zero.  
In Refs.\cite{Splettstoesser10-2, Splettstoesser12}, the line-width functions were energy-independent, namely $\Ga_{b\sig,s}(\al^\pr)=\dl_{\sig,s}\Ga_{b}$=constant. 
Hence $\f{\partial \Ga_{b\sig,s}(\al^\pr) }{\partial \om_0}=0=\f{\partial \Ga_{b\sig,s}(\al^\pr) }{\partial V}$ and $F_{\om_0,V}^{b\sig}(\al)=0$ hold consistently.

To calculate $F_{mn}^{b\sig}(\al)$, we need to assume the energy dependences of $\Ga_{b\sig,s}$. 
For the simplicity, we assume that
\bea
\Ga_{b\sig,s}\aeq  \dl_{\sig,s}[ \Ga_{b}+\Ga_{b}^\pr \cdot(sg_S B_S-\sig g_b B_b) ]\no\\
\aeq \dl_{\sig,s}\Dl_b[ \ga_{b}+\ga_{b}^\pr \cdot(sg_S B_S-\sig g_b B_b) ] , \la{LW}
\eea
where $\Gamma_{b}^\pr$ are energy differential coefficients of the line-width functions at $B_b=B_S=0$. 
Namely, we disregard spin flips induced by tunneling between the QD and the leads. 
(\ref{LW}) is always appropriate when $\abs{\Ga_{b}^\pr (g_S B_S-g_bB_b)} \ll \Ga_b $ is satisfied. 
Additionally, we fix $\al^{\pr\pr}$ to zero bias ($\be_b=\be$, $\mu_b=\mu$), 
in which $E_s(\al)$ is given by $E_s(\al)=f(\om_0+sg_SB_S )$ with $f(\om)=[e^{\be(\om-\mu)}+1]^{-1}$. 
In the following this subsection, we suppose two leads ($b=L,R$) case. 
 $(\al^m,\al^n)=(B_L,B_S),(\Dl_L,B_S)$ components of the charge and spin BSN curvatures of the lead $L$ are 
\bea
&&\hs{-10mm}F_{B_L,B_S}^{L\up}\pm F_{B_L,B_S}^{L\dw}  \no\\
\aeq  g_S g_L \Gamma_{L}^\pr  [ f^\pr(\om_0+g_S B_S)\pm f^\pr(\om_0-g_S B_S)]    \f{\Gamma_R }{ \Ga_\tot^2}\no\\
&&-g_S g_L \Gamma_{L}^\pr  [ f^\pr(\om_0+g_S B_S)\mp f^\pr(\om_0-g_S B_S)] \no\\
&&\hs{3mm} \times \Big( \Gamma_{L}^\pr(g_S B_S- g_L B_L) \f{2\Gamma_R }{ \Ga_\tot^3} +\Gamma_{R}^\pr(g_S B_S- g_R B_R)\f{\Ga_R-\Ga_L}{\Ga_\tot^3} \Big) \la{BL,BS,c,0} ,\\
&&\hs{-10mm}F_{\Dl_L,B_S}^{L\up} \pm F_{\Dl_L,B_S}^{L\dw} \no\\
\aeq   -g_S [f^\pr(\om_0+g_SB_S)\mp f^\pr(\om_0-g_SB_S)]   \f{ \ga_L \ga_R \Dl_R}{ (\ga_L \Dl_L+\ga_R \Dl_R)^2} \no\\
&&\hs{-10mm}-g_S [f^\pr(\om_0+g_SB_S)\pm f^\pr(\om_0-g_SB_S)] \ga_{L}^\pr (g_S B_S- g_LB_L) 
 \f{ \ga_R \Dl_R-  \ga_L \Dl_L}{ (\ga_L \Dl_L+\ga_R \Dl_R)^2} . \no\\
 \la{DL,BS,c,0} 
\eea
Here $f^\pr(\om)=\f{\partial f(\om)}{\partial  \om}$ and $\Ga_\tot=\Ga_L+\Ga_R$. 
The pumped charge (spin) induced by a slow cycle modulation of $(\al^n,B_S)$ ($\al^n=B_L, \Dl_L$) are given by
\bea
\bra \Dl N_{L\up} \ket\pm \bra \Dl N_{L\dw} \ket 
\aeq \int_{S^n} d\al^n dB_S \ (F_{\al^n,B_S}^{L\up}\pm F_{\al^n,B_S}^{L\dw} ), \la{Pump}
\eea
where $S^n$ are areas enclosed by the trajectories of $(\al^n,B_S)$. 
$F_{\al^n,B_S}^{L\up}\pm F_{\al^n,B_S}^{L\dw}$ ($\al^n=B_L, \Dl_L$) are invariant under the transformation $\ga_b \to c \ga$, $\ga_b^\pr \to c \ga_b^\pr $ (for any $c>0$). 
Hence relevant quantities are $\ga_b^\pr/\Ga_\tot$. The coupling strength $\Ga_\tot$ itself is not important. 
$F_{B_L,B_S}^{L\up}\pm F_{B_L,B_S}^{L\dw}$ are proportional to $g_Sg_L$ and $F_{\Dl_L,B_S}^{L\up} \pm F_{\Dl_L,B_S}^{L\dw}$ are proportional to $g_S$.  
The first terms of the right side of \re{BL,BS,c,0} and \re{DL,BS,c,0} are dominant terms. 
In the limit $\ga_L^\pr \to 0$, $F_{B_L,B_S}^{L\up}\pm F_{B_L,B_S}^{L\dw}$ and the second term of \re{DL,BS,c,0} vanish; however,  
the dominant term of \re{DL,BS,c,0} remains. At $\om_0=\mu$, $f^\pr(\om_0+g_SB_S)- f^\pr(\om_0-g_SB_S)$ vanish. 
Hence, at $\om_0=\mu$, the dominant terms of the spin BSN curvature of $(B_L,B_S)$ pump and 
the charge BSN of $(\Dl_L,B_S)$ pump vanish. 
The contour plots of these BSN curvatures are shown in Figs. \ref{Bs,Bl}(a) and 1(b) and Figs. \ref{Bs,Dl}(a) and 2(b). The details are explained in \res{pump}.

It is important to remark that $(\al^m,\al^n)=(B_L,B_R),(\Dl_L,\Dl_R)$ components of the charge and spin BSN curvatures are zero at zero bias 
because, in \re{F_mn}, $E_s(\al)=f(\om_0+sg_SB_S)$ are independent of $B_{L/R}$ and $\Dl_{L/R}$. 
As we showed in \res{Expansion}, for general model, the pumping dose not occur for all orders of the pumping frequency when $\al_S$ are fixed. 

\subsection{Interacting system} \la{Interacting}

In this section, we study the interacting system \re{H_S,1}.
First, we explain the Liouvillian for $0 \le U \le \infty$ (\res{Liouvillian}). 
Next, the instantaneous steady charge and spin currents are calculated at $U=\infty$ (\res{current,inf}). 
In \res{pump}, we confirm the consistency between our results and \Refe{Splettstoesser12} for $0 \le U \le \infty$.  
The BSN curvatures corresponding to \re{BL,BS,c,0} and \re{DL,BS,c,0} are calculated at $U=\infty$ and differences of the results 
between $U=0$ and $U=\infty$ are discussed (\res{pump}). 
Finally, in \res{pump_finite}, we study the pumping for $0 \le U \le \infty$ in the wide-band limit (i.e., \re{LW} with $\Ga_b^\pr=0$).

\subsubsection{Liouvillian} \la{Liouvillian}

We explain the Liouvillian for $k_{\rm{B}}T > \Ga$, in which the Born-Markov approximation is appropriate. 
The matrix representation of the Liouvillian of the RWA by the number states $\{\ke{n_\up n_\dw} \}$ ($n_s=0,1$ are the numbers of an electron with spin $s=\up,\dw$) 
is a $16 \times 16$ matrix which is block diagonalized
 to the ``diagonal"  space (spanned by  $\{\ke{n_\up n_\dw}\br{n_\up n_\dw}\}_{n_\up,n_\dw=0,1}$) and the ``off-diagonal" space 
(spanned by \\
$\{\ke{n_\up n_\dw}\br{m_\up m_\dw} \}_{(n_\up,n_\dw)\ne (m_\up,m_\dw)}$). The ``diagonal" block is given by
\bea
K^\chi(\al) \aeq  \begin{pmatrix} 
-[\Phi_{\up}^++\Phi_{\dw}^+] &\Phi_{\up}^{-,\chi}&\Phi_{\dw}^{-,\chi}&0\\
\Phi_{\up}^{+,\chi} &-[\Phi_{\up}^-+\phi_{\dw}^+]&0&\phi_{\dw}^{-,\chi}\\ 
\Phi_{\dw}^{+,\chi} &0&-[\Phi_{\dw}^-+\phi_{\up}^+]&\phi_{\up}^{-,\chi}\\ 
0 &\phi_\dw^{+,\chi}&\phi_\up^{+,\chi}&-[\phi_{\up}^-+\phi_{\dw}^-]  \\ 
\end{pmatrix}\begin{matrix} 
\dke{0000} \\
\dke{1010} \\
\dke{0101} \\
\dke{1111} \\
\end{matrix} \hs{2mm},  \la{K_int}
\eea
with 
\bea
\phi_{s}^{\pm,\chi} \aeq 2\pi\sum_b \Dl_b  \sum_{k,\sig}\abs{v_{bk\sig,s}}^2 f_b^\pm(\om_0+sg_SB_S+U) \no\\
&& \times e^{\mp i\chi_{b\sig}} \dl(\ep_{bk}+\sig g_bB_b-\om_0-sg_SB_S-U), 
\eea
and $\Phi_{s}^{\pm,\chi}=\phi_{s}^{\pm,\chi} \vert_{U=0}$. 
$\phi_{s}^{\pm,\chi}$ satisfies 
\bea 
\phi_{s}^{\pm}\aeq \phi_{s}^{\pm,\chi}\bv{\chi=0}=\sum_{b,\sig}\phi_{b\sig,s}^{\pm}, \la{Phi_s} 
\eea
and 
\bea
\f{\phi_{s}^{\pm,\chi}}{\p(i\chi_{b\sig})}\Bv{\chi=0}=\mp \phi_{b\sig,s}^{\pm},
\eea
with 
\bea
\phi_{b\sig,s}^{\pm} \aeq  2\pi \Dl_b  \sum_{k}\abs{v_{bk\sig,s}}^2 f_b^\pm(\om_0+sg_SB_S+U) \no\\
&&\times e^{\mp i\chi_{b\sig}} \dl(\ep_{bk}+\sig g_bB_b-\om_0-sg_SB_S-U).
\eea
The off-diagonal block is a $(12\times12)$-diagonal matrix, which dose not relate to the instantaneous steady state.
At $U=0$, $K^\chi(\al)$ becomes $K_\up^\chi(\al)\otimes 1_\dw+1_\up \otimes K_\dw^\chi(\al)$, where $K_s^\chi(\al)(s=\up,\dw)$ are given by \re{K_U=0} and $1_s$ are identity matrices. 
In the opposite limit $U \to \infty$, $K^\chi(\al)$ reduces to
\bea
K^{\chi(\infty)}(\al)
\aeq  \begin{pmatrix} 
-[\Phi_{\up}^++\Phi_{\dw}^+] &\Phi_{\up}^{-,\chi}&\Phi_{\dw}^{-,\chi}\\
\Phi_{\up}^{+,\chi} &-\Phi_{\up}^-&0\\ 
\Phi_{\dw}^{+,\chi} &0&-\Phi_{\dw}^-\\ 
\end{pmatrix} \begin{matrix} 
\dke{0000} \\
\dke{1010} \\
\dke{0101} \\
\end{matrix} , \la{K_int_inf}
\eea
because the density of states of the leads vanish at high energy ($\phi_{s}^\pm \to 0).$

\subsubsection{Instantaneous steady currents for $U\to \infty$} \la{current,inf}

In this subsection, we set $U=\infty$. The characteristic polynomial of $K^{\chi(\infty)}$ is denoted as
\bea
C_3(\chi,\lm)=\det(K^{\chi(\infty)}-\lm )=\sum_{n=0}^2 C_n(\chi)\lm^n-\lm^3.
\eea 
Because of $C_0(0)=0$, $\lm=0$ is one of the solutions at $\chi=0$.
Now we set $\chi_{b\sig}$ as infinitesimal and other counting fields are zero. 
Then, the eigenvalue corresponding to the instantaneous steady state is given by 
$\lm=\lm_0^\chi=i \chi_{b\sig} \cdot i_{b\sig}^\st+\mathcal{O}(\chi_{b\sig}^2)$. 
It leads to $0=C_3(\chi,\lm_0^\chi)=C_1(0)i\chi_{b\sig} i_{b\sig}^\st + i\chi_{b\sig} C_0^{b\sig}$ with
$C_0^{b\sig}=\f{\partial C_0(\chi)}{\partial (i\chi_{b\sig})} \big \vert _{\chi=0}$, and we obtain
\bea
i_{b\sig}^\st \aeq -\f{C_0^{b\sig}}{C_1(0)} , \la{key_tec}
\eea
with $C_1(0) = -[\Phi_{\up}^+\Phi_{\dw}^-+\Phi_{\up}^-\Phi_{\dw}^++\Phi_{\up}^-\Phi_{\dw}^-]$. 
From $C_0(\chi) = -[\Phi_{\up}^++\Phi_{\dw}^+]\Phi_{\up}^-\Phi_{\dw}^- 
+\Phi_{\dw}^{-,\chi}\Phi_{\up}^-\Phi_{\dw}^{+,\chi} +\Phi_{\dw}^-\Phi_{\up}^{-,\chi}\Phi_{\up}^{+,\chi}$, we have
\bea
i_{b\sig}^\st \aeq \f{\Phi_{\up}^-(\Phi_{b\sig,\dw}^{-}\Phi_{\dw}^{+}-\Phi_{\dw}^{-}\Phi_{b\sig,\dw}^{+} )
+\Phi_{\dw}^-(\Phi_{b\sig,\up}^{-}\Phi_{\up}^{+}-\Phi_{\up}^{-}\Phi_{b\sig,\up}^{+})}
{\Phi_{\up}^+\Phi_{\dw}^-+\Phi_{\up}^-\Phi_{\dw}^++\Phi_{\up}^-\Phi_{\dw}^-}.
\eea
The total instantaneous steady current vanishes:
\bea
\sum_{b,\sig}i_{b\sig}^\st=0. \la{N_hozon_inf}
\eea
$i_{b\sig}^\st$ can be rewritten as
\bea
i_{b\sig}^\st=\f{\sum_{s=\up,\dw}  \Phi_{-s}^-\Gamma_{b\sig,s}\sum_{b^\pr(\ne b)}\Ga_{b^\pr,s}[f_{b^\pr}( \om_s)-f_{b}( \om_s)]}
{\Phi_{\up}^+\Phi_{\dw}^-+\Phi_{\up}^-\Phi_{\dw}^++\Phi_{\up}^-\Phi_{\dw}^-} .
\eea
Here, $\Phi_{-s}^-$ $(s=\up,\dw)$ describes $\Phi_\dw^-$ for $s=\up$ and $\Phi_\dw^-$ for $s=\dw$. 
At zero bias, the instantaneous steady currents vanish. Similar to \res{SC}, 
$\bra \Dl N_{b\sig} \ket^\st_\tau$ are generally nonzero when $\al^{\pr\pr}$ is not fixed at zero bias.

\subsubsection{BSN curvatures for $U\to \infty$} \la{pump}

The instantaneous steady state $\rho_0(\al) $ and corresponding left eigenvector $l_0^\chi(\al)$ are written as 
\bea
\rho_0 = \rho_{0}\ke{00} \br{00} +\rho_{\up}\ke{10} \br{10} +\rho_{\dw}\ke{01} \br{01} +\rho_{2}\ke{11} \br{11} ,
\eea
and
\bea
l_0^\chi=  \ke{00} \br{00} +l_{\up}^\chi \ke{10} \br{10} +l_{\dw}^\chi \ke{01} \br{01} +l_2^\chi \ke{11} \br{11}.
\eea 
The BSN vectors are given by
\bea
A_n^{b\sig}(\al) 
\aeq -\sum_{c=\up,\dw,2}  l_c^{b\sig}(\al) \f{\partial \rho_c(\al) }{\partial \al^n} ,
\eea
where $l_c^{b\sig}(\al) = \f{\partial [l_c^{\chi}(\al)]^\ast} {\partial (i\chi_{b\sig})} \big \vert_{\chi=0}  $. 
It leads to the BSN curvatures
\bea
F_{mn}^{b\sig}(\al) 
\aeq -\sum_{c=\up,\dw,2}  \f{\partial l_c^{b\sig}(\al) }{\partial \al^m}   \f{\partial \rho_c(\al) }{\partial \al^n}  -(m \leftrightarrow n) . \la{F_mn_U_fin}
\eea
We confirmed the consistency between our results and \Refe{Splettstoesser12}, which studied the similar system for $0\le U \le\infty$ using the wide-band limit. 
As we explained in \rec{sFCS-QME}, $\varphi_\ka^{O_\mu}(\al)$ of \re{varphi} corresponds to $-\dbr{l_0^{O_\mu}(\al)}$, namely $-l_c^{b\sig}(\al)$. 
In the condition of the wide-band limit, we calculated $l_c^{b\sig}(\al)$ $(c=\up,\dw,2)$ for $0\le U \le \infty$ and
confirmed numerically the correspondence between $\varphi_c^{O_\mu}(\al)$ ($c=\up,\dw,2$) and $-[l_c^{b\up}(\al)\pm l_c^{b\dw}(\al) ]$ for the charge and spin pump.

Particularly, in the limit $U \to \infty$, $\rho_{2}$ vanishes and $F_{mn}^{b\sig}(\al) $ reduces to
\bea
F_{mn}^{b\sig(\infty)}(\al) 
\aeq -\sum_{s=\up,\dw} \f{\partial l_s^{b\sig(\infty)}(\al) }{\partial \al^m}   \f{\partial \rho_s^{(\infty)}(\al) }{\partial \al^n} -(m \leftrightarrow n) , \la{F_mn_U_inf}
\eea
where $\rho_s^{(\infty)}(\al)$ and $l_s^{b\sig(\infty)}(\al)$ are the limits $U \to \infty$ of $\rho_s(\al)$ and $l_s^{b\sig}(\al)$, respectively. From \re{K_int_inf} we obtain
\bea
\rho_s^{(\infty)}(\al) \aeq \f{ \Phi_s^+\Phi_{-s}^-}{\Phi_{\up}^- \Phi_{\dw}^-+\Phi_{\up}^-\Phi_{\dw}^++\Phi_{\up}^+\Phi_{\dw}^-} ,\\
\{l_s^{(\infty)}(\al)\}^\ast \aeq \f{\Phi_s^{-,\chi}}{\Phi_s^- +\lm_0^\chi} ,
\eea
and
\bea
l_s^{b\sig(\infty)}(\al)
\aeq \f{\Phi_{b\sig,s}^--i_{b\sig}^\st(\al)}{\Phi_s^-} .
\eea
\re{N_hozon_inf} leads $\sum_{b,\sig}l_s^{b\sig(\infty)}(\al)=1$. 
Then, we obtain
\bea
\sum_{b,\sig} A_n^{b\sig(\infty)}(\al) \aeq -\sum_{c=\up,\dw,2}  \f{\partial \rho_c^{(\infty)}(\al) }{\partial \al^n} .
\eea
This equation and \re{N_hozon_inf} lead
\bea
\sum_{b,\sig} \bra \Dl N_{b\sig} \ket \aeq -\sum_{s=\up,\dw}[\rho_c^{(\infty)}(\al_\tau)-\rho_c^{(\infty)}(\al_0)].
\eea
The RHS is $(-1)$ times the change of the  total electron average number of the QD. 
The above equation describes the conservation of the total electron number. 
In the following of this subsection, we fix $\al^{\pr\pr}$ to zero bias ($\be_b=\be$, $\mu_b=\mu$) and suppose \re{LW}. 
Then, $l_s^{b\sig(\infty)}(\al)$ equals $v_s^{b\sig}(\al^\pr)$ given by \re{v_s^bsig} and  $\rho_s^{(\infty)}(\al)$ are given by
\bea
\rho(sB_S) \aeq \f{ e^{-\be(\om_s-\mu)}}{1+ e^{-\be(\om_\dw-\mu)}+e^{-\be(\om_\up-\mu)} }.
\eea
We emphasize that $F_{mn}^{b\sig(\infty)}(\al)$ can be obtained by just a replacement,
\bea
E_s(\al)=f(\om_s) \to \rho(sB_S),
\eea
in \re{F_mn}. 
In the following this subsection, we suppose two leads ($b=L,R$) case. 
The charge and spin BSN curvatures of $(B_L,B_S),(\Dl_L,B_S)$ pump are given by a replacement $f^\pr(\om_0\pm g_S B_S) \to \rho^\pr(\pm B_S)$ in (\ref{BL,BS,c,0}) 
and (\ref{DL,BS,c,0}), where
$\rho^\pr(B_S) \defe \f{1}{g_S }\f{\partial \rho(B_S)}{\partial B_S}  $. 
Similar to $U=0$, the charge and spin BSN curvatures of $(B_L,B_R),(\Dl_L,\Dl_R)$ pump are zero.

\begin{figure*}
\includegraphics[width=16cm]{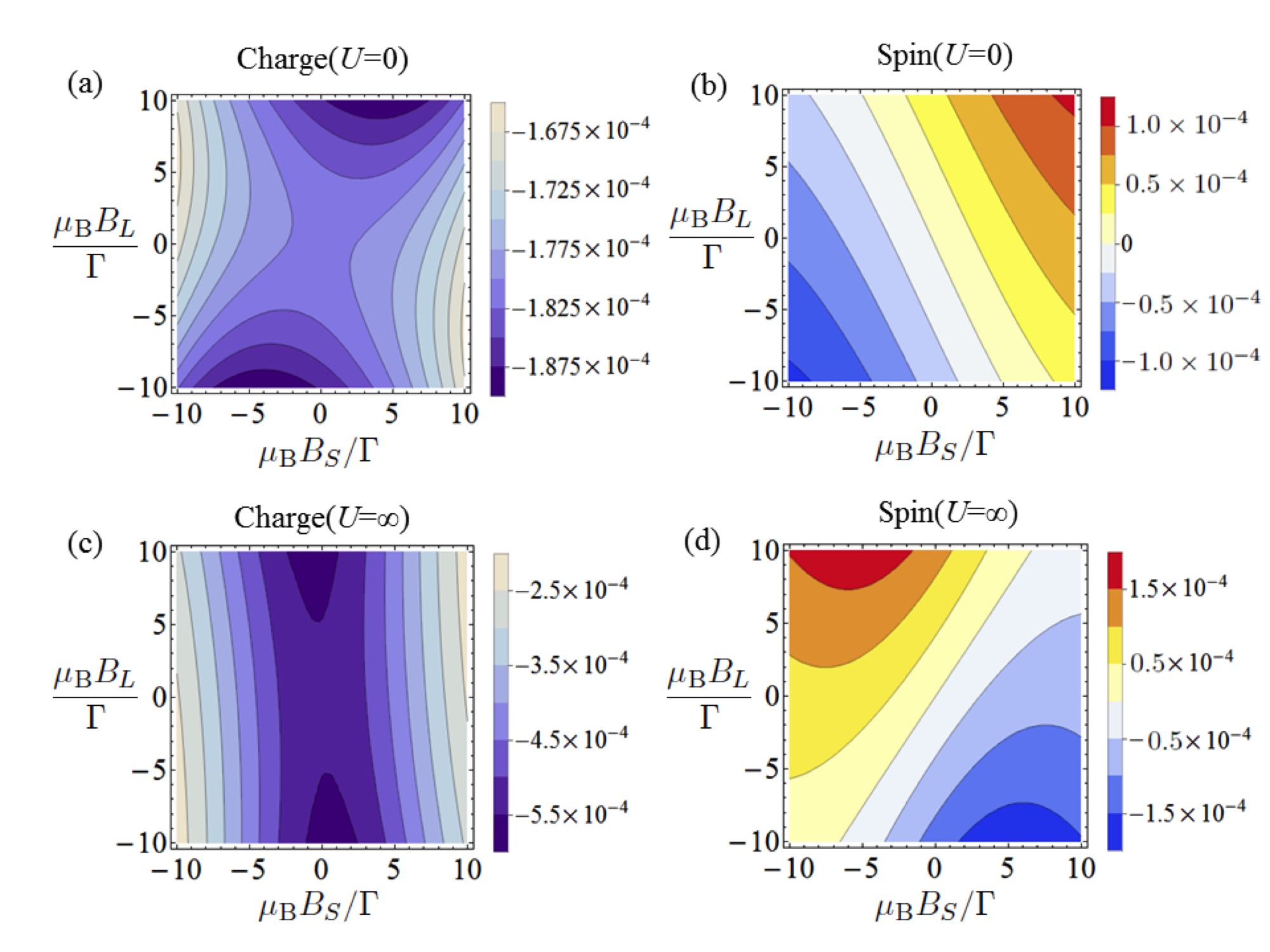}
\caption{\label{Bs,Bl}(a) BSN curvature of charge of $(B_L,B_S)$ pump, $[F_{B_L,B_S}^{L\up}+F_{B_L,B_S}^{L\dw}]/\big(\f{\mu_{\rm{B}}}{\Ga}\big)^2$ at $U=0$, 
(b) the BSN curvature of spin, $[F_{B_L,B_S}^{L\up}-F_{B_L,B_S}^{L\dw}]/\big(\f{\mu_{\rm{B}}}{\Ga}\big)^2$ at $U=0$, 
(c) $[F_{B_L,B_S}^{L\up}+F_{B_L,B_S}^{L\dw}]/\big(\f{\mu_{\rm{B}}}{\Ga}\big)^2$ at $U=\infty$, and
(d) $[F_{B_L,B_S}^{L\up}-F_{B_L,B_S}^{L\dw}]/\big(\f{\mu_{\rm{B}}}{\Ga}\big)^2$ at $U=\infty$. 
The values of the parameters used for these plots are $\Ga_L=\Ga_R=\Ga$, $\Ga_L^\pr=\Ga_R^\pr=0.1$, $\beta=0.5/\Ga$, $\om_0=\mu-3\Ga$, and $B_R=0$, 
and all $g$ factors ($g_L^\ast$, $g_R^\ast$, $g_S^\ast$) are $-0.44$ (bulk GaAs).}
\end{figure*}
\begin{figure*}
\includegraphics[width=16cm]{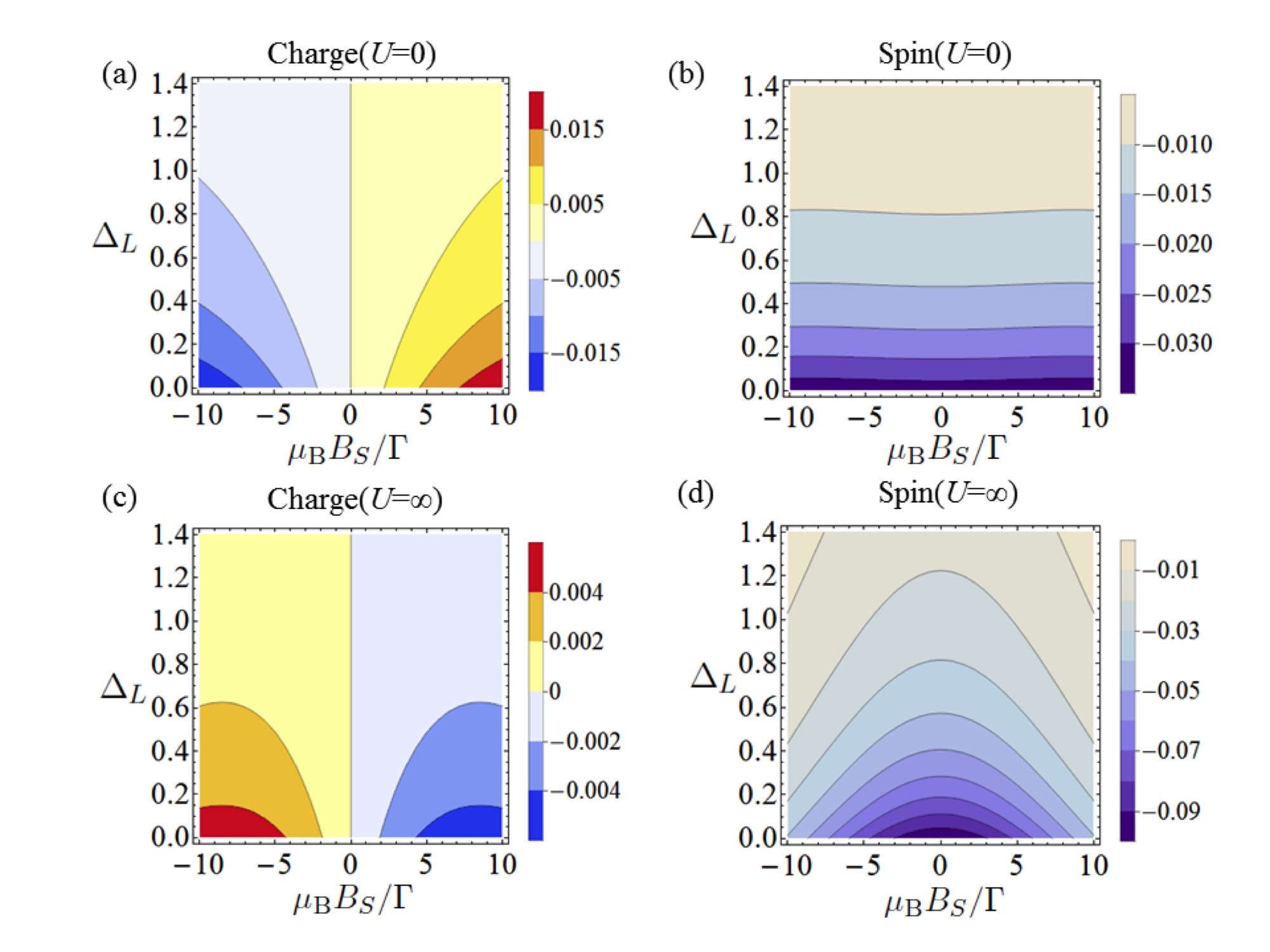}
\caption{\label{Bs,Dl}(a) BSN curvature of charge of $(\Dl_L,B_S)$ pump, $[F_{\Dl_L,B_S}^{L\up}+F_{\Dl_L,B_S}^{L\dw}]/\f{\mu_{\rm{B}}}{\Ga}$ at $U=0$, 
(b) the BSN curvature of spin, $[F_{\Dl_L,B_S}^{L\up}-F_{\Dl_L,B_S}^{L\dw}]/\f{\mu_{\rm{B}}}{\Ga}$ at $U=0$, 
(c) $[F_{\Dl_L,B_S}^{L\up}+F_{\Dl_L,B_S}^{L\dw}]/\f{\mu_{\rm{B}}}{\Ga}$ at $U=\infty$, and 
(d) $[F_{\Dl_L,B_S}^{L\up}-F_{\Dl_L,B_S}^{L\dw}]/\f{\mu_{\rm{B}}}{\Ga}$ at $U=\infty$. 
The values of the parameters used for these plots are $\ga_L=\Ga_R=\Ga$, $\ga_L^\pr=\Ga_R^\pr=0.1$, and $B_L=0$ and other conditions are the same as Fig. \ref{Bs,Bl}.}
\end{figure*}

\begin{figure}
\begin{center}
\includegraphics[width=16cm]{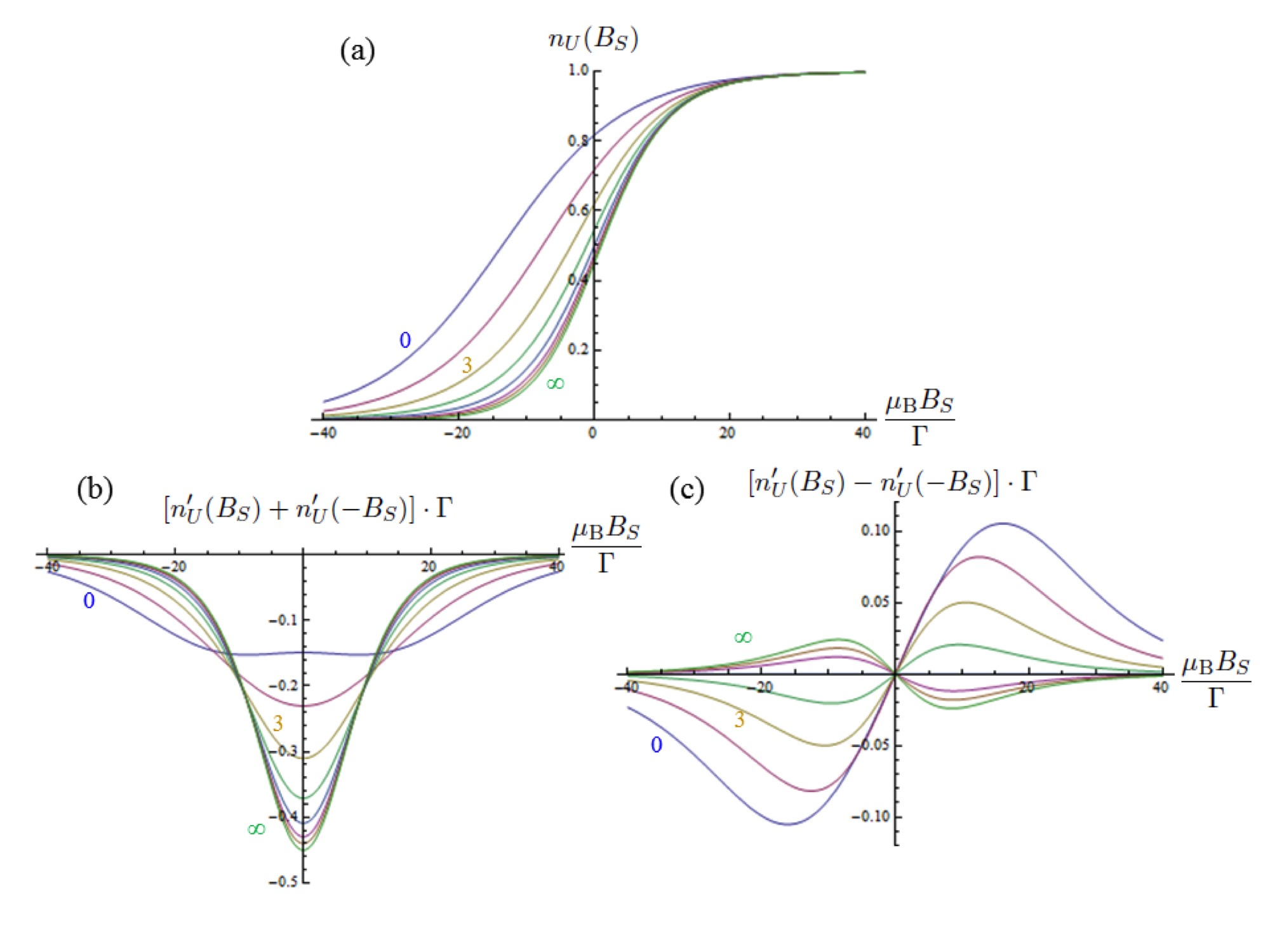}
 \caption{(a)$n_S(B)$, (b)$[n_S^\pr(B)+n_S^\pr(-B)]\cdot\Ga$, (c)$[n_S^\pr(B)-n_S^\pr(-B)]\cdot\Ga$ for $U/\Ga=$0, 1.5, 3, 4,5, 6, 7.5, 9, $\infty$. 
Here, $n^\pr_U(\pm B_S) =\f{1}{g_S}\f{\partial n_U(B)}{\partial B} \vert_{B=\pm B_S}$ and $n_U(B_S)$ is defined by \re{def_n_U}. 
The conditions are the same as Fig. \ref{Bs,Bl}.}
    \label{n_U}
  \end{center}
\end{figure}

In Figs. \ref{Bs,Bl}(a)-\ref{Bs,Bl}(d), we plot the BSN curvatures of $(B_L,B_S)$ pump normalized by $(\mu_{\rm{B}}/\Ga)^2$, where $\Ga=\Ga_L=\Ga_R$ 
and $\mu_{\rm{B}}=57.88$ $\mu$eV/T is the Bohr magneton.
For $U=0$, the charge and spin BSN curvatures  are shown in Fig. \ref{Bs,Bl}(a) and Fig. \ref{Bs,Bl}(b), and for $U=\infty$ these are shown in Figs. \ref{Bs,Bl}(c) and \ref{Bs,Bl}(d).
The horizontal and vertical axes of these plots are the strength of the magnetic fields $B_S$ and $B_L$ normalized by $\Ga/\mu_{\rm{B}}$.
The values of the parameters used for these plots are $\Ga_L=\Ga_R=\Ga$, $\Ga_L^\pr=\Ga_R^\pr=0.1$, $\beta=0.5/\Ga$, $\om_0=\mu-3\Ga$, $B_R=0$, 
and $g_L^\ast=g_R^\ast=g_S^\ast=-0.44$ (bulk GaAs).
The BSN curvatures of $(\Dl_L,B_S)$ pump normalized by $\mu_{\rm{B}}/\Ga$ are shown similarly in Figs. \ref{Bs,Dl}(a)-\ref{Bs,Dl}(d). 
In all plots, $\ga_L=\Ga_R=\Ga$, $\ga_L^\pr=\Ga_R^\pr=0.1$, $B_L=0$, and other conditions are the same as in Fig. \ref{Bs,Bl}. 
In Figs. \ref{Bs,Bl} and \ref{Bs,Dl}, the maximum values of $\abs{\Ga_{b}^\pr (g_S B_S-g_bB_b)}/\Ga_b$ are 0.44 and 0.22 ($<$1), respectively.
The pumped charges and spins are given by \re{Pump}. 

Figure \ref{n_U}(a) shows the instantaneous average numbers of the up spin electron of the QD, $n_U(B_S)$ defined by \re{def_n_U} at $U/\Ga=$0, 1.5, 3, 4.5, 6, 7.5, 9, $\infty$ for 
$\beta=0.5/\Ga$, $\om_0=\mu-3\Ga$, and $g_S=-0.44 \times \mu_{\rm{B}}/2$. In particular, $n_0=f(\om_0+g_SB_S)$ and $n_\infty(B_S)=\rho(B_S)$ hold.  
Because two electrons cannot occupy a QD at $U=\infty$, the magnetic field dependence of $\rho(B_S)$ is more sensitive than $f(\om_0+g_SB_S)$.
Figures \ref{n_U}(b) and \ref{n_U}(c) show $n_U^\pr(B_S)\pm n_U^\pr(-B_S)$ normalized by $1/\Ga$, where $n_U^\pr(\pm B_S) =\f{1}{g_S}\f{\partial n_U(B)}{\partial B} \vert_{B=\pm B_S}$.

In Figs. \ref{Bs,Dl}(a) and \ref{Bs,Dl}(c), the charge BSN curvatures of $(\Dl_L,B_S)$ pump vanish at $B_S=0$. 
This is because the first term of \re{DL,BS,c,0} vanishes since $n^\pr(B_S)-n^\pr(-B_S)=0$ ($n$ denotes $n_0$ or $n_\infty$) for $B_S=0$ and 
the second term vanishes since $g_SB_S-g_bB_b=0$ for $B_S=0=B_L$. Similarly, in Figs. \ref{Bs,Bl}(b) and \ref{Bs,Bl}(d), 
the spin BSN curvatures of $(B_L,B_S)$ pump vanish at $B_S=0=B_L$. 
The zero lines in these plots relate to the cancellation between the first and second terms of \re{BL,BS,c,0}. 
Figures. \ref{Bs,Bl}(a), \ref{Bs,Bl}(c) and Figs. \ref{Bs,Bl}(b), \ref{Bs,Bl}(d) are respectively symmetric and antisymmetric under the transformation $(B_S,B_L) \to (-B_S,-B_L)$. 
Similarly, Figs. \ref{Bs,Dl}(b), \ref{Bs,Dl}(d) and Figs. \ref{Bs,Dl}(a), \ref{Bs,Dl}(c) are respectively symmetric and antisymmetric under the transformation $B_S \to -B_S$. 
We emphasize that pure charge and pure spin pumps are respectively realized for $(B_L,B_S)$ pump and $(\Dl_L,B_S)$ pump such that the areas $S^n$ in \re{Pump} 
are symmetric under the above transformations. 
An instance of symmetric area of $(B_L,B_S)$ pump is a disk of which the center is $B_S=0=B_L$.
 
In $\om_0>\mu$ region, the larger $\om_0-\mu$, the less difference between $U=0$ and $U=\infty$ becomes.
The Coulomb interaction prevents two electrons from occupying the QD. 
This effect is conspicuous in the $\om_0<\mu$ region, although it is not important in the $\om_0>\mu$ region.

As shown in Figs. \ref{Bs,Bl}(a), \ref{Bs,Bl}(c) and Figs. \ref{Bs,Dl}(b), \ref{Bs,Dl}(d), the $B_S$ dependence of the charge BSN curvature of $(B_L,B_S)$ pump 
and the spin BSN curvature of $(\Dl_L,B_S)$ pump at $U=0$
are more gentle than those at $U=\infty$. It results from the behavior of $n^\pr(B_S)+n^\pr(-B_S)$ as shown in Fig. \ref{n_U}(b).

As shown in Figs. \ref{Bs,Bl}(b), \ref{Bs,Bl}(d) and in Figs. \ref{Bs,Dl}(a), \ref{Bs,Dl}(c), the $B_S$ dependence of the spin BSN curvature of $(B_L,B_S)$ pump 
and the charge BSN curvature of $(\Dl_L,B_S)$ pump 
are opposite. This is because the leading term (in weak magnetic field region) of these are proportional to $n^\pr(B_S)- n^\pr(-B_S)$ and 
its $B_S$ dependence is opposite in $U=0$ and $U=\infty$ for $\om_0-\mu<0$ as indicated in Fig. \ref{n_U}(b).
This inversion is realized for only $\om_0-\mu<0$ region. At $\om_0=\mu$, $f^\pr(\om_0+g_SB_S)-f^\pr(\om_0-g_SB_S)$ vanish. 
In $\om_0>\mu$ region, the signs of $f^\pr(\om_0+g_SB_S)-f^\pr(\om_0-g_SB_S)$ and $\rho^\pr(B_S)-\rho^\pr(-B_S)$ are the same. 

In Figs. \ref{Bs,Bl} and \ref{Bs,Dl}, absolute values of the normalized BSN curvatures are smaller than unity. However, we can improve this problem by tuning $g$ factors. 
The first and second terms of the right side of \re{BL,BS,c,0} are the second and third order in the $g$ factors, 
and the first and second terms of the right side of \re{DL,BS,c,0} are the first and second order in the $g$ factors.
If all $g$ factors change to $-20$ (for example for the materials like InAs, InSb), the first, second, and third order terms 
become about $45$, $2\ 000$, and $90\ 000$ times.
In fact, for these values of $g$ factors, the assumption \re{LW} is not appropriate for magnetic fields that are not small; 
we need concrete energy dependence of the line-width functions.

\subsubsection{Instantaneous steady currents} \la{current_finite}

The characteristic polynomial of \re{K_int} is 
\bea
\det(K^{\chi}-\lm )=\sum_{n=0}^3 c_n(\chi)\lm^n+\lm^4.
\eea
Similar to \re{current,inf}, we obtain
\bea
i_{b\sig}^\st \aeq -\f{c_0^{b\sig}}{c_1(0)} \la{i_finite},
\eea
with $c_0^{b\sig}=\f{\partial c_0(\chi)}{\partial (i\chi_{b\sig})} \big \vert _{\chi=0}$. 
$c_0(\chi)$ is given by 
\bea
c_0(\chi) 
 \aeq   K_{00}  K_{\up \up}K_{\dw \dw}K_{22 }-K_{00}K_{\dw \dw} \phi_{\dw}^{-,\chi}\phi_\dw^{+,\chi} -K_{00} K_{\up \up}  \phi_{\up}^{-,\chi} \phi_\up^{+,\chi} \no\\
&&-K_{\dw \dw}K_{22 } \Phi_{\up}^{-,\chi} \Phi_{\up}^{+,\chi}-K_{22 }K_{\up \up}\Phi_{\dw}^{-,\chi} \Phi_{\dw}^{+,\chi} \no\\
&&-\Phi_{\up}^{-,\chi}\Phi_{\dw}^{+,\chi}\phi_{\dw}^{-,\chi}\phi_\up^{+,\chi}+\Phi_{\up}^{-,\chi}\Phi_{\up}^{+,\chi}\phi_{\up}^{-,\chi} \phi_\up^{+,\chi}   \no\\
&&+\Phi_{\dw}^{-,\chi} \Phi_{\dw}^{+,\chi} \phi_{\dw}^{-,\chi}\phi_\dw^{+,\chi} - \Phi_{\dw}^{-,\chi}\Phi_{\up}^{+,\chi}  \phi_{\up}^{-,\chi}\phi_\dw^{+,\chi}.
\eea
Then, we obtain 
\bea
c_0^{b\sig} 
\aeq -K_{00}K_{\dw \dw} [\phi_{b\sig,\dw}^{-}\phi_\dw^+ -\phi_{\dw}^- \phi_{b\sig,\dw}^{+} ] 
-K_{00} K_{\up \up}  [\phi_{b\sig,\up}^{-}  \phi_\up^+ -\phi_{\up}^-  \phi_{b\sig,\up}^{+}] \no\\
&&-K_{\dw \dw}K_{22 } [\Phi_{b\sig,\up}^{-}  \Phi_{\up}^+ -\Phi_{\up}^-  \Phi_{b\sig,\up}^{+}] 
-K_{22 }K_{\up \up} [ \Phi_{b\sig,\dw}^{-}  \Phi_{\dw}^+ -\Phi_{\dw}^-  \Phi_{b\sig,\dw}^{+} ]  \no\\
&&-[\Phi_{b\sig,\up}^{-} \Phi_{\dw}^+-\Phi_{\up}^- \Phi_{b\sig,\dw}^{+}]  \phi_{\dw}^-\phi_\up^+ 
-\Phi_{\up}^- \Phi_{\dw}^+ [\phi_{b\sig,\dw}^{-} \phi_{\up}^+ -\phi_\dw^- \phi_{b\sig,\up}^{+} ] \no\\
&&+[\Phi_{b\sig,\up}^{-} \Phi_{\up}^+ -\Phi_{\up}^- \Phi_{b\sig,\up}^{+}  ]\phi_{\up}^-  \phi_\up^+   
+\Phi_{\up}^- \Phi_{\up}^+[ \phi_{b\sig,\up}^{-}  \phi_\up^+  -\phi_{\up}^-  \phi_{b\sig,\up}^{+}]\no\\
&&+[\Phi_{b\sig,\dw}^{-}  \Phi_{\dw}^+ -\Phi_{\dw}^-  \Phi_{b\sig,\dw}^{+} ]   \phi_{\dw}^-\phi_\dw^+  
+\Phi_{\dw}^-  \Phi_{\dw}^+  [\phi_{b\sig,\dw}^{-} \phi_{\dw}^- -\phi_\dw^+ \phi_{b\sig,\dw}^{+}]  \no\\
&&-[\Phi_{b\sig,\dw}^{-} \Phi_{\up}^+ -\Phi_{\dw}^- \Phi_{b\sig,\up}^{+} ]  \phi_{\up}^- \phi_\dw^+ 
- \Phi_{\dw}^- \Phi_{\up}^+  [\phi_{b\sig,\up}^{-} \phi_\dw^+ -\phi_{\up}^- \phi_{b\sig,\dw}^{+}]. \la{C_0^bsig}
\eea
Here, 
\bea
\phi_{b\sig,s}^{ \pm} \aeq 2\pi \Dl_b  \sum_{k}\abs{v_{bk\sig,s}}^2 f_b^\pm(\om_s+U) \dl(\ep_{bk}+\sig g_bB_b-\om_s-U), 
\eea
and $\Phi_{b\sig,s}^{\pm}=\phi_{b\sig,s}^{\pm}\bv{U=0}$. 
$c_1(0)$ is given by
\bea
c_1(0) 
\aeq -K_{00} K_{\up\up} K_{\dw\dw} -K_{00} K_{\up\up}  K_{22}-K_{00}  K_{\dw\dw} K_{22}- K_{\up\up} K_{\dw\dw} K_{22} \no\\
&&+K_{\dw \dw} (\phi_{\dw}^-\phi_\dw^++\Phi_{\up}^- \Phi_{\up}^+) +K_{00}(\phi_{\dw}^- \phi_\dw^+ +\phi_{\up}^- \phi_\up^+  ) \no\\
&&+K_{\up \up} (\phi_{\up}^-\phi_\up^+ +\Phi_{\dw}^-\Phi_{\dw}^+) +K_{22 }(\Phi_{\up}^- \Phi_{\up}^++\Phi_{\dw}^-\Phi_{\dw}^+) .
\eea
$\sum_{b,\sig}c_0^{b\sig}=0$ leads $\sum_{b,\sig}i_{b\sig}^\st=0$. 
At zero-bias, $i_{b\sig}^\st$ vanishes.  

\subsubsection{BSN curvatures} \la{pump_finite}

$(l_c^\chi)^\ast$ are given by
\bea
(l_\up^\chi)^\ast \aeq \f{[-K_{\dw \dw}+\lm_0^\chi][-K_{00}+\lm_0^\chi]\phi_{\dw}^{+,\chi}+\phi_{\up}^{+,\chi}\Phi_{\up}^{-,\chi}\Phi_{\dw}^{+,\chi}
-\phi_{\dw}^{+,\chi}\Phi_{\dw}^{-,\chi}\Phi_{\dw}^{+,\chi} }
{[-K_{\dw \dw}+\lm_0^\chi]\Phi_{\up}^{+,\chi}\phi_{\dw}^{+,\chi}+[-K_{\up \up}+\lm_0^\chi]\Phi_{\dw}^{+,\chi}\phi_{\up}^{+,\chi} } ,\\
(l_\dw^\chi)^\ast \aeq \f{[-K_{\up \up}+\lm_0^\chi][-K_{00}+\lm_0^\chi]\phi_{\up}^{+,\chi}+\phi_{\dw}^{+,\chi}\Phi_{\dw}^{-,\chi}\Phi_{\up}^{+,\chi}
-\phi_{\up}^{+,\chi}\Phi_{\up}^{-,\chi}\Phi_{\up}^{+,\chi} }
{[-K_{\up \up}+\lm_0^\chi]\Phi_{\dw}^{+,\chi}\phi_{\up}^{+,\chi}+[-K_{\dw \dw}+\lm_0^\chi]\Phi_{\up}^{+,\chi}\phi_{\dw}^{+,\chi} } ,\\
(l_{2}^\chi)^\ast \aeq \f{-\Phi_{\up}^{-,\chi}+[-K_{\up \up}+\lm_0^\chi](l_\up^\chi)^\ast}{\phi_{\dw}^{+,\chi}}. \la{l_2}
\eea
Similarly, we obtain $\rho_c=\rho_0r_c$ $(c=\up,\dw,2)$ with 
\bea
r_\up \aeq \f{K_{\dw \dw}K_{00}\phi_{\dw}^{-}+\phi_{\up}^{-}\Phi_{\up}^{+}\Phi_{\dw}^{-}
-\phi_{\dw}^{-}\Phi_{\dw}^{+}\Phi_{\dw}^{-} }
{-K_{\dw \dw}\Phi_{\up}^{-}\phi_{\dw}^{-}-K_{\up \up}\Phi_{\dw}^{-}\phi_{\up}^{-} } ,\\
r_\dw \aeq \f{K_{\up \up}K_{00}\phi_{\up}^{-}+\phi_{\dw}^{-}\Phi_{\dw}^{+}\Phi_{\up}^{-}
-\phi_{\up}^{-}\Phi_{\up}^{+}\Phi_{\up}^{-} }
{-K_{\up \up}\Phi_{\dw}^{-}\phi_{\up}^{-}-K_{\dw \dw}\Phi_{\up}^{-}\phi_{\dw}^{-} } ,\\
r_2\aeq \f{-\Phi_{\up}^{+}-K_{\up \up}r_\up}{\phi_{\dw}^{-}} ,\\
\rho_0 \aeq \f{1}{1+\sum_{c=\up,\dw,2} r_c}.
\eea
In the following of this subsection, we suppose zero-bias. 
Then, $\rho_0$ and $\rho_c$ become 
\bea
\rho_0 \aeq \f{1}{\Xi},\  \rho_s =\f{e^{-\be(\om_s-\mu)}}{\Xi},\ \rho_2 =\f{e^{-\be(\om_\up+\om_\dw+U-2\mu)}}{\Xi} .
\eea
Here, 
\bea
\Xi \aeq 1+e^{-\be(\om_\up-\mu)}+e^{-\be(\om_\dw-\mu)}+e^{-\be(\om_\up+\om_\dw+U-2\mu)}.
\eea

In the following of this subsection, we suppose that the line-width functions do not depend on the energy. 
Then, we obtain
\bea
l_s^{b\sig} \aeq \f{\Ga_{b\sig,s}}{\Ga}=v_s^{b\sig} ,\\
l_2^{b\sig} \aeq \f{\Ga_{b\sig,\up}+\Ga_{b\sig,\dw}}{\Ga}=l_\up^{b\sig}+l_\dw^{b\sig}.
\eea
Substituting these two equations to \re{F_mn_U_fin}, we obtain
\bea
F_{mn}^{b\sig}(\al) 
\aeq -\sum_{s=\up,\dw}  \Big( \f{\partial  }{\partial \al^m} \f{\Ga_{b\sig,s}}{\Ga} \Big)  \f{\partial n_s(\al) }{\partial \al^n}  -(m \leftrightarrow n). 
\eea
Here,
\bea
n_s \aeq \rho_s+\rho_2 =\f{e^{-\be(\om_s-\mu)}+e^{-\be(2\om_0+U-2\mu)}}{\Xi},
\eea
is the average number of the electrons in the QD with spin $s$.  
Because the line-width functions are energy-independent, the BSN curvatures of $(B_b,B_S)$-pump vanish. 
In the following this subsection, we suppose two leads ($b=L,R$) case. 
If we suppose
\bea
\Ga_{b\sig,s}(\om)=\dl_{\sig,s}\Ga_b=\dl_{\sig,s}\Dl_b\ga_b,
\eea
the BSN curvatures of $(\Dl_L,B_S)$-pump are given by
\bea
F_{\Dl_L,B_S}^{L\up} \pm F_{\Dl_L,B_S}^{L\dw}  \aeq -g_S[n_U^\pr(B_S)\mp n_U^\pr(-B_S)]\f{ \ga_L \ga_R \Dl_R}{ (\ga_L \Dl_L+\ga_R \Dl_R)^2} , \la{Dl_B_finite}
\eea
where 
\bea
n_U(B_S) \aeqd n_\up=\f{e^{-\be(\om_\up-\mu)}+e^{-\be(2\om_0+U-2\mu)}}{1+e^{-\be(\om_\up-\mu)}+e^{-\be(\om_\dw-\mu)}+e^{-\be(2\om_0+U-2\mu)}} \la{def_n_U},\\
n_U^\pr(B_S) \aeqd \f{1}{g_S}\f{\partial n_U(B)}{\partial B} \Bv{B=\pm B_S}.
\eea
Because $n_0=f(\om_0+g_SB_S)$ and $n_\infty(B_S)=\rho(B_S)$, \re{Dl_B_finite} confirms with the results of \res{pump,U=0} and \res{pump}. 
$n_U^\pr(sB_S)$ and $n_U^\pr(B_S) \mp n_U^\pr(-B_S)$ are given by 
\bea
&& n_U^\pr(sB_S) \no\\
\aeq -\be\f{ e^{-\be (\om_0-\mu)} [e^{ -s\be g_SB_S}+e^{ s\be g_SB_S}e^{-\be[2(\om_0-\mu)+ U]} +2e^{-\be (\om_0-\mu)} ]}
{\{1+e^{-\be (\om_0-\mu)}[e^{ \be g_SB_S}+e^{ -\be g_SB_S}]+e^{-\be[2(\om_0-\mu)+ U]} \}^2} ,\\
&&\hs{-10mm}n_U^\pr(B_S) - n_U^\pr(-B_S) \no\\
\aeq \be[1-e^{-\be[2(\om_0-\mu)+ U]} ]\f{ e^{-\be (\om_0-\mu)} (e^{ \be g_SB_S}-e^{ -\be g_SB_S})}
{\{1+e^{-\be (\om_0-\mu)} [e^{ \be g_SB_S}+e^{ -\be g_SB_S}]+e^{-\be[2(\om_0-\mu)+ U]} \}^2} , \no\\
\\
&&\hs{-10mm}n_U^\pr(B_S) +n_U^\pr(-B_S) \no\\
 \aeq -\be \f{ e^{-\be (\om_0-\mu)}[1+e^{-\be[2(\om_0-\mu)+ U]} ] (e^{ \be g_SB_S}+e^{ -\be g_SB_S}) +4e^{-2\be (\om_0-\mu)} }
{\{1+e^{-\be (\om_0-\mu)}[e^{ \be g_SB_S}+e^{ -\be g_SB_S}]+e^{-\be[2(\om_0-\mu)+ U]}\}^2} .
 \eea
In particular, at
\bea
\om_0-\mu \aeq -\f{U}{2}, \la{HFC}
\eea
$n_U^\pr(B_S) - n_U^\pr(-B_S)=0$ and $F_{\Dl_L,B_S}^{L\up}+F_{\Dl_L,B_S}^{L\dw}=0 $ hold. 
\re{HFC} is called the half-filling condition.  
Under this condition, pure spin pump is realized. 
$n_U^\pr(B_S) - n_U^\pr(-B_S)$ is proportional to $F_U\defe 1-e^{-\be[2(\om_0-\mu)+ U]}$.
This factor becomes $F_0=1-e^{-\be[2(\om_0-\mu)]}$ at $U=0$ and $F_\infty=1$ at $U \to \infty$. 
If $\om_0-\mu<0$, $F_0<0$ holds and $n_U^\pr(B_S) - n_U^\pr(-B_S)$ is negative for $0 \le U< -2(\om_0-\mu)$ and 
0 for $U=-2(\om_0-\mu)$ and positive for $U>-2(\om_0-\mu)$.
If $\om_0-\mu>0$, $n_U^\pr(B_S) - n_U^\pr(-B_S)$ is always positive.

We focus on a cyclic pump of an area $\Dl_L^-\le \Dl \le \Dl_L^+$, $B_S^-\le B_S \le B_S^+$. 
The pumped charge and spin are given by 
\bea
&&\hs{-10mm}\bra \Dl N_{L\up} \ket\pm \bra \Dl N_{L\dw} \ket \no\\
\aeq \int_{\Dl_L^-}^{\Dl_L^+}d\Dl_L\int_{B_S^-}^{B_S^+} dB_S \ (F_{\Dl_L,B_S}^{L\up}\pm F_{\Dl_L,B_S}^{L\dw} ) \no\\
\aeq -\int_{\Dl_L^-}^{\Dl_L^+}d\Dl_L\ \f{ \ga_L \ga_R \Dl_R}{ (\ga_L \Dl_L+\ga_R \Dl_R)^2} \int_{B_S^-}^{B_S^+} dB_S \ g_S[n_U^\pr(B_S)\mp n_U^\pr(-B_S)] \no\\
\aeq -\ga_R\Dl_R\Big[ \f{1}{\ga_L\Dl_L^-+\ga_R\Dl_R}-\f{1}{\ga_L\Dl_L^++\ga_R\Dl_R} \Big] \no\\
&&\times[n_U(B_S^+)-n_U(B_S^-)\pm \{n_U(-B_S^+)-n_U(-B_S^-) \} ].
\eea
In particular, if $\ga_L\Dl_L^- \ll \ga_R\Dl_R$, $\ga_L\Dl_L^+ \gg \ga_R\Dl_R$,
\bea
\bra \Dl N_{L\up} \ket\pm \bra \Dl N_{L\dw} \ket \aeq -[n_U(B_S^+)-n_U(B_S^-) \mp \{n_U(-B_S^-)-n_U(-B_S^+) \}],
\eea
holds. 
For instance, if the $g$-factor of the system is negative and $B_S^\pm =\pm \infty$, 
\bea
\bra \Dl N_{L\up} \ket+ \bra \Dl N_{L\dw} \ket\aeq 0, \\
\bra \Dl N_{L\up} \ket - \bra \Dl N_{L\dw} \ket\aeq -2, 
\eea
hold.

\newpage

\section{Quantum diabatic pump} \la{diabatic}

\subsection{Spinless one level quantum dot}

In this section, we consider spinless one level QD coupled to two leads ($b=L,R$).  
$\ke{0}$ ($\ke{1}$) denotes the state that the QD is empty (occupied).  
The diagonal components $p_n=\br{n}\rho\ke{n}$ ($n=0,1$) of the system state $\rho$ are governed by the master equation:
\bea
\f{d}{dt}\begin{pmatrix} 
p_0(t) \\
p_1(t) \\ 
\end{pmatrix} \aeq K(\al_t)\begin{pmatrix} 
p_0(t) \\
p_1(t) \\ 
\end{pmatrix}.
\eea
The Liouvillian is given by
\bea
K \aeq \sum_b \Ga_b \begin{pmatrix} 
-f_b &&1-f_b \\
f_b &&-(1-f_b) 
\end{pmatrix}.
\eea
Here, $\Ga_b$ is the line-width function of the lead $b$, 
$f_b=[e^{\be_b(\ep-\mu_b)}+1]^{-1}$ is the Fermi distribution function, 
$\be_b$ and $\mu_b$ are inverse temperature and chemical potential of the lead $b$, 
$\ep$ is the energy level of the QD. 
The right eigenvectors of the Liouvillian are the instantaneous steady state 
\bea
p^\st(\al)=\begin{pmatrix} 
p_0^\st(\al) \\
p_1^\st(\al) \\ 
\end{pmatrix}=\begin{pmatrix} 
1-F(\al) \\
F(\al) \\ 
\end{pmatrix} ,
\eea
and 
\bea
\begin{pmatrix} 
-1 \\
1 \\ 
\end{pmatrix} , \la{rev2}
\eea
 with the eigenvalue $(-\Ga)$. Here, 
\bea
F(\al) \defe \f{\sum_b \Ga_b f_b}{\Ga},\ \Ga\defe \sum_b \Ga_b.
\eea
As a specialty of this model, \re{rev2} is time-independent. 
We introduce 
$p_n^{(m)}=\br{n} \rho^{(m)}\ke{n}$ ($m=1,2,\cdots$) and 
$\tl p_n^{(m)}=\br{n}\tl \rho^{(m)}\ke{n}$ ($m=0,1,\cdots$). 
$\tl p_n^{(0)}$ are given by
\bea
\begin{pmatrix} 
\tl p_0^{(0)}(\al_t) \\
\tl p_1^{(1)}(\al_t)
\end{pmatrix} \aeq e^{-\int_0^t ds \ \Ga(s)}[p(0)-p^\st(\al_0)]=e^{-\int_0^t ds \ \Ga(s)}\begin{pmatrix} 
-p_1(0)+F(\al_0) \\
p_1(0)-F(\al_0)
\end{pmatrix}. 
\eea
We suppose $p(0)={}^t(p_0(0),p_1(0))=p^\st(\al_0)$.
Then, $\tl p_n^{(0)}=0$ holds. 
We choose the $\tl K^{-1}(\al)$ of \re{RR} as the pseudo-inverse of $K(\al)$.
We have
\bea
\tl K^{-1}=\f{1}{\Ga}\begin{pmatrix} 
0&1\\
1&0\\ 
\end{pmatrix} .
\eea
From $\rho^{(n)}(t)=[ \mR(\al_t) \f{d}{dt} ]^n \rho_0(\al_t)$, 
we obtain $p^{(n)}={}^t(p_0^{(n)},p_1^{(n)})$ as
\bea
p^{(1)}(\al_t) \aeq \tl K^{-1}(\al_t)\f{d}{dt}p^{\st}(\al_t) \no\\
\aeq \f{1}{\Ga(t)} \begin{pmatrix} 
0&1\\
1&0\\ 
\end{pmatrix} \f{d}{dt}\begin{pmatrix} 
1-F(\al_t)\\
F(\al_t)\\ 
\end{pmatrix} \no\\
\aeq -\f{1}{\Ga(t)}\f{d}{dt} \begin{pmatrix} 
-F(\al_t) \\
F(\al_t)
\end{pmatrix} ,
\eea
and 
\bea
p^{(n+1)}(t) \aeq \tl K^{-1}(\al_t)\f{d}{dt}p^{(n)}(t) \no\\
\aeq -\f{1}{\Ga(t)} \f{d}{dt}\begin{pmatrix} 
-p_1^{(n)}(t) \\
p_1^{(n)}(t)
\end{pmatrix} .
\eea
$\tl p^{(n)}={}^t(-\tl p_1^{(n)},\tl p_1^{(n)})$ ($n=1,2,\cdots$) is given by 
\bea
\tl p_1^{(n)}(t) \aeq - e^{-\int_0^t ds \ \Ga(s)}p_1^{(n)}(0) .
\eea
For by only modulating $\Ga_b$ at zero-bias, the pump dose not occur for all orders ($p^{(n)}(t)=0$)
because $p^\st(\al)$ dose not change.

We consider the particle current to the lead $b$.
From discussion of \res{pump,U=0}, we obtain
\bea
\dbr{l_0^{N_b}}=(0,\f{\Ga_b}{\Ga}).
\eea
Then, we get
\bea
i_{N_b}^{(1)} \aeq -\dbr{l_0^b}\f{d}{dt}p^\st(\al_t) = -\f{\Ga_b(t)}{\Ga(t)} \f{d}{dt}F(t) ,
\eea
and 
\bea
i_{N_b}^{(n+1)} \aeq -\f{\Ga_b(t)}{\Ga(t)} \f{d}{dt}p_1^{(n)}(t) .
\eea
$\tl i_{N_b}^{(n)}$ is given by
\bea
\tl i_{N_b}^{(n)} \aeq -\f{\Ga_b(t)}{\Ga(t)} \f{d}{dt}\tl p_1^{(n)}(t) \no\\
\aeq -\f{\Ga_b(t)}{\Ga(t)} [K(\al_t)\tl p^{(n)}(t)]_1 \no\\
\aeq -\f{\Ga_b(t)}{\Ga(t)} \Big[ \Ga(t) \begin{pmatrix} 
-F(\al_t)&1-F(\al_t)\\
F(\al_t)&-1+F(\al_t) \\ 
\end{pmatrix} \begin{pmatrix} 
-\tl p_1^{(n)}(t)\\
\tl p_1^{(n)}(t) \\ 
\end{pmatrix} \Big]_1 \no\\
\aeq -\f{\Ga_b(t)}{\Ga(t)} \Big[ -\Ga(t) \begin{pmatrix} 
-\tl p_1^{(n)}(t)\\
\tl p_1^{(n)}(t) \\ 
\end{pmatrix} \Big]_1 \no\\
\aeq \Ga_b(t)\tl p_1^{(n)}(\al_t) \no\\
\aeq -\Ga_b(t)e^{-\int_0^t ds \ \Ga(s)} p_1^{(n)}(0). 
\eea

\subsection{Numerical calculation}

We set the time-dependence of the control parameters as 
\bea
\Ga(t)\aeq \Ga_L(t)+\Ga_R ,\ \Ga_L(t)=\ga[1+g\sin \om (t+\dl))],\ \Ga_R=\ga ,\\
f_L(t) \aeq f_R(t)=f(t)=\f{1}{e^{\be(\ep(t)-\mu)}+1} ,\ \ep(t)-\mu=\ep_0 \sin \om t.
\eea
For the numerical calculation, we set
\bea
g=0.5, \ \om=0.3\ga,\ \be \ep_0=1 ,\ \dl=0,\f{\pi}{2}.
\eea
$\Ga$ of \re{Ga_G} is given by $\ga(2-g)=1.5\ga$. 
Then, 
\bea
\f{\om}{\Ga}=\f{0.3}{1.5}=0.2,
\eea
holds. 

\begin{figure}
\begin{center}
\includegraphics[width=16cm]{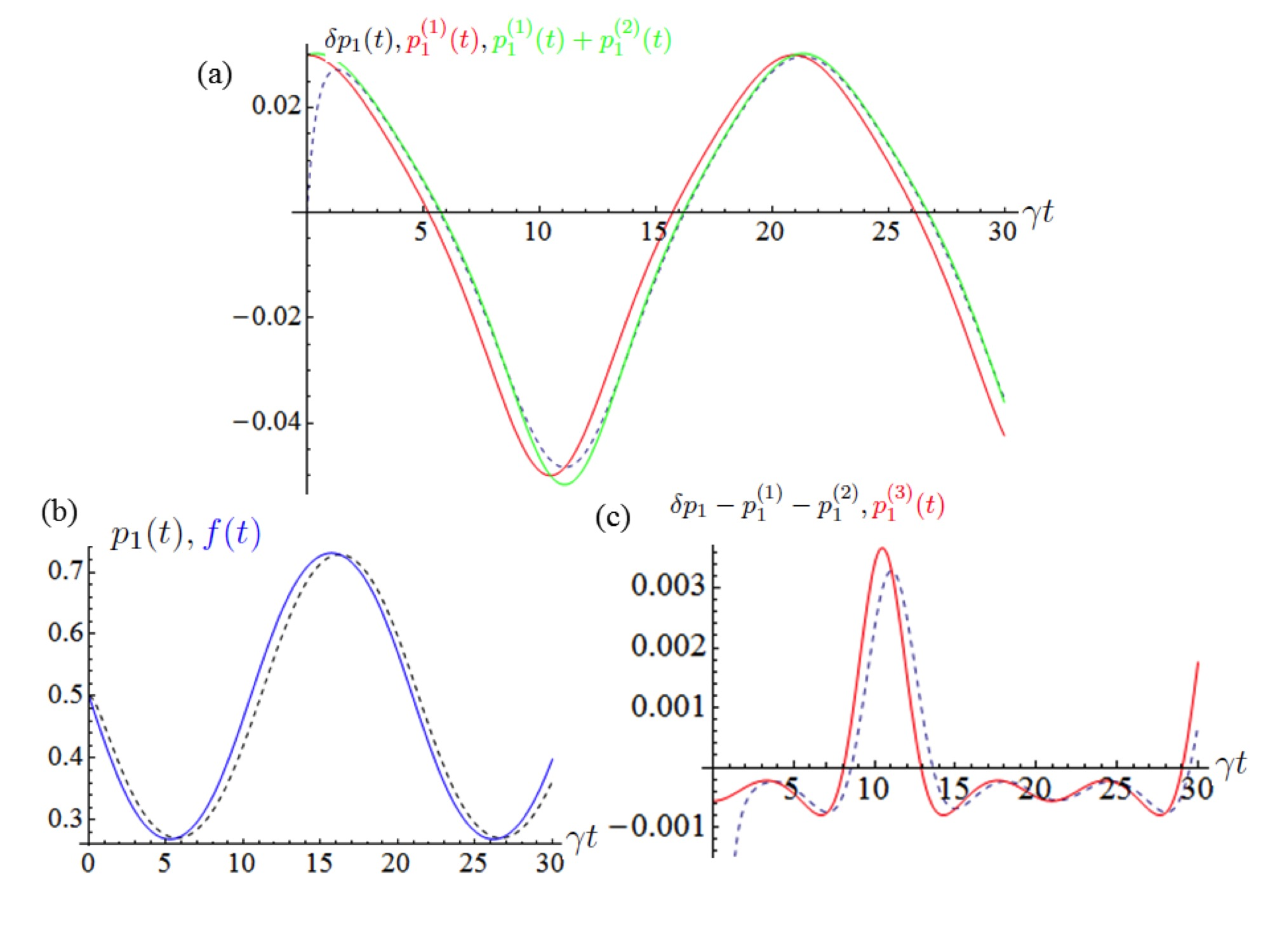}
 \caption{(a)$\dl p_1(t)\defe p_1(t)-f(t)$(dashed line), $p_1^{(1)}(t)$(red line) and $p_1^{(1)}(t)+p_1^{(2)}(t)$, 
(b)$p_1(t)$(dashed line) and $f(t)$, (c)$\dl p_1- p_1^{(1)}-p_1^{(2)}$(dashed line) and $p_1^{(3)}(t)$ for $\dl=\pi/2$. }
    \label{napcs}
  \end{center}
\end{figure}

For $\dl=\pi/2$, the pumped particle numbers of the first one cyclic are given by
\bea
\bra \Dl N_L \ket \aeq 7.69464 \times 10^{-2} ,\no\\
\bra \Dl N_L \ket^{\rm{BSN}}+\sum_{n=1}^5[\widetilde{ \bra \Dl N_L \ket}^{(n)}+\bra \Dl N_L \ket^{(n+1)}] \aeq 7.69583 \times 10^{-2} ,\no\\
\bra \Dl N_L \ket^{\rm{BSN}} \aeq 9.71762 \times 10^{-2},\no\\
\widetilde{ \bra \Dl N_L \ket}^{(1)} \aeq -1.79649\times 10^{-2} ,\no\\
\bra \Dl N_L \ket^{(2)} \aeq 0  ,\no\\
\widetilde{ \bra \Dl N_L \ket}^{(2)} \aeq 0 ,\no\\
\bra \Dl N_L \ket^{(3)} \aeq -0.270724\times 10^{-2},\no\\
\widetilde{ \bra \Dl N_L \ket}^{(3)} \aeq 0.0336304 \times 10^{-2},\no\\
\bra \Dl N_L \ket^{(4)} \aeq 0 ,\no\\
\widetilde{ \bra \Dl N_L \ket}^{(4)} \aeq 0 ,\no\\
\bra \Dl N_L \ket^{(5)} \aeq 0.0133644\times 10^{-2},\no\\
\widetilde{ \bra \Dl N_L \ket}^{(5)} \aeq -0.00156459 \times 10^{-2} ,\no\\
\bra \Dl N_L \ket^{(6)} \aeq 0.
\eea
Figure \ref{napcs}(b) shows that $p_1(t)$ and $f(t)$, 
Fig.\ref{napcs}(a) shows that $\dl p_1(t)\defe p_1(t)-f(t)$, $p_1^{(1)}(t)$ and $p_1^{(1)}(t)+p_1^{(2)}(t)$,
and Fig.\ref{napcs}(c) shows that $\dl p_1- p_1^{(1)}-p_1^{(2)}$ and $p_1^{(3)}(t)$.

\begin{figure}
\begin{center}
\includegraphics[width=16cm]{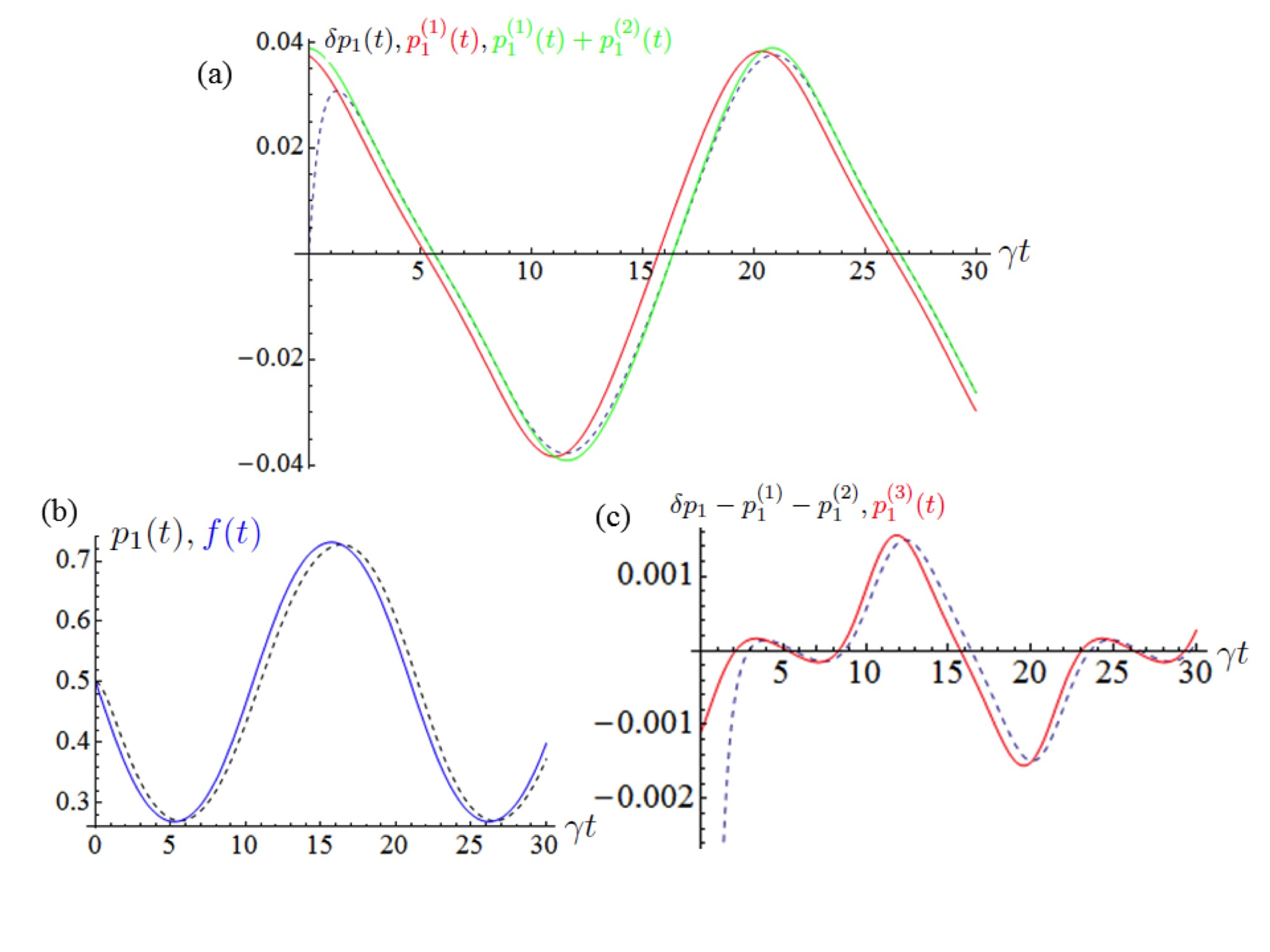}
 \caption{(a)$\dl p_1(t)= p_1(t)-f(t)$(dashed line), $p_1^{(1)}(t)$(red line) and $p_1^{(1)}(t)+p_1^{(2)}(t)$, 
(b)$p_1(t)$(dashed line) and $f(t)$, (c)$\dl p_1- p_1^{(1)}-p_1^{(2)}$(dashed line) and $p_1^{(3)}(t)$ for $\dl=0$. }
    \label{napss}
  \end{center}
\end{figure}

For $\dl=0$, the pumped particle numbers of the first one cyclic are given by
\bea
\bra \Dl N_L \ket \aeq -0.466997 \times 10^{-2} ,\no\\
\bra \Dl N_L \ket^{\rm{BSN}}+\sum_{n=1}^5[\widetilde{ \bra \Dl N_L \ket}^{(n)}+\bra \Dl N_L \ket^{(n+1)}] \aeq -0.464558 \times 10^{-2} ,\no\\
\bra \Dl N_L \ket^{\rm{BSN}} \aeq 0,\no\\
\widetilde{ \bra \Dl N_L \ket}^{(1)} \aeq -1.9376 \times 10^{-2} ,\no\\
\bra \Dl N_L \ket^{(2)} \aeq 1.52006 \times 10^{-2} ,\no\\
\widetilde{ \bra \Dl N_L \ket}^{(2)} \aeq -0.0726599 \times 10^{-2} ,\no\\
\bra \Dl N_L \ket^{(3)} \aeq 0,\no\\
\widetilde{ \bra \Dl N_L \ket}^{(3)} \aeq 0.0572197 \times 10^{-2},\no\\
\bra \Dl N_L \ket^{(4)} \aeq -0.0462914 \times 10^{-2} ,\no\\
\widetilde{ \bra \Dl N_L \ket}^{(4)} \aeq 0.0148158 \times 10^{-2} ,\no\\
\bra \Dl N_L \ket^{(5)} \aeq 0,\no\\
\widetilde{ \bra \Dl N_L \ket}^{(5)} \aeq -0.00221088 \times 10^{-2} ,\no\\
\bra \Dl N_L \ket^{(6)} \aeq 0.00210926 \times 10^{-2} .
\eea
Figure \ref{napss}(b) shows that $p_1(t)$ and $f(t)$, 
Fig.\ref{napss}(a) shows that $\dl p_1(t)= p_1(t)-f(t)$, $p_1^{(1)}(t)$ and $p_1^{(1)}(t)+p_1^{(2)}(t)$,
and Fig.\ref{napss}(c) shows that $\dl p_1- p_1^{(1)}-p_1^{(2)}$ and $p_1^{(3)}(t)$.

\newpage

\section{Generalized quantum master equation for entropy production} \la{sGQME}

\subsection{Definition of entropy production}

In this chapter and \rec{GEEE} and \rec{Other_def}, 
we suppose that $\{H_b\}_b$ are time-independent. 

It is natural to identify the average entropy production rate with
\bea
\dot{\sig}(t)\defe \sum_{b\in\RM{\bm{\mC}}} \be_b(t)[-i_{H_b}(t)]+\sum_{b\in\RM{\bm{\mG}}} \be_b(t)[-i_{H_b}(t)-\mu_b(t)\{ -i_{N_b}(t)\}] .
\eea
This is given by $\dot{\sig}(t)=\tr_S[W^\sig (\al_t)\rho(t)]$ with 
\bea
W^\sig (\al) \defe \sum_{b\in\RM{\bm{\mC}}} \be_b[-W^{H_b}(\al)] + \sum_{b\in\RM{\bm{\mG}}} \be_b[-W^{H_b}(\al)-\mu_b\{ -W^{N_b}(\al)\}]  \la{def_W^sig} .
\eea
The average entropy production is given by
\bea
\sig \aeqd \int_0^\tau dt\ \dot{\sig}(t) \no\\
\aeq \int_0^\tau dt \ J_\sig^\st(\al_t)+\int_C \ d\al^n \ A_n^\sig(\al) +\mO\big(\f{\om}{\Ga}\big)\la{def_sig},
\eea
where 
\bea
J_\sig^\st(\al)\defe \sum_{b\in\RM{\bm{\mC}}}\be_b[-i^\st_{H_b}(\al)]+\sum_{b\in\RM{\bm{\mG}}}\be_b[-i^\st_{H_b}(\al)-\mu_b\{-i^\st_{N_b}(\al)\}] ,
\eea
and 
\bea
A_n^\sig(\al)\defe \sum_{b\in\RM{\bm{\mC}}} \be_b[-A_n^{H_b}(\al)]+\sum_{b\in\RM{\bm{\mG}}} \be_b[-A_n^{H_b}(\al)-\mu_b\{-A_n^{N_b}(\al)\}] .
\eea
Here, we used \re{i^O} for $\{O_\mu\}=\{H_b\}_b+\{N_b\}_{b\in \RM{\bm{\mG}}}$.  
The excess entropy production is defined by 
\bea
\sig_\RM{ex} \defe \sig-\int_0^\tau dt \ J_\sig^\st(\al_t) = \int_C \ d\al^n \ A_n^\sig(\al)+\mO\big(\f{\om}{\Ga}\big) .
\eea
While we can calculate the average of the entropy production, 
our formalism is not compatible to discuss the higher moments of the entropy production.  
Although \re{Dl o} is the average of the difference between outcomes at $t=\tau$ and $t=0$ of $O_\mu$, 
$\sig$ is not that of some bath's operator if $\al^{\pr\pr}$ are modulated.

\subsection{Introduction of generalized QME} 

We consider a kind of generalized quantum master equation (GQME) 
\bea
\f{d}{dt}\rho^\lm(t) = \mK^\lm(\al_t)\rho^\lm(t) ,\la{GQME}
\eea
with the initial condition $\rho^\lm(0)=\rho(0)$. 
Here, $\lm$ is a single real parameter.
We suppose that the Liouvillian is given by 
\bea
\mK^\lm(\al)\bu=-i[H_S(\al_S),\bu]+\sum_b \mL_b^\lm(\al) \bu,
\eea
 with 
\bea
\mL_b^\lm (\al)\bu=\sum_a c_{ba}^\lm(\al) A_a \bu B_a ,
\eea
and 
\bea
c_{ba}^\lm(\al)\bv{\lm=0}=c_{ba}.
\eea 
While $c_{ba}^\chi(\al)$ of \re{mL_b} depend on $\chi$ if and only if $A_a, B_a \ne 1$, 
$c_{ba}^\lm(\al)$ can depend on $\lm$ for all $a$. 
We suppose that the solution of \re{GQME} satisfies
\bea
\tr_S[\rho^\pr(\tau)] = \sig ,\la{G_def} 
\eea
where $X^\pr \defe \f{\p X^\lm}{\p (i\lm)}\Bv{\lm=0}$. 
This condition is equivalent to 
\bea
\dbr{1}\mK^\pr(\al) = \dbr{1}W^\sig(\al). \la{G_con}
\eea
Let's consider 
\bea 
\dbr{l_0^\lm(\al)}\mK^\lm(\al)=\lm_0^\lm(\al)\dbr{l_0^\lm(\al)} \la{G_l_0} ,
\eea
corresponding to \re{left} for $n=0$.
Similar to \re{touka_sc} and \re{BSN}, 
\bea
\lm_0^\pr(\al) \aeq \dbr{1}W^\sig(\al)\dke{\rho_0(\al)}=J^\st_\sig(\al), \la{lm_0^pr}
\eea
and 
\bea
A_n^\sig(\al) \aeq -\dbr{l_0^\pr (\al)}\f{\p }{\p \al^n}\dke{\rho_0(\al)}=\dbr{1}W^\sig(\al)\mR(\al)\f{\p }{\p \al^n}\dke{\rho_0(\al)}, \la{A_n^sig_F}
\eea
hold. 
Although $\lm_0^\lm(\al)$ and $l_0^\lm(\al)$ depend on the choice of $\mK^\lm(\al)$, $\lm_0^\pr(\al)$ and $A_n^\sig(\al)$ do not depend, 
as can be seen in the RHS of the \re{lm_0^pr} and \re{A_n^sig_F}.
The LHS of \re{G_con} is given by
\bea
\dbr{1}\mK^\pr(\al) = \dbr{1} \sum_{b,a} c_{ba}^\pr(\al) B_a A_a .
\eea
Using this and \re{W^O}, \re{G_con} becomes
\bea
\sum_{b,a} c_{ba}^\pr(\al) B_a A_a = \sum_{a} [-\sum_b \be_b c_{ba}^{H_b}(\al)+\sum_{b\in\RM{\bm{\mG}}} \be_b \mu_b c_{ba}^{N_b}(\al) ] B_a A_a .
\eea
Infinite solutions of this equation exist. 
One choice of $\mK^\lm(\al)$ satisfying this relation is $\chi_{H_b} \to -\be_b \lm$ (for all $b$) and 
$\chi_{N_b} \to \be_b\mu_b \lm$ (for $b \in \RM{\bm{\mG}}$) limit of $\hat{K}^\chi(\al)$. 

``Higher moments'' $\f{\p^n}{\p (i\lm)^n}\tr_S[\rho^\lm(\tau)]\bv{\lm=0}$ ($n=2,3,\cdots$) depend on the choice of $\mK^\lm(\al)$ and seems have no physical meening.
In contrast, the higher moments of the entropy production could be considered for the classical Markov jump process.
In Appendix \ref{Appendix B}, we review the entropy production of the Markov jump process \cite{Komatsu15, Jarzynski}, and in \rec{Other_def}, we compare that and \re{def_sig}. 

\subsection{Current operators}

The particle and energy current operators from the system into bath $b$, $w^{N_b}(\al)$ and $w^{H_b}(\al)$, 
are usually defined by 
\bea
w^{X_b}(\al) \defe -[\mL_b\dg(\al) X_S]\dg=-\mL_b\dg(\al) X_S \ (X=N,H). \la{def_w}
\eea
For a super-operator $\mJ$, $\mJ\dg$ is defined by $\dbr{\mJ\dg X}Y \dket=\dbra X \dke{\mJ Y} $ ($X,Y \in \RM{\bm{B}}$).
\bea
\mL_b\dg(\al) \bu =\sum_a c_{ba}^\ast(\al) A_a\dg \bu B_a\dg ,
\eea
holds. 
$w^{X_b}(\al)$ is a Hermitian operator and is given by 
\bea
w^{X_b}(\al) =- \sum_a c_{ba}(\al) B_a X_S A_a \ (X=N,H) . \la{W_A_B}
\eea
In general, for the RWA, 
\bea
w^{H_b}=W^{H_b}(\al)= \sum_{\om } \sum_{\mu,\nu} \om \Phi_{b,\mu \nu}(\om) [s_{b\mu}  (\om)]\dg s_{b\nu}(\om),
\eea
holds (Appendix \ref{Current_op}).
For the Born-Markov approximation and the CGA, $w^{H_b}(\al)\ne W^{H_b}(\al)$. 
From \re{s_N_S}, \re{h_bN_S} and \re{PidgN_S}, 
\bea
w^{N_b}(\al)=0 \ (b \in \RM{\bm{\mC}}),
\eea
holds for the RWA, the Born-Markov approximation, and the CGA.  
In the following, we set
\bea
W^{N_b}(\al) \defe w^{N_b}(\al)=0 \ (b \in \RM{\bm{\mC}}),
\eea
and 
\bea
i_{N_b}^\st(\al) \defe \tr_S[w^{N_b}(\al)\rho_0(\al)]=0 \ (b \in \RM{\bm{\mC}}).
\eea
Here, we suppose \re{H_Sb} for $b \in \RM{\bm{\mG}}$. 
The generalization to \re{H_SbG} case is straightforward. 
For $\{ O_\mu \}=\{N_b\}_{b \in \RM{\bm{\mG}}}+\{H_b \}_b$, \re{Phi_chi} holds in \re{Pi_RWA}. 
For the Born-Markov approximation and the CGA, $w^{N_b}(\al)=W^{N_b}$, however, $w^{H_b}(\al)\ne W^{H_b}(\al)$. 
For the RWA, 
\bea
&&\hs{-10mm}w^{N_b}(\al)= W^{N_b}(\al) \no\\
\aeq \sum_{\om}\sum_{\al,\beta}  \Big\{   \Phi_{b,\al\beta }^-(\om) [a_{\al }(\om)] \dg a_{\beta }(\om) 
-   \Phi_{b,\al \beta}^+(\om)a_{\al }  (\om)  [a_{\beta } (\om)]\dg  \Big\} \ (b \in \RM{\bm{\mG}}),\la{w=W} \\
&&\hs{-10mm} w^{H_b}(\al)= W^{H_b}(\al) \no\\
\aeq \sum_{\om}\sum_{\al,\beta}  \Big\{  \om \Phi_{b,\al \beta }^-(\om) [a_{\al }(\om)] \dg a_{\beta }(\om) 
-  \om \Phi_{b,\al \beta }^+(\om)a_{\al }  (\om)  [a_{\beta } (\om)]\dg  \Big\}\ (b \in \RM{\bm{\mG}}), \la{w=W2}
\eea
hold. 
Therefore, \re{def_W^sig} and \re{def_w} imply that
$W^\sig(\al)$ is given by 
\bea
W^\sig(\al) = \sum_b \mL_b\dg(\al) (\be_b H_S-\be_b \nu_b N_S)=\sum_b \Pi_b\dg(\al) (\be_b H_S-\be_b \nu_b N_S) .\la{W^sig}
\eea
Here, 
\bea
\nu_b \defe  \left \{ \begin{array}{ll}
\mu_b^\pr \hs{3mm} b \in \RM{\bm{\mC}} \\
\mu_b \hs{3.2mm} b \in \RM{\bm{\mG}}
\end{array} \right. .
\eea
$\mu_b^\pr$ is an arbitrary real number.

\newpage

\section{Geometrical expression of excess entropy production} \la{GEEE}

In this chapter and \rec{Other_def}, we focus on the RWA. 
We use
\bea
W^{H_b}\aeq w^{H_b}\ (b \in \RM{\bm{\mC}}) ,\\
W^{H_b}\aeq w^{H_b},\  W^{N_b}= w^{N_b}\ (b \in \RM{\bm{\mG}}), \la{rqG}
\eea
and 
\bea
\Pi_b(\bu e^{-\be_b(H_S-\nu_bN_S)})=(\Pi_b\dg \bu)e^{-\be_b(H_S-\nu_bN_S)}, \la{KMS2}
\eea
and
\bea
[h_b(\al),N_S] =0 .\la{H_L_N_S=0} 
\eea
If we suppose \re{s_N_S} for $b \in \RM{\bm{\mC}}$ and \re{H_Sb} or \re{H_SbG} for 
$b \in \RM{\bm{\mG}}$, these relations hold. 
If $n_\RM{GC}=0$, existence of $N_S$, \re{s_N_S}, \re{rqG} and \re{H_L_N_S=0} are not required and the system $S$ 
does not have to be described by the annihilation and creation operators 
($S$ can be spin chain or few level system, etc.). 
Using $\mL_b \dg 1=\Pi_b \dg 1=0$ (see \re{Pidg1=0}) for \re{KMS2} with $\bu=1$, we obtain
\bea
\Pi_b e^{-\be_b(H_S-\nu_bN_S)}=\mL_b e^{-\be_b(H_S-\nu_bN_S)}=0. \la{rho_ss}
\eea
Here, we used \re{H_L_H_S=0} and \re{H_L_N_S=0}.

\subsection{Equilibrium state} \la{eq}

In this section, we consider equilibrium state $\be_b=\be$ (for all $b$) and $\mu_b=\mu$ $(b \in \RM{\bm{\mG}})$, 
and $\al$ denotes the set of $(\al_S$, $\{\al_{Sb}\}_b$, $\be$, $\be \mu)$.
We show that $A_n^\sig(\al)$ is a total derivative of the von Neumann entropy of the instantaneous steady state.
Differentiating \re{G_l_0} by $\lm$, we obtain
\bea
\dbr{l_0^\pr(\al)}\hat{K}(\al)+\dbr{1}\mK^\pr(\al) = \lm_0^\pr(\al)\dbr{1}. \la{ee}
\eea
In the RHS, $\lm_0^\pr(\al)=J^\st_\sig(\al)=0$ holds. 
The second term of the LHS is $\dbr{1}W^\sig(\al)$.
\re{W^sig} leads 
\bea
W^\sig(\al) = \be \sum_b  \mL_b\dg(\al)[ H_S-\mu N_S]=\be \hat{K}\dg(\al)[H_S-\mu N_S],
\eea
i.e., 
\bea
\dbr{\be[H_S-\mu N_S]} \hat{K}(\al) = \dbr{1}W^\sig(\al).
\eea
Then, \re{ee} leads 
\bea
\big[\dbr{l_0^\pr(\al)}+\dbr{\be[H_S-\mu N_S]} \big] \hat{K}(\al) = 0.
\eea
This implies 
\bea
\dbr{l_0^\pr(\al)} = -\dbr{\be[H_S-\mu N_S]} +c(\al)\dbr{1} ,
\eea
i.e., $\{l_0^\pr(\al)\}\dg = -\be[H_S-\mu N_S]+c(\al)$ where $c(\al)$ is unimportant complex number. 
By the way, $\rho_0(\al)$ is given by
\bea
\rho_0(\al) = \rho_\RM{gc}(\al_S;\be,\be \mu) \defe \f{e^{-\be(H_S(\al_S)- \mu N_S)}}{\Xi(\al_S;\be,\be \mu)} ,
\eea
with $\Xi(\al_S;\be,\be \mu)\defe\tr_S[e^{-\be(H_S(\al_S)-\mu N_S)}]$.
This is derived from \re{rho_ss} (Cf.\re{BM_rho_ss}).  
Then, 
\bea
\{l_0^\pr(\al)\}\dg = \ln\rho_\RM{gc}(\al_S;\be,\be \mu) +c^\pr(\al) 1 ,\ c^\pr(\al)=c(\al)+\ln\Xi(\al_S;\be,\be \mu),
\eea
holds. 
Substituting this equation into \re{A_n^sig_F}, we obtain 
\bea
A_n^\sig(\al) = \f{\partial }{\partial \al^n} S_\RM{vN}(\rho_\RM{gc}(\al_S;\be,\be \mu)) ,
\eea
using \re{dS}.

\subsection{Weakly nonequilibrium regime} \la{noneq}

We introduce
\bea
\ep_{1,b}\defe \be_b -\overline{\be},\ \ep_{2,b}\defe \left \{ \begin{array}{ll}
0 \hs{18mm} b \in \RM{\bm{\mC}} \\
\be_b\mu_b -\overline{\be\mu} \hs{3mm} b \in \RM{\bm{\mG}}
\end{array} \right. , \ \ep \defe \max_{b}\big\{\f{\abs{\ep_{1,b}}}{\overline{\be}},
\f{\abs{\ep_{2,b}}}{\abs{\overline{\be\mu}}} \big\}, 
\eea
where $\overline{\be}$ and $\overline{\be \mu}$ are the reference values, 
which satisfy 
\bea
\min_b \be_b \le \overline{\be} \le \max_b \be_b ,\\ 
\min_{b \in \RM{\bm{\mG}}} \be_b\mu_b \le \overline{\be\mu} \le \max_{b \in \RM{\bm{\mG}}} \be_b\mu_b .
\eea 
$\ep$ is a measure of degree of nonequilibrium.
We consider $\ep \ll 1$ regime. 
Now, we introduce
\bea
 \hat{K}_\ka (\al)\bu \defe -i[H_S(\al_S)+\ka H_\RM{L}(\al),\bu]+\sum_b \Pi_b(\al)\bu ,
\eea
and corresponding instantaneous steady state $\rho_0^{(\ka)}(\al)$:
\bea
 \hat{K}_\ka (\al)\rho_0^{(\ka)}(\al)=0. \la{rho_0^ka}
\eea
Here, $\ka$ is a real parameter satisfying $-1 \le \ka \le 1$. 
$\dbr{1} \hat{K}_\ka (\al) =0$ holds.
In the following, we show
\bea
A_n^\sig(\al) = -\tr_S\Big[\ln \rho_0^{(-1)}(\al)\f{\partial \rho_0(\al)}{\partial \al^n}\Big]+\mO(\ep^2). \la{goal}
\eea
We use the following notations:
\bea
\al_{1,b} \defe \be_b, \ \al_{2,b} \defe \be_b\nu_b,\ \overline{X} \defe X \bv{\al_{i,b}=\overline{\al_{i}}} .
\eea
Here, $\overline{\al_1}=\overline{\be}$ and $\overline{\al_2}=\overline{\be\mu}$. 

We expand $\rho_0^{(\ka)}$ and $l_0^\pr$ as
\bea
\rho_0^{(\ka)}(\al) \aeq \overline{\rho_0^{(\ka)}}+\sum_b(\ep_{1,b}\rho_{1,b}^{(\ka)}+\ep_{2,b}\rho_{2,b}^{(\ka)})+\mO(\ep^2) ,\\
l_0^\pr(\al) \aeq \overline{l_0^\pr(\al)}+\sum_b(\ep_{1,b}k_{1,b}+\ep_{2,b}k_{2,b})+\mO(\ep^2), \la{l_ep}
\eea
with 
\bea
\overline{\rho_0^{(\ka)}} = \rho_\RM{gc},\
\overline{l_0^\pr(\al) } =-\overline{\be}H_S+\overline{\be\mu}N_S+\overline{c}^\ast1 = \ln \rho_\RM{gc}+\overline{c^\pr}^\ast 1.
\eea
Here, $\rho_\RM{gc}\defe \rho_\RM{gc}(\al_S;\overline{\be},\overline{\be \mu})$, $\overline{c}$ and $\overline{c^\pr}$ are the same with $c(\al)$ and $c^\pr(\al)$ in \res{eq}.

First, we investigate $k_{i,b}$ in \re{l_ep}.
\re{ee} can be rewritten as
\bea
\hat{K}\dg(\al)l_0^\pr(\al)+[\mK^\pr(\al)]\dg1 = J_\sig^\st(\al). \la{eer}
\eea
Here, 
\bea
J_\sig^\st(\al) = \mO(\ep^2),
\eea
holds
because $i^\st_{H_b}(\al),i^\st_{N_b}(\al)=\mO(\ep)$ and 
\bea
J^\st_\sig(\al)=\sum_b (-i^\st_{H_b}(\al)\ep_{1,b}+i^\st_{N_b}(\al)\ep_{2,b}) ,
\eea
 since
\bea
\sum_b i^\st_{X_b}(\al)= -\tr_S[X_S \sum_b \mL_b(\al) \rho_0(\al)]=0 \hs{3mm}(X=N,H).
\eea
Then we obtain 
\bea
\overline{\partial_{i,b}\mK^\pr} \dg 1+\overline{K}\dg k_{i,b}+\overline{\partial_{i,b}\mL_{b}} \dg \overline{l_0^\pr} \aeq 0, \la{ee,1}
\eea
in $\mO(\ep_{i,b})$. 
Here, $\partial_{i,b}X \defe \p X/\p \al_{i,b}$ and $\overline{K}\defe \overline{\hat{K}}$.
The first term of the LHS is 
\bea
\overline{\partial_{i,b}\mK^\pr} \dg 1 \aeq  \f{\partial [\mK^\pr]\dg 1}{\partial \al_{i,b}}\Bv{\al_{i,b}=\overline{\al_i}} \no\\
\aeq \f{\partial \mL_b\dg [\al_{1,b}H_S-\al_{2,b}N_S]}{\partial \al_{i,b}}\Bv{\al_{i,b}=\overline{\al_i}} \no\\
\aeq \overline{\partial_{i,b} \mL_b} \dg[\overline{\be}H_S-\overline{\be \mu}N_S]+\overline{\Pi_b}\dg \f{\partial[\al_{1,b}H_S-\al_{2,b}N_S]}{\partial \al_{i,b}}.
\eea
The third term of the LHS becomes
\bea
\overline{\partial_{i,b}\mL_{b}} \dg \overline{l_0^\pr} \aeq \overline{\partial_{i,b}\mL_{b}} \dg ( -\overline{\be}H_S+\overline{\be\mu}N_S+c1) \no\\
\aeq -\overline{\partial_{i,b}\mL_{b}} \dg(\overline{\be}H_S-\overline{\be\mu}N_S).
\eea
Here, we used $\overline{\partial_{i,b}\mL_{b}}\dg1=0$ derived from $\hat{K}\dg 1=0$. 
Then, \re{ee,1} becomes
\bea
\overline{K}\dg k_{1,b}+\overline{\Pi_b}\dg H_S \aeq 0,\la{ee,2} \\
\overline{K}\dg k_{2,b}-\overline{\Pi_b}\dg N_S \aeq 0 \la{ee,3}.
\eea

Next, we show the relation between $k_{i,b}$ and $\rho_{i,b}^{(-1)}$. 
\re{rho_0^ka} leads 
\bea
\overline{K_\ka}\rho_{i,b}^{(\ka)}+\overline{\partial_{i,b}\mL_b}\rho_\RM{gc} = 0 , \la{ss,1}
\eea
in $\mO(\ep_{i,b})$. 
Here, $\overline{K_\ka}\defe \overline{\hat{K}}_\ka$.
By the way, \re{rho_ss} is
\bea
\mL_b \rho_\RM{gc}(\al_S;\be_b,\be_b \nu_b)=0.
\eea
Differentiating this equation by $\al_{i,b}$, we obtain
\bea
\overline{\partial_{i,b}\mL_b}\rho_\RM{gc} \aeq -\overline{\mL_b}\overline{\f{\rho_\RM{gc}(\al_S;\be_b,\be_b \nu_b)}{\partial \al_{i,b}}} =
 \overline{\mL_b}\f{\partial [\al_{1,b} H_S-\al_{2,b}N_S]}{\partial \al_{i,b}}\rho_\RM{gc}(\al_S;\overline{\be},\overline{\be \mu}).
\eea
Substituting these equations into \re{ss,1}, we obtain
\bea
\overline{K}_\ka \rho_{1,b}^{(\ka)}+\overline{\Pi_b}(H_S \rho_\RM{gc}) \aeq 0, \la{ss,1c}\\
\overline{K}_\ka \rho_{2,b}^{(\ka)}-\overline{\Pi_b}(N_S \rho_\RM{gc}) \aeq 0.\la{ss,1d}
\eea
Now, we use \re{KMS2}, namely, 
\bea
\overline{\Pi_b}(\bu\rho_\RM{gc})=(\overline{\Pi_b}\dg \bu)\rho_\RM{gc} \la{DB2m}.
\eea
Using this relation, we rewire \re{ss,1c} and \re{ss,1d} as
\bea
\overline{K}_\ka\rho_{1,b}^{(\ka)}+(\overline{\Pi_b}\dg H_S) \rho_\RM{gc} \aeq 0, \la{ss,1e}\\
\overline{K}_\ka\rho_{2,b}^{(\ka)}-(\overline{\Pi_b}\dg N_S) \rho_\RM{gc} \aeq 0 \la{ss,1f}.
\eea
Multiplying $\rho_\RM{gc}^{-1}$ from the right, we obtain
\bea
(\overline{K}_\ka\rho_{1,b}^{(\ka)})\rho_\RM{gc}^{-1}+\overline{\Pi_b}\dg H_S\aeq 0, \la{ss,1g}\\
(\overline{K}_\ka\rho_{2,b}^{(\ka)})\rho_\RM{gc}^{-1}-\overline{\Pi_b}\dg N_S\aeq 0 \la{ss,1h}.
\eea
 \re{DB2m} can be rewritten as
\bea
(\overline{\Pi_b}Y)\rho_\RM{gc}^{-1}=\overline{\Pi_b}\dg(Y\rho_\RM{gc}^{-1}), \la{DB2r}
\eea
for any $Y=\bu\rho_\RM{gc} \in \RM{\bm{B}}$ by multiplying $\rho_\RM{gc}^{-1}$ from the right.
\re{DB2r} leads
\bea
(\overline{\Pi}\rho_{i,b}^{(\ka)})\rho_\RM{gc}^{-1} =\overline{\Pi}\dg(\rho_{i,b}^{(\ka)}\rho_\RM{gc}^{-1}) ,\la{K1}
\eea
where $\overline{\Pi}\defe \sum_b \overline{\Pi_b}$.
By the way, $[H_S(\al_S),\rho_0^{(\ka)}(\al)]= 0$ holds similarly to \re{H_S,rho_0}. 
Differentiating this equation by $\al_{i,b}$, we obtain
\bea
[H_S(\al_S),\rho_{i,b}^{(\ka)}] = 0. \la{Key}
\eea
This relation leads 
\bea
(\overline{H_\ka^{\times}}\rho_{i,b}^{(\ka)})\rho_\RM{gc}^{-1}=\overline{H_\ka^{\times}}(\rho_{i,b}^{(\ka)}\rho_\RM{gc}^{-1})
= \overline{H_{-\ka}^{\times}}\dg(\rho_{i,b}^{(\ka)}\rho_\RM{gc}^{-1}), \la{K2}
\eea
where $H_\ka^{\times}\bu \defe -i[H_S(\al_S)+\ka H_\RM{L}(\al),\bu]$. 
We used $(H_\ka^{\times})\dg=-H_\ka^{\times}$. 
In the first equality, we used that $\rho_\RM{gc}$ commutes with $H_S$ and $H_\RM{L}$.
\re{K1} and \re{K2} lead
\bea
(\overline{K}_\ka\rho_{i,b}^{(\ka)})\rho_\RM{gc}^{-1} = \overline{K}_{-\ka}\dg(\rho_{i,b}^{(\ka)}\rho_\RM{gc}^{-1}). \la{imp}
\eea
Substituting this into \re{ss,1g} and \re{ss,1h}, we obtain
\bea
 \overline{K}_{-\ka}\dg (\rho_{1,b}^{(\ka)}\rho_\RM{gc}^{-1})+\overline{\Pi_b}\dg H_S\aeq 0 ,\la{ss,1i}\\
  \overline{K}_{-\ka}\dg (\rho_{2,b}^{(\ka)}\rho_\RM{gc}^{-1})-\overline{\Pi_b}\dg N_S\aeq 0 \la{ss,1j}.
\eea
Subtracting \re{ss,1i} (\re{ss,1j}) for $\ka=-1$ from \re{ee,2} (\re{ee,3}), we obtain
\bea
\overline{K}\dg(k_{i,b}-\rho_{i,b}^{(-1)}\rho_\RM{gc}^{-1}) = 0.
\eea
This means
\bea
k_{i,b} = \rho_{i,b}^{(-1)}\rho_\RM{gc}^{-1}+\overline{c_{i,b}}1,
\eea
where $\overline{c_{i,b}}$ is unknown complex number.
Using this relation, \re{l_ep} becomes
\bea
l_0^\pr(\al) 
\aeq \ln \rho_\RM{gc}(\al_S;\overline{\be},\overline{\be \mu})+C(\al)1+\sum_b \sum_{i=1}^2\ep_{i,b}\rho_{i,b}^{(-1)}\rho_\RM{gc}^{-1}+\mO(\ep^2) \no\\
\aeq \ln \rho_0^{(-1)}(\al)+C(\al)1+\mO(\ep^2). \la{Goal-} 
\eea
Substituting this equation into \re{A_n^sig_F}, we obtain  \re{goal}.
Here, $C(\al)\defe\overline{c^\pr}^\ast+\sum_{b,i}\overline{c_{i,b}}\ep_{i,b}$.
We supposed $[\rho_\RM{gc},\rho_{i,b}^{(-1)}] =0$, which leads $\ln \rho_0^{(-1)}(\al) =\ln\rho_\RM{gc}+\sum_{i,b}\ep_{i,b}\rho_{i,b}^{(-1)}\rho_\RM{gc}^{-1}+\mO(\ep^2) $.
This supposition is satisfied if $[N_S,\rho_0^{(-1)}(\al)]=\mO(\ep^2)$ (which leads $[N_S,\rho_{i,b}^{(-1)}] =0$) or $\overline{\be\mu}=0$ holds. 
If $H_S$ is non-degenerate, $[N_S,\rho_0^{(-1)}(\al)]=0$ holds, then $[N_S,\rho_{i,b}^{(-1)}] =0$, $[\rho_\RM{gc},\rho_{i,b}^{(-1)}] =0$ and \re{Goal-} hold.
If $n_\RM{GC}=0$, $\rho_\RM{gc}$ is replaced by the canonical distribution and \re{Goal-} holds without any assumption. 

If 
\bea
[H_\RM{L}(\al), \rho_0^{(\ka)}(\al)]=0, \la{Key2a}
\eea
holds, $\rho_0^{(\ka)}(\al)$ is independent of $\ka$ ($\rho_0^{(\ka)}(\al)=\rho_0(\al)$), 
then \re{goal} becomes
\bea
A_n^\sig(\al) = \f{\partial }{\partial \al^n} S_\RM{vN}(\rho_0(\al))+\mO(\ep^2), \la{A_S}
\eea
using \re{dS}. 
\re{Key2a} holds if $H_S$ is non-degenerate.
\re{A_S} can be shown from $[\overline{H_\RM{L}},\rho_{i,b}^{(1)}]=0$, which is weaker assumption than \re{Key2a} and is derived from \re{Key2a} for $\ka=1$.
If we neglect the Lamb shift Hamiltonian, namely we consider the QME for $\hat{K}_0(\al)$, \re{A_S} holds (with a replacement $\rho_0 \to \rho_0^{(0)}$).
From \re{A_S}, we obtain 
\bea
\sig_\RM{ex}= S_\RM{vN}(\rho_0(\al_\tau))-S_\RM{vN}(\rho_0(\al_0))+\mO(\ep^2\dl), \la{ex=}
\eea
with $\dl=\max_{n,\al \in C} \f{\abs{\al^n-\al^n_0}}{\abs{\bar{\al}^n}}$.
$\bar{\al}^n$ is typical value of the $n$-th control parameter.

Yuge {\it et al.} \cite{Yuge13} considered the outputs of $A(t)=-\sum_b \be_b(t)[H_b-\mu_b(t)N_b]$ (for $n_\RM{C}=0$) at $t=0$ and $t=\tau$ as $a(0)$ and $a(\tau)$, 
and errorneously identified $ a(\tau)-a(0)$ with the entropy production.
To analyze $\sig^\pr \defe \bra a(\tau)-a(0)\ket$, improperly, they took $\chi_{H_b} \to -\be_b \lm$ and $\chi_{N_b} \to \be_b\mu_b \lm$ limit of the FCS-QME \re{FCS-QME} 
only valid for time independent observables.
The obtained Liouvillian (of which the Lamb shift Hamiltonian is neglected) incidentally satisfy \re{G_con}. 
Using that Liouvillian, for the time-reversal symmetric system, 
Yuge {\it et al.} studied the relation between $A_n^\sig(\al)$ and the symmetrized von Neumann entropy.
In contrast, up to here, we do not suppose the time-reversal symmetry. 
In \res{Discussion}, we consider the time-reversal operations and show that the potential $\mS(\al)$ such that 
$A_n^\sig(\al)=\p \mS/\p \al^n+\mO(\ep^2)$ dose not exist if the time-reversal symmetry is broken.

\subsection{Time-reversal operations} \la{Discussion}

We define the time-reversal operation. 
We denote the time-reversal operator of the system by $\theta$. 
We also define 
\bea
\tl Y \defe \theta Y \theta^{-1},
\eea
for all $Y \in \RM{\bm{B}}$ and 
\bea
\tl \mJ \tl Y \defe \theta ( \mJ Y) \theta^{-1},
\eea
for a super-operator $\mJ$ of the system. 
The time-reversal of $\hat{K}(\al)\rho_0(\al)=0$ is given by
\bea
i[\tl H_\RM{L}(\al),\tl \rho_0(\al)]+\sum_b \tl \Pi_b(\al) \tl \rho_0(\al)=0,
\eea
using \re{H_S,rho_0}. 
If 
\bea
\tl H_\RM{L}(\al) =H_\RM{L}(\al),\ \ \tl \Pi_b(\al)= \Pi_b(\al), \la{TRS}
\eea
hold, the above equation coincides with the equation of $\rho_0^{(-1)}(\al)$ since $[H_S,\rho_0^{(\ka)}]=0$,
then 
\bea
\tl \rho_0(\al)=\rho_0^{(-1)}(\al), \la{tl=-1}
\eea
holds. 
If the total Hamiltonian is time-reversal invariant, \re{TRS} holds \cite{trs}. 
If \re{TRS} holds and we neglect the Lamb shift Hamiltonian, the instantaneous steady state is time-reversal invariant: $\tl \rho_0^{(0)}=\rho_0^{(0)}$.

As we will show, for time-reversal symmetric system, 
\bea
\f{\p}{\p \al^n}S_\RM{sym}(\rho_0(\al)) =-\tr_S\Big[\ln \tl \rho_0(\al)\f{\partial \rho_0(\al)}{\partial \al^n}\Big]+\mO(\ep^2), \la{giro}
\eea
holds. 
Here,
\bea
S_\RM{sym}(\rho) \defe - \tr_S\big[\rho\half(\ln \rho+\ln \tl \rho)\big],
\eea
is the symmetrized von Neumann entropy. 
Combining \re{goal} with \re{tl=-1}, we obtain 
\bea
A_n^\sig(\al)=\f{\p}{\p \al^n}S_\RM{sym}(\rho_0(\al))+\mO(\ep^2),
\eea
then, the equation \re{ex=} with $S_\RM{vN} \to S_\RM{sym}$ holds.
As analogy, we consider
\bea
S^\pr(\al) \defe - \tr_S\big[\rho_0(\al)\half(\ln \rho_0(\al)+\ln \rho_0^{(-1)}(\al))\big],
\eea
for generally non-time-reversal symmetric system.
The difference between $\p S^\pr(\al)/\p \al^n$ and the first term of the RHS of \re{goal} is 
\bea  
&&\hs{-10mm}\f{\p S^\pr(\al)}{\p \al^n}-\Big(-\tr_S\Big[\ln \rho_0^{(-1)}(\al)\f{\partial \rho_0(\al)}{\partial \al^n}\Big]\Big) \no\\
\aeq-\half \tr_S\big[\f{\p \rho_0}{\p \al^n}(\ln \rho_0-\ln \rho_0^{(-1)}) \big]-\half  \tr_S\big[\rho_0\f{\p }{\p \al^n}\ln \rho_0^{(-1)} \big]. \la{giron}
\eea
To calculate the RHS of this equation, we use formulas
\bea
\ln(A+\dl B) \aeq \ln A+\int_0^\infty ds \ \Big(\dl \f{1}{A+s}B\f{1}{A+s} \no\\
&&- \dl^2 \f{1}{A+s}B\f{1}{A+s}B\f{1}{A+s}+\mO(\dl^3) \Big) ,\la{ln A+B}\\
\f{\p}{\p \al^n}\ln A(\al) \aeq \int_0^\infty ds \  \f{1}{A(\al)+s}\f{\p A(\al)}{\p \al^n}\f{1}{A(\al)+s} \la{d ln A},
\eea
where $A,B,A(\al)\in \RM{\bm{B}}$ and $\dl$ is small real number. 
We proof \re{ln A+B} in Appendix \ref{proof_ln A+B}. 
\re{d ln A} is derived from \re{ln A+B}. 
$\rho_0-\rho_0^{(-1)}=\ep \psi+\mO(\ep^2)$ holds because $\overline{\rho_0^{(\ka)}} = \rho_\RM{gc}(\al_S;\overline{\be},\overline{\be\mu})$. 
Then, the first term of the RHS of \re{giron} is given by
\bea
&&\hs{-10mm}-\half \tr_S\big[\f{\p \rho_0}{\p \al^n}(\ln \rho_0-\ln \rho_0^{(-1)}) \big] \no\\ 
\aeq-\f{\ep}{2}\int_0^\infty ds \  \tr_S\Big[\f{\p \rho_0}{\p \al^n}\f{1}{\rho_0^{(-1)}+s}\psi\f{1}{\rho_0^{(-1)}+s} \Big]+\mO(\ep^2) .
\eea
The second term of the RHS of \re{giron} is given by
\bea
&&\hs{-10mm}-\half  \tr_S\big[\rho_0\f{\p }{\p \al^n}\ln \rho_0^{(-1)} \big] \no\\
 \aeq -\half \int_0^\infty ds \  \tr_S\Big[\f{\p \rho_0^{(-1)}}{\p \al^n}
\f{1}{\rho_0^{(-1)}+s}(\rho_0^{(-1)}+\ep \psi)\f{1}{\rho_0^{(-1)}+s}  \Big] +\mO(\ep^2)\no\\
\aeq -\half  \tr_S\big[\f{\p \rho_0^{(-1)}}{\p \al^n}\big]-\f{\ep}{2} \int_0^\infty ds \  \tr_S\Big[\f{\p \rho_0^{(-1)}}{\p \al^n}
\f{1}{\rho_0^{(-1)}+s} \psi\f{1}{\rho_0^{(-1)}+s}  \Big]+\mO(\ep^2) \no\\
\aeq -\f{\ep}{2} \int_0^\infty ds \  \tr_S\Big[\f{\p \rho_0^{(-1)}}{\p \al^n}
\f{1}{\rho_0+s} \psi\f{1}{\rho_0+s}  \Big]+\mO(\ep^2) \no\\
\aeq  -\f{\ep}{2} \int_0^\infty ds \  \tr_S\Big[\f{\p (\theta \rho_0^{(-1)}\theta^{-1})}{\p \al^n}
\f{1}{\tl \rho_0+s} \tl \psi\f{1}{\tl \rho_0+s}  \Big]+\mO(\ep^2).
\eea
Here, we used $\ep(\rho_0^{(-1)}+s)^{-1}=\ep(\rho_0+s)^{-1}+\mO(\ep^2)$ and $\tr_S\bu=\tr_S \tl \bu$ if $\tr_S \bu$ is real.
In general, the RHS of \re{giron} is not $\mO(\ep^2)$. 
However, if $\tl \rho_0=\rho_0^{(-1)}$ holds, the RHS of \re{giron} becomes $\mO(\ep^2)$ since $\tl \psi=-\psi$, then \re{giro} holds.
In the proof of \re{giro}, Yuge {\it et al.} \cite{Yuge13} used incorrect equations
$\f{\p}{\p \al^n}\ln \tl \rho_0=\tl \rho_0^{-1}\f{\p\tl \rho_0}{\p \al^n}$ and $\ln \rho_0-\ln \tl \rho_0=\ep \psi \tl \rho_0^{-1}+\mO(\ep^2)$. 

We introduce the BSN curvature
\bea
F_{mn}^\sig(\al)=\f{\p A_n^\sig}{\p \al^m}-\f{\p A_m^\sig}{\p \al^n}.
\eea
$F_{mn}^\sig(\al)=\mO(\ep^2)$ and the existence of $\mS(\al)$ such that $A_n^\sig(\al)=\p \mS(\al)/\p \al^n+\mO(\ep^2)$ are equivalent.  
If $F_{mn}^\sig(\al)=\mO(\ep)$ holds, $\mS(\al)$ does not exist. 
$F_{mn}^\sig(\al)$ is given by 
\bea
F_{mn}^\sig(\al)\aeq f_{mn}(\al)-f_{nm}(\al)+\mO(\ep^2),
\eea
where 
\bea
f_{mn}(\al)\aeqd  -\tr_S\Big(\f{\p \ln \rho_0^{(-1)}}{\p \al^m}\f{\p  \rho_0}{\p \al^n} \Big).
\eea
$f_{mn}(\al)$ is given by 
\bea
f_{mn}(\al)\aeq -\int_0^\infty ds \ \tr_S\Big(\f{1}{\rho_0^{(-1)}+s}\f{\p \rho_0^{(-1)}}{\p \al^m}\f{1}{\rho_0^{(-1)}+s}\f{\p  \rho_0}{\p \al^n} \Big) \no\\
\aeq  \int_0^\infty ds \ [\mF_{mn}^{(0)}(s)+\mF_{mn}^{(1)}(s)+\mO(\ep^2)],
\eea
with
\bea
\mF_{mn}^{(0)}(s)=-\tr_S \Big(\sig_s \f{\p \rho_\RM{gc}}{\p \al^m} \sig_s \f{\p \rho_\RM{gc}}{\p \al^n} \Big),
\eea
and  
\bea
\sig_s \defe \f{1}{\rho_\RM{gc}+s}. 
\eea
$\mF_{mn}^{(1)}(s)$ is given by 
\bea
\mF_{mn}^{(1)}(s) \aeq \mF_{mn}^{(1,0)}(s)+\mF_{mn}^{(1,1)}(s),\\
\mF_{mn}^{(1,0)}(s) \aeq \tr_S \Big(\sig_s\eta^{(-1)}\sig_s \f{\p \rho_\RM{gc}}{\p \al^m}\sig_s\f{\p \rho_\RM{gc}}{\p \al^n}
+\sig_s \f{\p \rho_\RM{gc}}{\p \al^m}\sig_s\eta^{(-1)}\sig_s\f{\p \rho_\RM{gc}}{\p \al^n}\Big), \\
\mF_{mn}^{(1,1)}(s) \aeq \tr_S \Big(-\sig_s \f{\p\eta^{(-1)}}{\p \al^m}\sig_s\f{\p \rho_\RM{gc}}{\p \al^n} -\sig_s\f{\p \rho_\RM{gc}}{\p \al^m} \sig_s\f{\p\eta^{(1)}}{\p \al^n}\Big),
\eea
where 
\bea
 \eta^{(\ka)}(\al) \defe \sum_b \sum_{i=1,2} \ep_{i,b}\rho_{i,b}^{(\ka)}.
\eea
$\rho_0=\rho_\RM{gc}+ \eta^{(1)}+\mO(\ep^2)$ and $\rho_0^{(-1)}=\rho_\RM{gc}+ \eta^{(-1)}+\mO(\ep^2)$ hold.
Because of $\mF_{mn}^{(0)}(s)-\mF_{nm}^{(0)}(s)=0$ and $\mF_{mn}^{(1,0)}(s)-\mF_{nm}^{(1,0)}(s)=0$, we obtain 
\bea
F_{mn}^\sig(\al)\aeq \int_0^\infty ds \ [\mF_{mn}^{(1,1)}(s)-\mF_{nm}^{(1,1)}(s)]+\mO(\ep^2) \no\\
\aeq \zeta_{mn}-\zeta_{nm}+\mO(\ep^2),
\eea
with
\bea
\zeta_{mn} \aeq \int_0^\infty ds \ \tr_S \Big(\sig_s \f{\p[\eta^{(1)}-\eta^{(-1)}]}{\p \al^m}\sig_s\f{\p \rho_\RM{gc}}{\p \al^n}\Big) \no\\
\aeq \tr_S \Big(\f{\p[\eta^{(1)}-\eta^{(-1)}]}{\p \al^m}\f{\p \ln \rho_\RM{gc}}{\p \al^n}\Big) \no\\
\aeq -\tr_S \Big(\f{\p[\eta^{(1)}-\eta^{(-1)}]}{\p \al^m}\f{\p [\overline{\be}H_S(\al_S)-\overline{\be\mu}N_S]}{\p \al^n}\Big).
\eea
Here, we used $\tr_S \Big(\f{\p[\eta^{(1)}-\eta^{(-1)}]}{\p \al^m}\f{\p \ln \Xi}{\p \al^n}\Big)=0$ because $\tr_S\eta^{(1)}=0=\tr_S\eta^{(-1)}$. 
$\overline{\be}$ and $\overline{\be\mu}$ are functions of $\al^{\pr\pr}$. 
Using $\tr_S\bu=\tr_S \tl \bu$ if $\tr_S \bu$ is real, we obtain 
\bea
\zeta_{mn} \aeq -\tr_S \Big(\f{\p[\theta \eta^{(1)}\theta^{-1}-\theta \eta^{(-1)}\theta^{-1}]}{\p \al^m}
\f{\p [\overline{\be}\tl H_S(\al_S)-\overline{\be\mu} \tl N_S]}{\p \al^n}\Big) .
\eea
For time-reversal symmetric system, $\tl H_S=H_S$, $\tl N_S=N_S$ and $\theta \eta^{(1)}\theta^{-1}=\eta^{(-1)}$ hold. 
Then, the above equation becomes $\zeta_{mn}=-\zeta_{mn}$, namely, $\zeta_{mn}=0$ and $F_{mn}^\sig(\al)=\mO(\ep^2)$ hold. 
However, if the time-reversal symmetry is broken, $\theta \eta^{(1)}\theta^{-1} \ne \eta^{(-1)}$ holds in general. 
Then, $\zeta_{mn}\ne -\zeta_{mn}$ namely $\zeta_{mn}\ne 0$ hold.
$\zeta_{mn}$ is not symmetric for $m$ and $n$. 
Then, if the time-reversal symmetry is broken and $H_S$ is degenerated,  $\mS(\al)$ dose not exist in general. 
This is the most important result of this thesis. 

\subsection{Born-Markov approximation} \la{Born_Markov}

We denote the BSN vector for the entropy production and instantaneous steady state of the Born-Markov approximation by 
$A_n^{\sig,\rm{BM}}(\al)$ and $\rho_0^\RM{BM}(\al)$. 
Then, 
\bea
A_n^{\sig,\rm{BM}}(\al) \aeq A_n^{\sig}(\al)+\mO(v^2),\\
S_\RM{vN}(\rho_0^\RM{BM}(\al)) \aeq S_\RM{vN}(\rho_0(\al)) +\mO(v^2), \\
S_\RM{sym}(\rho_0^\RM{BM}(\al)) \aeq S_\RM{sym}(\rho_0(\al)) +\mO(v^2),
\eea
hold \cite{Yuge13}. 
Here, $v=u^2$ and $u(\ll 1)$ describes the order of $H_{Sb}$.
Then, if \re{A_S} holds, we obtain
\bea
A_n^{\sig,\rm{BM}}(\al) = \f{\p }{\p \al^n}S_\RM{vN}(\rho_0^\RM{BM}(\al))+\mO(\ep^2)+\mO(v^2).
\eea
For time-reversal symmetric system,
\bea
A_n^{\sig,\rm{BM}}(\al) = \f{\p }{\p \al^n}S_\RM{sym}(\rho_0^\RM{BM}(\al))+\mO(\ep^2)+\mO(v^2),
\eea
holds.

\newpage

\section{Comparison of two definitions of entropy production} \la{Other_def}

In this chapter, we compare preceding study on of the entropy production in the classical Markov jump process \cite{Komatsu15, Jarzynski} with ours.
We consider the Markov jump process on the states $n=1,2,\cdots,\mN$, where the definitions are explained in Appendix \ref{Appendix B}.
The probability to find the system in a state $n$ is $p_n(t)$ and it obeys the master equation:
\bea
\f{dp_n(t)}{dt} = \sum_{m=1}^\mN K_{nm}(\al_t)p_m(t) .\la{M}
\eea
The Liouvillian is given by 
\bea
K_{nm}(\al)=\sum_b K_{nm}^{(b)}(\al),
\eea
where $K_{nm}^{(b)}$ originates the couping between the system and the bath $b$.
$ \sum_n K_{nm}^{(b)}(\al)=0$ holds. 
We suppose that $K_{mn}^{(b)}(\al)\ne 0(=0)$ holds if $K_{nm}^{(b)}(\al)\ne0(=0)$ for all $n \ne m$.
The definition of the entropy production for each Markov jump process \re{MJ} is \re{def_EP}. 
The average entropy production $\sig^\RM{C}$ is given by (see \re{sig^C0})
\bea
\sig^\RM{C}  = \int_0^\tau dt \ \sum_{n,m} \sig_{nm}^\RM{C}(\al_t)p_m(t), \la{sig^C}
\eea
where
\bea
\sig_{nm}^\RM{C}(\al)  =  -K_{nm}(\al) \ln \f{K_{nm}(\al)}{K_{mn}(\al)}.
\eea

We denote the solution of the QME with RWA by $\rho(t)$.
We suppose $p_n(t)\defe \br{n}\rho(t)\ke{n}$ is governed by \re{M} with 
\bea
K_{nm}^{(b)}(\al)=(\Pi_b(\al))_{nn,mm}.
\eea 
Here, $\ke{n}$ is the energy eigenstate of $H_S(\al_S)$, 
\bea
(\Pi_b(\al) \bu)_{nm}=\sum_{k,l}(\Pi_b(\al))_{nm,kl} (\bu)_{kl},\ (\bu)_{kl}\defe \br{k}\bu \ke{n}.
\eea 
This supposition implies \re{Key2a}. 
A sufficient condition by which $p_n(t)$ obeys \re{M} is below: 
(1) $H_S(\al_S)$ is non-degenerate and (2) $ \{\al^n \in \al_S \vert \ \f{\p }{\p \al^n}\ke{n} \ne 0 \}$ are fixed. 
The eigenenergy can depend on $\{\al^n \in \al_S \vert \ \f{\p }{\p \al^n}\ke{n} =0 \}$.
We show that our average entropy production \re{def_sig} is given by a similar expression of \re{sig^C}:
\bea
\sig = \int_0^\tau dt \ \sum_{n,m} \sig_{nm}(\al_t)p_m(t). \la{sig^Q}
\eea
Here, 
\bea
 \sig_{nm}(\al) \defe \sum_b K_{nm}^{(b)}(\al)\theta_{nm}^{(b)}(\al) = -\sum_b K_{nm}^{(b)}(\al)\ln \f{K_{nm}^{(b)}(\al)}{K_{mn}^{(b)}(\al)},
\eea
with 
\bea
\theta_{nm}^{(b)}(\al)\defe  \left \{ \begin{array}{ll}
- \ln\f{K_{nm}^{(b)}(\al)}{K_{mn}^{(b)}(\al)} \hs{3mm} K_{nm}^{(b)}(\al) \ne 0 \\
0 \hs{20mm} K_{nm}^{(b)}(\al) = 0
\end{array} \right. .
\eea
Because of \re{def_w}, \re{w=W} and \re{w=W2}, the particle and energy currents are given by
 $i_{X_b}= \tr_S[W^{X_b}\rho(t)]$ with $W^{X_b} = -(\Pi_b \dg X_S)\dg \ (X=H,N) $. 
\re{W_A_B} leads 
\bea
(W^{X_b})_{nm} = -\sum_{k,l} (\Pi_b)_{lk,mn} (X_S)_{kl}.
\eea
We suppose $(X_S)_{nm}=(X_S)_{nn}\dl_{nm}$ for $X=N,H$. 
Since $(X_S)_{kl}$ is a diagonal matrix, $(W^{X_b})_{nm}$ is also a diagonal matrix. 
Then, 
\bea
i_{X_b} = \sum_m (W^{X_b})_{mm}p_m(t), \la{i^X_b3}
\eea
holds. 
Substituting $(W^{X_b})_{mm} =-\sum_n K^{(b)}_{nm} (X_S)_{nn}$ into \re{i^X_b3}, we obtain
\bea
i_{X_b}\aeq -\sum_{n,m}  K^{(b)}_{nm} (X_S)_{nn}p_m(t) \no\\
\aeq \sum_{n,m}  K^{(b)}_{nm} [(X_S)_{mm}-(X_S)_{nn}]p_m(t). \la{i_X_b}
\eea
This equation leads
\bea
\dot{\sig}(t) = -\sum_{n,m}\sum_b K^{(b)}_{nm} \be_b(t) \{[(H_S)_{mm}-(H_S)_{nn}]-\nu_b(t)[(N_S)_{mm}-(N_S)_{nn}] \}p_m(t).
\eea
Using the local detailed balance condition derived from \re{KMS2}
\bea
\ln \f{K_{nm}^{(b)}(\al)}{K_{mn}^{(b)}(\al)} = \be_b \{[(H_S)_{mm}-(H_S)_{nn}]-\nu_b[(N_S)_{mm}-(N_S)_{nn}]\}, \la{LDBC}
\eea
we obtain \re{sig^Q}. For $b \in \RM{\bm{\mC}}$, $(N_S)_{mm}-(N_S)_{nn}=0$ holds for $n$ and $m$ such that $K_{nm}^{(b)}(\al) \ne 0$. 

\re{i_X_b} can be rewritten as 
\bea
i_{X_b}\aeq \sum_{n,m}w_{nm}^{X_b}(\al_t)p_m(t),\  w_{nm}^{X_b}(\al_t) \defe K^{(b)}_{nm} [(X_S)_{mm}-(X_S)_{nn}]. \la{i_X_b2}
\eea
This $w_{nm}^{X_b}(\al_t)$ corresponds to $w^{O_\mu}_{i j}(\al_t)$ of \re{I_mu_rtd}. 

Now we introduce $\RM{\bm{\mA}}_{nm}=\{b \vert K_{nm}^{(b)} \ne 0 \}$. 
From the assumption, $\RM{\bm{\mA}}_{nm}=\RM{\bm{\mA}}_{mn}$ holds. 
If we suppose \re{s_N_S} for $b \in \RM{\bm{\mC}}$ and \re{H_SbG} for $b \in \RM{\bm{\mG}}$, 
$\RM{\bm{\mA}}_{nm}=\RM{\bm{\mC}}$ for $(N_S)_{mm}=(N_S)_{nn}$ and $\RM{\bm{\mA}}_{nm}=\RM{\bm{\mG}}$ for $(N_S)_{mm}\ne(N_S)_{nn}$. 
Then, \re{LDBC} means 
\bea
\ln \f{K_{nm}^{(b)}(\al)}{K_{mn}^{(b)}(\al)} = \be_b [(H_S)_{mm}-(H_S)_{nn}]  \ (b \in \RM{\bm{\mC}}),
\eea
with $(N_S)_{mm}=(N_S)_{nn}$ and 
\bea
\ln \f{K_{nm}^{(b)}(\al)}{K_{mn}^{(b)}(\al)} =\be_b \{[(H_S)_{mm}-(H_S)_{nn}]-\mu_b[(N_S)_{mm}-(N_S)_{nn}]\} \ (b \in \RM{\bm{\mG}}).
\eea
with $(N_S)_{mm}\ne(N_S)_{nn}$. 

Now we introduce a matrix $\mK^\lm(\al)$ by
\bea
[\mK^\lm(\al)]_{nm} \defe \sum_b K_{nm}^{(b)}(\al)e^{i\lm \theta_{nm}^{(b)}(\al)}.
\eea
Then, we obtain
\bea
\f{\partial}{\partial (i\lm)}\Big \vert_{\lm=0} \sum_{n,m}  \Big[ {\rm{T}} \exp \big[ \int_0^\tau dt \ \mK^\lm(\al_t) \big] \Big]_{nm} p_{m}(0) 
= \int_0^\tau dt \ \sum_{n,m}\sig_{nm}(\al_t)p_m(t)=\sig .
\eea
$\mK^\lm$ was originally introduced by Sagawa and Hayakawa \cite{Sagawa}.
About averages, our entropy production is the same with Sagawa and Hayakawa. 

We show that the difference between $\sig^\RM{C}_{nm}(\al)$ and $\sig_{nm}(\al)$ is $\mO(\ep^2)$:
\bea
\sig^\RM{C}_{nm}(\al) = \sig_{nm}(\al)+\mO(\ep^2). \la{sig_sig^C}
\eea
In fact, $K^{(b)}_{nm}$ can be expanded as 
\bea
K_{nm}^{(b)}=\bar{K}_{nm}^{(b)}+\sum_{i=1,2} \ep_{i,b}K^{i,b}_{nm}+\mO(\ep^2) ,
\eea
then we obtain
\bea
\sig^\RM{C}_{nm}(\al) \aeq \sig_{nm}^{\RM{C}(0,1)}+\sig^{\rm{C}(2)}_{nm}(\al)+\mO(\ep^3) ,\\
 \sig_{nm}(\al) \aeq \sig_{nm}^{(0,1)}+\sig^{(2)}_{nm}(\al)+\mO(\ep^3) ,
\eea
with
\bea
\sig_{nm}^{\RM{C}(0,1)}\aeq-\bar{K}_{nm} \ln \f{\bar{K}_{nm}}{\bar{K}_{mn}}+\sum_{i,b}\ep_{i,b} \Big[K^{i,b}_{nm} \ln \f{\bar{K}_{nm}}{\bar{K}_{mn}}+
 K^{i,b}_{nm}-K^{i,b}_{mn}\f{\bar{K}_{nm}}{\bar{K}_{mn}}  \Big] ,\\
 \sig_{nm}^{(0,1)} \aeq \sum_{b \in \RM{\bm{\mA}}_{nm}}\Big( -\bar{K}_{nm}^{(b)} \ln \f{\bar{K}_{nm}^{(b)}}{\bar{K}_{mn}^{(b)}}
+\sum_{i}\ep_{i,b} \Big[K^{i,b}_{nm} \ln \f{\bar{K}_{nm}^{(b)}}{\bar{K}_{mn}^{(b)}}+
 K^{i,b}_{nm}-K^{i,b}_{mn}\f{\bar{K}_{nm}^{(b)}}{\bar{K}_{mn}^{(b)}}  \Big] \Big) .\no\\
\eea\
$\sig^{\rm{C}(2)}_{nm}(\al)$ and $\sig^{(2)}_{nm}(\al)$ are quadratic orders of $\ep_{i,b}$. 
While the former includes $\ep_{i,b}\ep_{i^\pr,b^\pr}$ ($b \ne b^\pr$) terms, the latter dose not. 
$\RM{\bm{\mA}}_{nm}=\RM{\bm{\mA}}_{mn}$ leads
\bea
\bar{K}_{nm} = \sum_{b \in \RM{\bm{\mA}}_{nm}} \bar{K}_{nm}^{(b)},\ \bar{K}_{mn} = \sum_{b \in \RM{\bm{\mA}}_{nm}} \bar{K}_{mn}^{(b)}. \la{K_K^b}
\eea
\re{LDBC} leads $\bar{K}_{nm}^{(b)}/\bar{K}_{mn}^{(b)}$ is independent of $b \in \RM{\bm{\mA}}_{nm}$. 
Then, we obtain
\bea
\f{\bar{K}_{nm}^{(b)}}{\bar{K}_{mn}^{(b)}}=\f{\bar{K}_{nm}}{\bar{K}_{mn}} \ (b \in \RM{\bm{\mA}}_{nm}).
\eea
The above relation and \re{K_K^b} lead 
\bea
\sig_{nm}^{\RM{C}(0,1)}=\sig_{nm}^{(0,1)},
\eea
and \re{sig_sig^C}. \re{sig_sig^C} leads 
\bea
\sig_\RM{ex}^\RM{C} \aeq \sig_\RM{ex}+\mO(\ep^2\dl).
\eea
Here, $\sig_\RM{ex}^\RM{C}$ is given by \re{sig_ex^C}. 
Then, \re{K15}, the result of Ref.\cite{Komatsu15}, coincides with \re{ex=} when $p_n(t)= \br{n}\rho(t)\ke{n}$ is governed by the master equation \re{M}.

\newpage

\section{Conclusion} \la{Conclusion}

\subsection{General conclusion} 

In this thesis, for open systems described by the quantum master equation (QME), we investigated the quantum pump and the excess entropy production.

First, we investigated quantum pump using the FCS-QME (full counting statistics with quantum master equation) approach. 
We studied the non-adiabatic effect and the showed that the general solution of the QME $\rho(t)$ is decomposed as 
$\rho(t)=\rho_0(\al_t)+\sum_{n=1}^\infty \rho^{(n)}(t)+\sum_{n=0}^\infty \tl \rho^{(n)}(t) $ (\rec{sFCS-QME}). 
Here, $\al_t$ is the value of the set of the control parameters at time $t$ and $\rho_0(\al_t)$ is the instantaneous steady state of the QME, 
$\rho^{(n)}(t)$ and $ \tl \rho^{(n)}(t) $ are calculable and order $(\om/\Ga)^n$ where $\om$ is the modulation frequency of the control parameters and 
$\Ga$ is the coupling strength between the system and the baths. 
$ \tl \rho^{(n)}(t) $ exponentially damps (like $e^{-\Ga t}$) as a function of time. 
We showed that the generalized mater equation (GME) approach provides $\bm{p}(t)=\bm{p}_\RM{(ss)}(t)+\dl \bm{p}(t)$ in the Born approximation (Appendix \res{A_GME}).
Here, $\bm{p}$ corresponds to the set of the diagonal components of $\rho$ in the matrix representation by the energy eigenstates,  
$\bm{p}_\RM{(ss)}(t)$ corresponds to $\rho_0(\al_t)+\sum_{n=1}^\infty \rho^{(n)}(t)$ and the 
the term $\dl \bm{p}(t)$ originates from non-Markovian effects. 
The FCS-QME picks out one higher order non-adiabatic piece of information from the solution of the QME, namely, 
if we have $\rho^{(n)}(t)$, the FCS-QME method provides $(n+1)$-th order pump currents. 
Moreover, we showed that the Berry-Sinitsyn-Nemenman (BSN) phase derived under the ``adiabatic" condition which makes the Berry phase like treatment appropriate 
has the non-adiabatic (first order of $\om$) information. 
We showed that the quantum pump dose not occur in all orders of the pumping frequency when the system control parameters and the
thermodynamic parameters (the temperatures and the chemical potentials of the baths) are fixed under the zero-bias condition. 

Next, we studied the quantum adiabatic pump of the quantum dot (QD) system weakly coupled to two leads ($L$ and $R$) in \res{Non-interacting} and \res{Interacting}
using the FCS-QME with the rotating wave approximation (RWA) defined as the long coarse-graining time limit of the coarse-graining approximation (CGA). 
We confirmed the consistency between the FCS-QME approach and the GME approach for a QD of one quantum level with finite Coulomb interaction (\res{pump,U=0} and \res{pump}). 
We showed that the pumped charge and spin coming from the instantaneous steady current are not negligible when the thermodynamic parameters are not fixed to zero bias
 (\res{SC} and \res{current,inf}). 
To observe the spin effects, we consider collinear magnetic fields, which affect the spins through the Zeeman effect, with different amplitudes applying to the QDs ($B_S$) 
and the leads ($B_L$ and $B_R$). 
We focused on the dynamic parameters ($B_S$, $B_{L/R}$ and the coupling strength between QDs and leads, $\Dl_{L/R}$) as control parameters. 
In one level QD with the Coulomb interaction $U$, we analytically calculated the BSN curvatures of spin and charge of $(B_L,B_S)$ pump 
and $(\Dl_L,B_S)$ pump for the noninteracting limit ($U=0$) and the strong interaction limit ($U=\infty$) at zero-bias.
The difference depending on $U$ appeared through $n_U(s B_S)$ which is the average number of the electrons with spin $s$ in the QD. 
For $(B_L, B_S)$ pump, the energy dependences of the line-width functions are essential. 
Moreover, we studied the $(\Dl_L,B_S)$ pump for finite $U$ at zero-bias (\res{pump_finite}). 
The effect of $U$ appeared through $n_U(s B_S)$. 
When half-filling condition is satisfied, the charge pump does not occur.

We studied the quantum diabatic pump for spinless one level QD coupled to two leads (\rec{diabatic}).  
We calculated $\{\rho^{(n)}(t)\}_{n=1}^5$, $\{\tl \rho^{(n)}(t)\}_{n=1}^5$ and particle current up to 6th order and 
pumped particle numbers. 
 
In \res{Introduction_EE}, we newly defined average entropy production rate $\dot{\sig}(t)$ using the average energy and particle currents,
which are calculated by using the FCS-QME. 
Next, we introduced the generalized QMEs (GQMEs) providing $\dot{\sig}(t)$ (\rec{sGQME}). 
The GQMEs do not relate the higher moments (thus and the FCS) of the entropy production. 
We can calculate only the average of the entropy production. 
In \res{noneq}, using the GQME, in weakly nonequilibrium regime, we analyzed the BSN vector for the entropy production, $A_n^\sig(\al)$, 
which provides the excess entropy production $\sig_\RM{ex}$ under quasistatic operations between nonequilibrium steady states
 as $\sig_\RM{ex}=\int_C \ d\al^n \ A_n^\sig(\al)$, and showed
$A_n^\sig(\al)=-\tr_S\big[\ln \rho_0^{(-1)}(\al)\f{\p \rho_0(\al)}{\p \al^n} \big]+\mO(\ep^2)$.  
Here, $\al$ is the set of the control parameters and $\al^n$ is $n$-th component of the control parameters, $C$ is the trajectory in the control parameter space, 
$\tr_S$ denotes the trace of the system, and $\ep$ is a measure of degree of nonequilibrium. 
$\rho_0^{(-1)}(\al)$ is the instantaneous steady state obtained from the QME with reversing the sign of the Lamb shift term. 
In general, the potential $\mS(\al)$ such that $A_n^\sig(\al)=\f{\p \mS(\al)}{\p \al^n} +\mO(\ep^2)$ dose not exist (\res{Discussion}). 
This is the most important result of this thesis. 
The origins of the non-existence of the potential $\mS(\al)$ are a quantum effect (the Lamb shift term) and the breaking of the time-reversal symmetry. 
The non-existence of the potential means that the excess entropy essentially depends on the path of the modulation. 
In this case, it is important to consider the generalization of the entropy concept. 
In contrast, if the system Hamiltonian is non-degenerate or the Lamb shift term is negligible, we obtain 
$\sig_\RM{ex} =S_\RM{vN}(\rho_0(\al_{t_f}))-S_\RM{vN}(\rho_0(\al_{t_i}))+\mO(\ep^2 \dl)$. 
Here, $S_\RM{vN}(\rho) = -\tr_S[\rho \ln\rho]$ is the von Neumann entropy, $t_i$ and $t_f$ are the initial and final times of the operation, 
and $\dl$ describes the amplitude of the change of the control parameters. 
For time-reversal symmetric system, we showed that $\mS(\al)$ is the symmetrized von Neumann entropy.  
Additionally, we pointed out that preceding expression of the entropy production in the classical Markov jump process is different from ours 
and showed that these are approximately equivalent in the weakly nonequilibrium regime. 
We also checked that the definition of the average entropy production in the classical Markov jump process by Ref.\cite{Sagawa} is equivalent to ours.

\subsection{Future perspective} 

 $\rho_0^{(-1)}$ and $A_n^\sig(\al)$ should be calculated for concrete model 
in which the system Hamiltonian is degenerated or/and the time-reversal symmetry is broken. 
For instance, multi-level QD system applying the magnetic field is a candidate. 

If $\mS(\al)$ does not exist, the path dependence of the excess entropy is essential. 
The path dependence and the path of which the excess entropy is minimized should be studied.

\newpage

\renewcommand{\abstractname}{Acknowledgements}
\begin{abstract}
I would express my sincere appreciation to the PhD adviser Prof. Yasuhiro Tokura, for invaluable discussions, his enthusiasm and patience.
I acknowledge helpful discussions with N. Taniguchi, S. Okada, S. Nomura, 
S. Ajisaka, K. Watanabe, H. Hayakawa, R. Yoshii and  Yu Watanabe. 
\end{abstract}

\newpage

\appendix

\section{Born-Markov approximation} \la{BMA}

We denote $\mL_b^\chi$ in the Born-Markov approximation by $\mL_{b(\RM{BM})}^\chi$. 
From \re{BM} and \re{tr_u_s}, we obtain
\bea
\mL_{b(\RM{BM})}^\chi \bu 
\aeq -\int_0^\infty ds \ \sum_{\mu,\nu}\Big( s_{b\nu}^{\dagger}s_{b\mu}^I(t-s,t)\bu C_{b,\nu\mu}(s) 
-s_{b\mu}^I(t-s,t)\bu s_{b\nu}^{\dagger}C_{b,\nu\mu}^\chi(s) \no\\
&&- s_{b\nu}\bu s_{b\mu}^{I\dagger}(t-s,t) C_{b,\mu\nu}^\chi(-s) 
+ \bu s_{b\mu}^{I\dagger}(t-s,t)s_{b\nu} C_{b,\mu\nu}(-s) \Big). 
\eea
$C_{b,\mu\nu}(s)$ damps exponentially as $e^{-\abs{s}/\tau_b}$ where $\tau_b$ is the relaxation time of the bath $b$. 
Then, in the calculations of $s_{b\mu}^I(t-s,t)$ and $s_{b\mu}^{I\dagger}(t-s,t)$, 
the values of the control parameters can be approximated by $\al_S(t)$. 
Then, we obtain
\bea
s_{b\mu}^I(t-s,t) = \sum_\om  e^{i\om s}s_{b\mu}(\om), \
s_{b\mu}^{I\dagger}(t-s,t) = \sum_\om  e^{-i\om s}[s_{b\mu}(\om)]\dg,
\eea
and 
\bea
\mL_{b(\RM{BM})}^\chi \bu 
\aeq -\int_0^\infty ds \  \sum_{\mu,\nu}\sum_\om \Big( \Big\{ s_{b\nu}^{\dagger}s_{b\mu}(\om)\bu C_{b,\nu\mu}(s) 
-s_{b\mu}(\om)\bu s_{b\nu}^{\dagger}C_{b,\nu\mu}^\chi(s) \Big\}e^{i\om s} \no\\
&&+\Big\{- s_{b\nu}\bu [s_{b\mu}(\om)]\dg C_{b,\mu\nu}^\chi(-s) 
+ \bu [s_{b\mu}(\om)]\dg s_{b\nu} C_{b,\mu\nu}(-s) \Big\}e^{-i\om s} \Big). 
\eea
Here, 
\bea
\int_0^\infty ds \ C_{b,\nu\mu}^\chi(s) e^{i\om s}\aeq \int_0^\infty ds\int_{-\infty}^\infty d\Om \ \f{1}{2\pi}\Phi_{b,\nu\mu}^\chi(\Om) e^{i(\om-\Om) s} \no\\
\aeq \int_{-\infty}^\infty d\Om \ \f{1}{2\pi}\Big[\pi\dl(\Om-\om)-i\RM{P}\f{1}{\Om-\om} \Big]\Phi_{b,\nu\mu}^\chi(\Om)\no\\
\aeq \Phi_{b,\nu\mu}^{(+)\chi}(\om),
\eea
and 
\bea
\int_0^\infty ds \ C_{b,\mu\nu}^\chi(-s) e^{-i\om s}\aeq \Phi_{b,\mu\nu}^{(-)\chi}(\om),
\eea
hold. 
Then, we get 
\bea
\mL_{b(\RM{BM})}^\chi \bu 
\aeq - \sum_{\mu,\nu}\sum_\om \Big(  s_{b\mu}^{\dagger}s_{b\nu}(\om)\bu \Phi_{b,\mu\nu}^{(+)}(\om) 
-s_{b\nu}(\om)\bu s_{b\mu}^{\dagger}\Phi_{b,\mu\nu}^{(+)\chi}(\om)  \no\\
&&- s_{b\nu}\bu [s_{b\mu}(\om)]\dg \Phi_{b,\mu\nu}^{(-)\chi}(\om)
+ \bu [s_{b\mu}(\om)]\dg s_{b\nu} \Phi_{b,\mu\nu}^{(-)}(\om)  \Big) \no\\
\aeq \mL_{b(\RM{BM})}^{\Phi, \chi}\bu+\mL_{b(\RM{BM})}^{\Phi, \chi}\bu.
\eea
Here, 
\bea
\mL_{b(\RM{BM})}^{\Phi, \chi}\bu \aeq
-\half \sum_{\mu,\nu}\sum_\om  \Big(  \Phi_{b,\mu\nu}(\om)s_{b\mu}^{\dagger}s_{b\nu}(\om)\bu  
-\Phi_{b,\mu\nu}^{\chi}(\om)s_{b\nu}(\om)\bu s_{b\mu}^{\dagger}  \no\\
&&-\Phi_{b,\mu\nu}^{\chi}(\om) s_{b\nu}\bu [s_{b\mu}(\om)]\dg 
+ \Phi_{b,\mu\nu}(\om)\bu [s_{b\mu}(\om)]\dg s_{b\nu}  \Big) ,\\
\mL_{b(\RM{BM})}^{\Psi, \chi}\bu \aeq \f{i}{2} \sum_{\mu,\nu}\sum_\om \half \Big(  \Psi_{b,\mu\nu}(\om)s_{b\mu}^{\dagger}s_{b\nu}(\om)\bu  
-\Psi_{b,\mu\nu}^{\chi}(\om)s_{b\nu}(\om)\bu s_{b\mu}^{\dagger}  \no\\
&&+ \Psi_{b,\mu\nu}^{\chi}(\om)s_{b\nu}\bu [s_{b\mu}(\om)]\dg 
- \bu \Psi_{b,\mu\nu}(\om) [s_{b\mu}(\om)]\dg s_{b\nu} \Big).
\eea
For \re{H_Sb}, we obtain 
\bea
\mL_{b(\RM{BM})}^{\Phi, \chi}\bu \aeq
-\half \sum_{\al,\be}\sum_\om  \Big(  \Phi_{b,\al\be}^-(\om)a_\al^{\dagger}a_\be(\om)\bu  
-\Phi_{b,\al\be}^{-,\chi}(\om)a_\be(\om)\bu a_\al^{\dagger}  \no\\
&&-\Phi_{b,\al\be}^{-,\chi}(\om) a_\be\bu [a_\al(\om)]\dg 
+  \Phi_{b,\al\be}^-(\om) \bu [a_\al(\om)]\dg a_\be \no\\
&&+\Phi_{b,\al\be}^+(\om)  a_\al[a_\be(\om)]\dg \bu 
-\Phi_{b,\al\be}^{+,\chi}(\om)[a_\be(\om)]\dg \bu a_\al   \no\\
&&-\Phi_{b,\al\be}^{+,\chi}(\om) a_\be\dg \bu a_\al(\om)
+\Phi_{b,\al\be}^+(\om)  \bu a_\al(\om)a_\be \dg \Big), \la{BM_phi}
\eea
and 
\bea
\mL_{b(\RM{BM})}^{\Psi, \chi}\bu \aeq
\f{i}{2} \sum_{\al,\be}\sum_\om  \Big(  \Psi_{b,\al\be}^-(\om)a_\al^{\dagger}a_\be(\om)\bu  
-\Psi_{b,\al\be}^{-,\chi}(\om)a_\be(\om)\bu a_\al^{\dagger}  \no\\
&&+\Psi_{b,\al\be}^{-,\chi}(\om) a_\be\bu [a_\al(\om)]\dg 
-  \Psi_{b,\al\be}^-(\om) \bu [a_\al(\om)]\dg a_\be \no\\
&&-\Psi_{b,\al\be}^+(\om)  a_\al[a_\be(\om)]\dg \bu 
+\Psi_{b,\al\be}^{+,\chi}(\om)[a_\be(\om)]\dg \bu a_\al   \no\\
&&-\Psi_{b,\al\be}^{+,\chi}(\om) a_\be\dg \bu a_\al(\om)
+\Psi_{b,\al\be}^+(\om)  \bu a_\al(\om)a_\be \dg \Big). \la{BM_psi}
\eea
By the way,
\bea
\mL_{b(\RM{BM})}^{\Phi} e^{-\be_b (H_S-\mu N_S)}=0, \la{BM_rho_ss}
\eea
holds. Here, $\mL_{b(\RM{BM})}^{\Phi}=\mL_{b(\RM{BM})}^{\Phi,\chi}\bv{\chi=0}$. 
Because of \re{KMS}, 1st and 7th terms of \re{BM_phi} cancel in the LHS of \re{BM_rho_ss}.
Similarly, 2nd and 8th, 3rd and 5th, 4th and 6th terms of \re{BM_phi} cancel.
If $\mL_{b(\RM{BM})}^{\Psi}$ is negligible, $\rho_0$ becomes \re{rho_gc} at zero-bias.

\newpage

\section{Liouville space} \la{Liouville space}

By following correspondence, an arbitrary linear operator (which operates to the Hilbert space) $\bu =\sum_{n,m} \br{n}\bu \ke{m} \ke{n}\br{m}$ is mapped to a vector of the 
Liouville space\cite{FCS-QME, Fano}, $\dke{\bu}=\sum_{n,m} \br{n}\bu \ke{m} \dke{nm} $:
\bea
 \ke{n}\br{m} \!\!\! &\longleftrightarrow& \!\!\! \dke{nm} ,\la{taiou1} \\
 \tr(\ke{m}\br{n} n^\pr \ket \br{m^\pr} )  \!\!\! &\longleftrightarrow& \!\!\! \dbra nm \vert n^\pr m^\pr \dket , \\
 \tr(A\dg B) \!\!\! &\longleftrightarrow& \!\!\! \dbra A \vert B \dket ,\la{taiou3}  \\
\tr(\bu)  \!\!\! &\longleftrightarrow& \!\!\! \dbra 1\dke{\bu} .
\eea
Here, $\{\ke{n}\}$ is an arbitrarily complete orthonormal basis. The inner product of the Liouville space is defined by the Hilbert-Schmidt product [\re{taiou3}]. The Hermitian conjugate of $\dke{\bu}$ is defined as $\dbr{\bu}=(\dke{\bu})\dg=\sum_{n,m} \br{n}\bu \ke{m}^\ast \dbr{nm} $.
An arbitrary linear super-operator $\Hat{J}$ which operates to any operator ($\bu$) is mapped to a corresponding operator of the Liouville space ($\check{J}$) as
\bea
\dke{\Hat{J}\bu} \aeq \check{J}\dke{\bu} .
\eea
The matrix representation of $\check{J}$ (or $\Hat{J}$) is defined by 
\bea
J_{nm,kl} \aeq \dbr{nm} \check{J} \dke{kl} .
\eea
In the main text of this thesis, both $\check{J}$ and $\Hat{J}$ are denoted by $\Hat{J}$.

Generally, the Liouvillian $\Hat{K}^\chi$ operates to an operator $\bu$ as
\bea
\Hat{K}^\chi\bu \aeq -i[H_S,\bu]+\Hat{\Pi}^\chi \bu ,\la{K}\\
\Hat{\Pi}^\chi \bu \aeq \sum_a c_a^\chi A_a \bu B_a ,
\eea
where $H_S$ is the system Hamiltonian, $\Hat{\Pi}^\chi$ is the dissipator, $A_a$, $B_a$ are operators, and $c_{a}^\chi(\al)$ is a complex number.
The matrix representation of \re{K} is given by
\bea
\sum_{k,l} K_{nm,kl}^\chi \bu_{kl}
\aeq \sum_{k,l} \big[-i\{ (H_S)_{nk} \dl_{lm} -\dl_{nk} (H_S)_{lm} \} \bu_{kl} \no\\
&&+\{ \sum_a c_a^\chi(A_a)_{nk} (B_a)_{lm} \} \bu_{kl} \big] ,
\eea
where $\bu_{kl}=\br{k}\bu \ke{l}$.  Hence the matrix representation of $\Hat{K}^\chi$ is given by
\bea
K_{nm,kl}^\chi \aeq -i H_{nm,kl}+\Pi_{nm,kl}^\chi , \la{mat_rp1}\\
H_{nm,kl} \aeq (H_S)_{nk} \dl_{lm} -\dl_{nk} (H_S)_{lm} ,\la{mat_rp2} \\
\Pi_{nm,kl}^\chi  \aeq  \sum_a c_a^\chi (A_a)_{nk} (B_a)_{lm} \la{mat_rp3} .
\eea

Finally, we consider the current operators defined by \re{defW}. $\Hat{K}^{O_\mu}=\f{\partial \Hat{K}^\chi(\al) }{\partial (i\chi_{O_\mu})} \big \vert_{\chi=0}$ is given by
\bea
\Hat{K}^{O_\mu} \bu \aeq \sum_a c_a^{O_\mu}A_a \bu B_a.
\eea
Hence the current operators defined by \re{defW} are given by
\bea
W^{O_\mu} \aeq \sum_a c_a^{O_\mu}B_a A_a. \la{W_AB}
\eea

\newpage

\section{The time evolutions of $c_n^\chi(t)$} \la{co}

In this chapter, we derive the time evolution equations of $c_n^\chi(t)$ of \re{exp}.
The LHS of the FCS-QME, $\f{d}{dt} \dke{\rho^\chi(t)}=\Hat{K}^\chi(\al_t)  \dke{\rho^\chi(t)}$, is
\bea 
\f{d}{dt} \dke{\rho^\chi(t)} \aeq \sum_n \Big\{ \f{dc_n^\chi(t)}{dt} e^{\Lm_n^\chi (t)}\dke{\rho_n^\chi(\al_t)}
+c_n^\chi(t) e^{\Lm_n^\chi (t)} \lm_n(\al_t) \dke{\rho_n^\chi(\al_t)}\no\\
&&+c_n^\chi(t) e^{\Lm_n^\chi (t)} \f{d}{dt}\dke{\rho_n^\chi(\al_t)}  \Big\} .
\eea
And the RHS  of the FCS-QME is
\bea
\Hat{K}^\chi(\al_t)  \dke{\rho^\chi(t)}
\aeq \sum_n c_n^\chi(t) e^{\Lm_n^\chi (t)}\Hat{K}^\chi(\al_t)  \dke{\rho_n^\chi(\al_t)} \no\\
\aeq \sum_n c_n^\chi(t) e^{\Lm_n^\chi (t)}\lm_n(\al_t)   \dke{\rho_n^\chi(\al_t)}.
\eea
Hence we obtain
\bea
\sum_n \Big\{ \f{dc_n^\chi(t)}{dt} e^{\Lm_n^\chi (t)}\dke{\rho_n^\chi(\al_t)}
+c_n^\chi(t) e^{\Lm_n^\chi (t)} \f{d\dke{\rho_n^\chi(\al_t)}}{dt}  \Big\} =0. \la{e.c}
\eea
Applying $\dbr{l_m^\chi(\al_t)}$ to \re{e.c}, and using $\dbra l_n^\chi(\al) \dke{\rho_m^\chi(\al)}=\dl_{nm}$, we obtain
\bea
\hs{-7mm}\f{d}{dt}c_m^\chi(t) \aeq -\sum_n c_n^\chi(t) e^{\Lm_n^\chi (t)-\Lm_m^\chi (t)} \dbr{l_m^\chi(\al_t)}\f{d\dke{\rho_n^\chi(\al_t)}}{dt} \la{berry_key}.
\eea

By the way, the time derivative of \re{rig}, $ \Hat{K}^\chi(\al_t) \dke{\rho_n^\chi(\al_t)} =\lm^\chi_n(\al_t) \dke{\rho_n^\chi(\al_t)}$, is 
\bea
\f{d \Hat{K}^\chi(\al_t) }{dt} \dke{\rho_n^\chi(\al_t)}+ \Hat{K}^\chi(\al_t) \f{d \dke{\rho_n^\chi(\al_t)} }{dt} 
\aeq \f{ d\lm^\chi_n(\al_t)}{dt} \dke{\rho_n^\chi(\al_t)} +\lm^\chi_n(\al_t) \f{d\dke{\rho_n^\chi(\al_t)} }{dt} .
\eea
Applying $\dbr{l_m^\chi(\al_t)}$ to this equation, we obtain
\bea
&&\hs{-10mm} \dbr{l_m^\chi(\al_t)} \f{d \Hat{K}^\chi(\al_t) }{dt} \dke{\rho_n^\chi(\al_t)}+ \lm_m^\chi(\al_t) \dbr{l_m^\chi(\al_t)} \f{d \dke{\rho_n^\chi(\al_t)} }{dt} \no\\
 \aeq \f{ d\lm^\chi_n(\al_t)}{dt} \dl_{mn}+\lm^\chi_n(\al_t) \dbr{l_m^\chi(\al_t)} \f{d\dke{\rho_n^\chi(\al_t)} }{dt} ,
\eea
and it leads to
\bea
\dbr{l_m^\chi(\al_t)} \f{d\dke{\rho_n^\chi(\al_t)}}{dt} \aeq -\f{\dbr{l_m^\chi(\al_t)} \f{d \Hat{K}^\chi(\al_t) }{dt} \dke{\rho_n^\chi(\al_t)}}{\lm^\chi_m(\al_t)-\lm^\chi_n(\al_t)},
 \la{at_hisyokutai}
\eea
for $\lm^\chi_m(\al_t)\ne \lm^\chi_n(\al_t)$. Substituting this to \re{berry_key}, we obtain
\bea
\f{dc^\chi_m(t)}{dt}\aeq-\dbr{l_m^\chi(\al_t)} \f{d}{dt}\dke{\rho_m^\chi(\al_t)}c^\chi_m(t) \no\\
&&+\sum_{n(\ne m)} c^\chi_n(t) e^{\Lm_n^\chi (t)-\Lm_m^\chi (t)} 
\f{\dbr{l_m^\chi(\al_t)} \f{d \Hat{K}^\chi(\al_t) }{dt} \dke{\rho_n^\chi(\al_t)}}{\lm^\chi_m(\al_t)-\lm^\chi_n(\al_t)} . \la{comment2}
\eea

The above equation can also be written as 
\bea
\f{d\tl c^\chi_m(t)}{dt}=\sum_{n(\ne m)} \tl c^\chi_n(t) e^{\Lm_n^\chi (t)-\Lm_m^\chi (t)+\eta_m^\chi(t)-\eta_n^\chi(t)} 
\f{\dbr{l_m^\chi(\al_t)} \f{d \Hat{K}^\chi(\al_t) }{dt} \dke{\rho_n^\chi(\al_t)}}{\lm^\chi_m(\al_t)-\lm^\chi_n(\al_t)} \la{comment3},
\eea
where $\tl c_m^\chi(t)=c_m^\chi(t)e^{\eta_m^\chi(t)}$ with 
\bea
\eta_m^\chi(t)\aeq \int_0^t ds \ \dbr{l_m^\chi(\al_s)} \f{d}{ds}\dke{\rho_m^\chi(\al_s)} \no\\
\aeq \sum_k \int_C d\al^k \  \dbr{l_m^\chi(\al)} \f{\partial}{\partial \al^k}\dke{\rho_m^\chi(\al)}.
\eea
Here, $C$ is the trajectory from $\al_0$ to $\al_t$, $\al^k$ are the $k$-th component of the control parameters, and  
$\eta_m^\chi(t)=\mO(1)$ since $\dbr{l_m^\chi(\al_t)} \f{d}{dt}\dke{\rho_m^\chi(\al_t)}=\mO(\om)$ with $\om=2\pi/\tau$. 
In the RHS of \re{comment3}, the dominant term is $n=0$ if $m\ne 0$ because Re$\lm_0^\chi(\al)>{\rm{Re}}\lm_n^\chi(\al)$. 
Using $\f{d \Hat{K}^\chi(\al_t)} {dt}=\mO(\Ga\om)$, $\lm_n^\chi(\al_t)=\mO(\Ga)$, $e^{\eta_n^\chi(t)}=\mO(1)$ and $c_0^\chi(t)e^{\Lm_0^\chi}=\mO(1)$, we obtain 
\bea
\f{d\tl c^\chi_m(t)}{dt}=\mO( e^{-\Lm_m^\chi (t)}\om ),
\eea 
and 
\bea
c_m^\chi(t)e^{\Lm_m^\chi(t)}\aeq \mO \big(\om \int_0^t ds \ e^{\Lm_m^\chi(t)-\Lm_m^\chi (s)} \big) = \mO\big(\f{\om}{\Ga}\big) \la{ada}.
\eea
For $\chi=0$, \re{ada} is also derived from 
\bea
\dl \rho(t) =\rho(t)-\rho_0(\al_t) =\sum_{m\ne 0}c_m(t)e^{\Lm_m(t)}\rho_m(\al_t),
\eea 
and  (\ref{key6}) and (\ref{order}).

\newpage

\section{The validity of the adiabatic expansion} \la{val}

In the derivation of the QME with CGA, when going from \re{def,CGA} to \re{B7}, we used the following type of approximation:
\bea
\int_t^{t+T}du \int_t^u ds \ G([\al]_s^u;s,u;t) \aeqap \int_t^{t+T}du \int_t^u ds \ G([\al_t];s,u;t).
\eea
Here, $G([\al]_s^u,s,u,t) \sim e^{-(u-s)/\tau_B}$ and $[\al]_s^u=(\al_{t^\pr})_{t^\pr=s}^u$ is the control parameter trajectory and $[\al_t]$ is the trajectory which 
$\al_{t^\pr}=\al_t \ (s \le t^\pr \le u)$. $\tau_B$ is  the relaxation time of the baths.
Similarly, in the Born-Markov approximation (BM), when going from 
\bea
\f{d\rho^{I,\chi}(t)}{dt} \aeq -\int_0^t du \ \tr_B\Big\{[H_{\rm{int}}^I(t),[H_{\rm{int}}^I(u),\rho^{I,\chi}(t) \rho_B (\al_t)]_\chi]_\chi \Big\}, \la{RD}
\eea
 to \re{FCS-QME}, we used
\bea
\int_0^t du \ G([\al]_u^t;u,t;t) \aeqap \int_{-\infty}^t du \ G([\al_t];u,t;t).
\eea
Considering the corrections of the above approximations, the QME is given by
\bea
\f{d\dke{\rho(t)}}{dt} \aeq \mathcal{K}(t)\dke{\rho(t)} ,\la{QM} \\ 
\mathcal{K}(t)\aeq \Hat{K}(\al_t)+\Hat{K}^{(1)}(t) ,\ \Hat{K}^{(1)}(t) =\mO(\Ga \om \tau_X) ,
\eea
with $\om=2\pi/\tau$ and $\tau_X=\tau_{\rm{CG}}$ for CGA; $\tau_X=\tau_B$ for BM. 
$\Hat{K}^{(1)}(t)$ corresponds to $\bm{K}_{[1]}^{(1)}(t)$ of Appendix \ref{A_GME}. 
The discussions of \res{Expansion} are correct after replacing ${K}(\al_t) \to \mathcal{K}(t)$, $\mR(\al_t) \to \bar{\mR}(t)$ and $\rho_0(\al_t) \to \bar{\rho}_0(t)$. 
Here, $\bar{\rho}_0(t)$ and $\bar{\mR}(t)$ are defined by $\mathcal{K}(t)\dke{\bar{\rho}_0(t)}=0$ and
$\bar{\mR}(t)\mathcal{K}(t)=1-\dke{\bar{\rho}_0(t)}\dbr{1}$, respectively. (\ref{key6}) is corrected to
\bea
 \dke{\bar{\dl} \rho_{(\RM{ss})}(t)} = \sum_{n=1}^\infty \Big[\bar{\mR}(t)\f{d}{dt} \Big]^n \dke{\bar{\rho}_0(t)}\e \sum_{n=1}^\infty \dke{\bar{\rho}^{(n)}(t)},
\eea
with $\bar{\dl} \rho(t) \defe \rho(t)-\bar{\rho}_0(t)$. The corrections are given by
\bea
\bar{\rho}_0=\rho_0[1+\mO(\om \tau_X)] ,\ \bar{\mR}=\mR[1+\mO(\om \tau_X)] \la{D1},
\eea
and
\bea
 \bar{\rho}^{(n)}(t)-\rho^{(n)}(t)=\mO\Big(\big(\f{\om}{\Ga}\big)^n \om \tau_X \Big).
\eea
Next, we consider the reasonable range of $n$ of $\rho^{(n)}(t)$. Because $\rho^{(n)}(t)=\mO(\f{\om}{\Ga})^n$ and $\bar{\rho}_0(t)-\rho_0(\al_t)=\mO(\om \tau_X)$, 
the reasonable range is $n \le n_{\rm{max}}$, where $n_{\rm{max}}$ is determined by
\bea
\big(\f{\om}{\Ga}\big)^{n_{\rm{max}}+1}<\om \tau_X \ll \big(\f{\om}{\Ga}\big)^{n_{\rm{max}}} . \la{n_max}
\eea
Let us consider that reasonable concrete values of the parameters in the model of \res{model}:
$\om=10^p $ MHz, $\Ga=10\ \mu$eV=0.116 K, $1/\Ga=65.8$ ps, $\tau_{\rm{CG}}=1$ ps, and $\tau_B=0.1$ ps. These values lead to
\bea
\hs{-2.5mm}\om \tau_{\rm{CG}}\aeq 10^{-6+p} ,\ \om \tau_B= 10^{-7+p},\ \f{\om}{\Ga}=10^{-4.18+p},
\eea
and $n_{\rm{max}}=[\tl n_{\rm{max}}]$ with
\bea
\tl n_{\rm{max}}=\f{-6+p}{-4.18+p} \ ({\rm{CGA}}),\ \f{-7+p}{-4.18+p} \ (\rm{BM}).
\eea
Here, $[n]$ means the biggest integer below $n$. At $p=0$, $\tl n_{\rm{max}}=1.44$ (CGA), 1.67 (BM) and at $p=3$, $\tl n_{\rm{max}}=2.54$ (CGA), 3.39 (BM).
The larger the non-adiabaticity ($\f{\om}{\Ga}$), the larger $n_{\rm{max}}$ becomes.

\newpage

\section{Proof of \re{W_R}} \la{proof}

First, using  (\ref{defR}) and (\ref{touka_sc}), 
we obtain 
\bea
\dbr{1}W^{O_\mu}(\al)\mR(\al)\Hat{K}(\al)=\dbr{1}W^{O_\mu}(\al)-\lm_0^{O_\mu}(\al)\dbr{1}.
\eea 
Next, $\dbr{l_0(\al)}=\dbr{1}$, $\lm_0(\al)=0$, and  (\ref{left}) and (\ref{defW}) lead to 
\bea
\dbr{l_0^{O_\mu}(\al)}\Hat{K}(\al)=\lm_0^{O_\mu}(\al)\dbr{1}-\dbr{1}W^{O_\mu}(\al).
\eea 
Hence, we obtain
\bea
\big[ \dbr{1}W^{O_\mu}(\al)\mR(\al)+\dbr{l_0^{O_\mu}(\al)} \big]\Hat{K}(\al)=0 ,
\eea
 and it leads to \re{W_R}. 
To prove \re{W_R} only \re{defR} is required and $\Hat{K}(\al)\mR(\al)=1-\dke{\rho_0(\al)}\dbr{1}$ is not necessary. 
Additionally, the pseudo-inverse of the GME approach \re{RR} satisfies 
\bea
\sum_{j}R_{ij}K_{j k}^{(0)} =\dl_{i k}-p_i^{(0)} \ne \sum_{j}K^{(0)} _{i j}R_{j k}, 
\eea
which corresponds to our
\bea
\mR(\al)\Hat{K}(\al)=1-\dke{\rho_0(\al)}\dbr{1}\ne \Hat{K}(\al)\mR(\al). 
\eea

(\ref{W_R}) is shown also as follows. \re{defR} and $\dbr{1}\Hat{K}(\al)=0$ lead to $\Hat{K}(\al)\mR(\al)\Hat{K}(\al)=\Hat{K}(\al)$, which implies
\bea
\Hat{K}(\al)\mR(\al) \aeq  1-\dke{\sig(\al)} \dbr{1}, \ \dbra 1\dke{\sig(\al)}=1 . \la{KR} 
\eea
Applying $\dbr{1}$ to \re{defR}, we obtain $\dbr{1}\mR(\al)\Hat{K}(\al)=0$, which is equivalent to
\bea
\dbr{1} \mR(\al)\aeq \mathcal{C}(\al)\dbr{1}. \la{1_R}
\eea
By the way, differentiating \re{left} for $n=0$ by $i\chi_{O_\mu}$, we obtain 
\bea
\dbr{l_0^{O_\mu}(\al)}\Hat{K} (\al)+\dbr{1} \Hat{K}^{O_\mu}(\al)  \aeq \dbr{1}\lm_0^{O_\mu}(\al). 
\eea
Applying $\mR(\al)$ to this equation and using  (\ref{KR}) and (\ref{1_R}), we obtain\cite{Sagawa, yuge2}
\bea
\dbr{l_0^{O_\mu}(\al)}\aeq -\dbr{1} \Hat{K}^{O_\mu}(\al)\mR(\al) +c^{O_\mu}(\al) \dbr{1} ,\la{Y}\\
c^{O_\mu}(\al) \aeq \mathcal{C}(\al)\lm_0^{O_\mu}(\al)+\dbr{l_0^{O_\mu}(\al)}\sig(\al)\dket \la{Y2}.
\eea
\re{Y} becomes \re{W_R} because of \re{defW}. 
Particularly,  Yuge\cite{yuge2} used
\bea
\mR(\al) \aeq -\lim_{s \to \infty}\int_0^s dt \ e^{\Hat{K}(\al)t}(1-\dke{\rho_{0}(\al)} \dbr{1}) \la{R_Yuge},
\eea
which satisfies  (\ref{KR}) and (\ref{1_R}) with $\sig(\al)=\rho_0(\al)$, $\mathcal{C}(\al) = 0$ and \re{defR} (in \Refe{yuge2}, $\mathcal{C}(\al)$ was incorrectly set to $-1$).

\newpage

\section{Generalized mater equation and frequency-expansion} \la{A_GME}

The first half of this chapter is based on \cite{RT09}. The GME is  
\bea
\f{d}{dt}\bm{p}(t) \aeq \int_{-\infty}^t dt^\pr \ \bm{W}(t,t^\pr)\bm{p}(t^\pr) , \la{1}
\eea
where $\bm{p}={}^t(p_1,p_2,\cdots,p_n)$.
$\bm{W}(t,t^\pr)$ functionally depends on $\al_t$.
We expand $\bm{p}$ and $\bm{W}$ by the modulation frequency $\om$ of the control parameters: 
\bea
\bm{p}(t) = \sum_{k=0}^\infty \bm{p}^{(k)}(t) ,\ \bm{W}(t,t^\pr) = \sum_{k=0}^\infty \bm{W}^{(k)}(t;t-t^\pr).
\eea
$\bm{p}^{(k)}$ and $\bm{W}^{(k)}$ are proportional to $\om^k$. 
In general, $\bm{p}(t)$ should contain a term which exponentially damps as $e^{-\Ga \cdot(t-t_0)}$. 
Here, $\Ga$ is the coupling strength between the system and the baths. 
However, this method suppose $t_0 \to -\infty$. 
The RHS of \re{1} becomes
\bea
\int_{-\infty}^t dt^\pr \ \bm{W}(t,t^\pr)\bm{p}(t^\pr) \aeq \sum_{p,q}\int_{-\infty}^t dt^\pr \ \bm{W}^{(p)}(t;t-t^\pr)\bm{p}^{(q)}(t^\pr) \no\\
\aeq \sum_{p,q,k}\int_{-\infty}^t dt^\pr \ \bm{W}^{(p)}(t;t-t^\pr)\f{(-1)^k(t-t^\pr)^k}{k!}\f{d^k\bm{p}^{(q)}(t)}{dt^k} \no\\
\aeq \sum_{p,q,k}\f{1}{k!} \f{\partial^k}{\partial z^k}\Bv{z=0}\int_{-\infty}^t dt^\pr \ \bm{W}^{(p)}(t;t-t^\pr)e^{-z(t-t^\pr)}\f{d^k\bm{p}^{(q)}(t)}{dt^k} \no\\
\aeqe \sum_{p,q,k}\f{1}{k!}  \partial^k \bm{K}^{(p)}(t)\f{d^k\bm{p}^{(q)}(t)}{dt^k}.
\eea
Then, \re{1} becomes 
\bea
\f{d}{dt}\bm{p}^{(n)}(t) \aeq \sum_{p,q,k}^{(n)}\f{1}{k!}  \partial^k \bm{K}^{(p)}(t)\f{d^k\bm{p}^{(q)}(t)}{dt^k}.
\eea
$\sum_{p,q,k}^{(n)}$ is the summation over terms which have the same order with the LHS. 
How to count the time derivatives of $\bm{p}$ in this
expansion depends on the considered frequency regime.
In this chapter, we consider the regime $\om \lesssim \Ga$, for which
the system quickly relaxes to an oscillatory steady state with the frequency of $\al_t$; i.e., each time derivative introduces one power in $\om$. 
Then,
\bea
n+1=p+q+k, \la{M1}
\eea
holds. 

Now, we expand $\bm{p}^{(q)}$ and $\bm{K}^{(p)}(t)$ by $\Ga$:
\bea 
\bm{K}^{(p)}(t) = \sum_{j_K=1}^\infty \bm{K}_{[j_K]}^{(p)}(t) ,\ \bm{p}^{(k)}(t) = \sum_{j_p^{(k)}=-k}^\infty \bm{p}_{[j_p^{(k)}]}^{(k)}(t).
\eea
${}_{[j]}$ indicates terms of order $\Ga^j$. 
This matching requires the expansion
for $\bm{p}^{(k)}(t)$ to start from $\Ga^{-k}$. 
The matching condition for $\Ga$ is  
\bea
j_p^{(n)}=j_K+j_p^{(q)}. \la{M2}
\eea

In the following, we consider the Born approximation: $\bm{K}^{(p)}(t) = \bm{K}_{[1]}^{(p)}(t)$. 
Then, \re{M2} becomes $j_p^{(n)}=1+j_p^{(q)}$. 
This can be rewritten as 
\bea
n+1=q+j_n-j_q ,
\eea
 where $j_p^{(n)}=-n+j_n$ and $j_p^{(q)}=-q+j_q$. 
The above equation and \re{M1} lead
\bea
p+k+j_q=j_n. \la{MC}
\eea
First, we consider $j_n=0$. 
Then, solution of \re{MC} is only $(p,k,j_q)=(0,0,0)$. 
Then, we obtain
\bea
\f{d\bm{p}_{[-k]}^{(k)}(t)}{dt} \aeq \bm{K}_{[1]}^{(0)}(\al_t) \bm{p}_{[-(k+1)]}^{(k+1)}(t) \hs{3mm}(k=0,1,\cdots). \la{L}
\eea
Where, $\bm{K}_{[1]}^{(0)}(t)$ is function of only $\al_t$. 
Because the LHS of \re{1} does not have terms of $\mO(\om^0)$, 
we get 
\bea
0=\bm{K}_{[1]}^{(0)}(\al_t)\bm{p}_{[0]}^{(0)}. \la{SS}
\eea
This is the definition of the instantaneous steady state. 

Reference\cite{RT09} considered only the solutions of $j_n=0$. 
However, the solutions of $j_n>0$ should also be considered.
We consider $j_n=1$. Then, the solutions of \re{MC} are $(p,k,j_q)=(1,0,0)$, $(0,1,0)$, $(0,0,1)$:
\bea
\f{d\bm{p}_{[-k+1]}^{(k)}(t)}{dt} \aeq \bm{K}_{[1]}^{(1)}(t) \bm{p}_{[-k]}^{(k)}(t)+\partial \bm{K}_{[1]}^{(0)}(\al_t) \f{d\bm{p}_{[-k]}^{(k)}(t)}{dt}
+\bm{K}_{[1]}^{(0)}(\al_t)\bm{p}_{[-k]}^{(k+1)}(t). \la{H}
\eea
Here, 
\bea
\bm{p}_{[-k]}^{(k)}(t) =\mO(\f{\om^k}{\Ga^k}) ,\ \f{d\bm{p}_{[-k]}^{(k)}(t)}{dt}=\mO(\f{\om^{k+1}}{\Ga^k}),
\eea
and 
\bea
\bm{K}_{[1]}^{(1)}(t)=\mO(\Ga \om\tau_B) ,\ \partial \bm{K}_{[1]}^{(0)}(\al_t)=\mO(\Ga \tau_B),
\eea
hold. $\tau_B$ is the relaxation time of the baths. Then, we obtain
\bea
\bm{K}_{[1]}^{(1)}(t) \bm{p}_{[-k]}^{(k)}(t)=\mO(\f{\om^{k+1}}{\Ga^{k-1}}\tau_B) ,\
\partial \bm{K}_{[1]}^{(0)}(\al_t) \f{d\bm{p}_{[-k]}^{(k)}(t)}{dt}=\mO(\f{\om^{k+1}}{\Ga^{k-1}}\tau_B).
\eea
\re{H} leads
\bea
\bm{p}_{[-k]}^{(k+1)}(t) \aeq \bm{R}(\al_t)\f{d\bm{p}_{[-k+1]}^{(k)}(t)}{dt}-\bm{R}(\al_t) \Big[\bm{K}_{[1]}^{(1)}(t) \bm{p}_{[-k]}^{(k)}(t)
+\partial \bm{K}_{[1]}^{(0)}(\al_t) \f{d\bm{p}_{[-k]}^{(k)}(t)}{dt} \Big]. \no\\
 \la{p_-k^k+1}
\eea
Here, $\bm{R}(\al)$ is the pseudo-inverse of $\bm{K}_{[1]}^{(0)}(\al)$:
\bea
\bm{R}(\al)\bm{K}_{[1]}^{(0)}(\al) \aeq 1-\bm{p}_{[0]}^{(0)}(\al)e ,\ e=(1,\cdots,1).
\eea
\re{p_-k^k+1} for $k=0$ leads
\bea
\bm{p}_{[0]}^{(1)}(t) \aeq \mO(\om\tau_B). \la{p_0^1}
\eea
Here, we used $\bm{p}_{[1]}^{(0)}(t)=0$. 
Then, considering $\bm{p}_{[-k]}^{(k)}$ smaller than $\mO(\om\tau_B)$ is meaningless. 
This result \re{p_0^1} is the same order with that discussed in Appendix \ref{val}. 

Under the Born approximation, the difference between the QME and the GME is 
\bea
\sum_{k=0}^\infty \sum_{j_p^{(k)}=-k+1}^\infty \bm{p}_{[j_p^{(k)}]}^{(k)}(t)=\bm{p}_{[0]}^{(1)}(t)+\cdots .
\eea
The origin of this is the non-Markovian property of the GME.

\newpage

\section{Energy current operator} \la{Current_op}

Similar to \re{def_B(om)}, we introduce 
\bea
[R_{b,\mu}\dg](\Om_b) \aeq \sum_{n,m,r,s} \dl_{\Om_{b,m n},\Om_b} \ke{E_{b,n},r}\br{E_{b,n},r} R_{b,\mu}\dg \ke{E_{b,m},s} \br{E_{b,m},s},
\eea
with $ \Om_{b,mn}= E_{b,m}-E_{b,n}$ and $H_b\ke{E_{b,n},r}=E_{b,n}\ke{E_{b,n},r}$. $r$ denotes the label of the degeneracy. 
$\Om_b$ is one of the elements of $\{ \Om_{b,mn} \vert \ {  \br{E_{b,n},r} R_{b\mu}\dg \ke{E_{b,m},s} \ne 0 \hs{2mm} }^\exists \mu \} $. 
We set $\{O_\mu\}=\{H_b\}_b$. 
Then, 
\bea
R_{b,\mu,-2\chi}^{I\dagger}(u)=\sum_{\Om_b} [R_{b,\mu}\dg](\Om_b) e^{-i\Om_b u+i\chi_{H_b}\Om_b} ,
\eea
holds. 
Using this, we obtain 
\bea
\Phi_{b,\al \be}^{\chi}(\Om)\aeq 2\pi \sum_{\Om_b} \dl(\Om-\Om_b)e^{i\chi_{H_b}\Om_b}\tr_b(\rho_b [R_{b,\mu}\dg](\Om_b) R_{b,\nu} ).
\eea
This means
\bea
\Phi_{b,\mu\nu}^\chi(\Om)= e^{i\chi_{H_b}\Om}\Phi_{b,\mu\nu}(\Om).
\eea
Using this, we obtain 
\bea
W^{H_b}(\al)= \sum_{\om } \sum_{\mu,\nu} \om \Phi_{b,\mu \nu}(\om) [s_{b\mu}  (\om)]\dg s_{b\nu}(\om),
\eea
for the RWA. 

Using \re{Pi_dg}, we obtain 
\bea
w^{H_b}(\al)\aeq - \sum_{\om }  \sum_{\mu,\nu} \Big[
 \Phi_{b,\mu \nu}(\om) [s_{b\mu}(\om)]\dg H_S s_{b\nu} (\om) \no\\
&&-\half \Phi_{b,\mu \nu}(\om) H_S [s_{b\mu}(\om)] \dg s_{b\nu}(\om)
-\half \Phi_{b,\mu \nu}(\om) [s_{b\mu}(\om)] \dg s_{b\nu}(\om) H_S \Big] \no\\
\aeq -  \sum_{\om }\sum_{\mu,\nu}  \Phi_{b,\mu \nu}(\om) [s_{b\mu}(\om)]\dg [H_S ,s_{b\nu} (\om)] \no\\
\aeq   \sum_{\om }\sum_{\mu,\nu} \om \Phi_{b,\mu \nu}(\om) [s_{b\mu}(\om)]\dg s_{b\nu} (\om).
\eea
Then, 
\bea
W^{H_b}(\al)=w^{H_b}(\al),
\eea
holds.

\newpage

\section{Derivative of the von Neumann entropy} \la{Appendix A}

We show that 
\bea
\f{\p S_\RM{vN}(\rho_0(\al))}{\p \al^n}=-\tr_S \Big[\ln \rho_0(\al) \f{\p \rho_0(\al)}{\p \al^n} \Big] . \la{dS}
\eea
From the definition of the von Neumann entropy, the LHS of the above equation is given by
\bea
\f{\p S_\RM{vN}(\rho_0(\al))}{\p \al^n}=-\tr_S \Big[\ln \rho_0(\al) \f{\p \rho_0(\al)}{\p \al^n} \Big]-\tr_S \Big[ \f{\p \ln \rho_0(\al)}{\p \al^n} \rho_0(\al) \Big].
\eea
Using \re{d ln A}, the second term of the RHS of the above equation becomes
\bea
-\tr_S \Big[ \f{\p \ln \rho_0(\al)}{\p \al^n} \rho_0(\al) \Big] \aeq 
-   \tr_S \Big[ \int_0^\infty ds \ \f{1}{\rho_0(\al)+s}\f{\p \rho_0(\al)}{\p \al^n}\f{1}{\rho_0(\al)+s} \rho_0(\al) \Big] \no\\
\aeq -  \tr_S \Big[ \int_0^\infty ds \ \f{\rho_0(\al)}{(\rho_0(\al)+s)^2}\f{\p \rho_0(\al)}{\p \al^n}  \Big] \no\\
\aeq - \tr_S \Big[ \f{\p \rho_0(\al)}{\p \al^n}  \Big]=0.
\eea
Then, we obtain \re{dS}.

\newpage

\section{Proof of \re{ln A+B}} \la{proof_ln A+B}

For an arbitrary operator $X$, 
\bea
\f{1}{1+\dl X}=(1+\dl X)^{-1}=1-\dl X+ \dl^2 X^2-\dl^3X^3+\cdots, \la{(1+X)^-1}
\eea
holds if the absolute value of a real number $\dl$ is sufficiently small.
Using this equation, 
\bea
\f{1}{A+\dl B}\aeq [A(1+\dl A^{-1}B)]^{-1}=(1+\dl A^{-1}B)^{-1}A^{-1} \no\\
\aeq \f{1}{A}-\dl \f{1}{A}B\f{1}{A}+\dl^2 \f{1}{A}B\f{1}{A}B\f{1}{A}-\dl^3 \f{1}{A}B\f{1}{A}B\f{1}{A}B\f{1}{A}+\cdots , \la{(A+B)^-1}
\eea
holds for an arbitrary operator $A$ which has $A^{-1}$. 
Here, we used \re{(1+X)^-1} for $X= A^{-1}B$. 
For an arbitrary operator $Y$ which has $Y^{-1}$, 
\bea
\int_0^a ds \ \f{1}{Y+s}=\ln(Y+a)-\ln Y,
\eea
holds for a real number $a$. 
Using this equation for $Y=A$ and $Y=A+\dl B$, we obtain
\bea
\ln(A+\dl B)\aeq \ln A+ \ln (A+\dl B+a)-\ln (A+a) \no\\
&&+\int_0^a ds \ \Big(\f{1}{A+s}-\f{1}{A+\dl B+s}\Big).
\eea
Using this equation and \re{(A+B)^-1}, we get
\bea
\ln(A+\dl B)\aeq \ln A+\ln (A+\dl B+a)-\ln (A+a) \no\\
&&+\int_0^a ds \ \Big(\dl \f{1}{A+s}B\f{1}{A+s}-\dl^2 \f{1}{A+s}B\f{1}{A+s}B\f{1}{A+s}+\cdots \Big). \la{ln A+B+a_1}
\eea
Because the second and third terms of the RHS are 
\bea
\ln (A+B+a)-\ln (A+a) \aeq \ln \Big(1+\f{A+B}{a}\Big)-\ln \Big(1+\f{A}{a}\Big)=\mO(\f{1}{a}), \la{ln A+B+a_2}
\eea
we obtain
\bea
\ln(A+\dl B)\aeq \ln A+\int_0^\infty ds \ \Big(\dl \f{1}{A+s}B\f{1}{A+s}-\dl^2 \f{1}{A+s}B\f{1}{A+s}B\f{1}{A+s}+\cdots \Big), \no\\
 \la{ln A+B_2}
\eea
for $a \to \infty$. 
The above equation is \re{ln A+B}.

We show \re{ln A+B+a_2}. 
Substituting $A=1$ to \re{ln A+B+a_1}, we get
\bea
\ln(1+\dl B)\aeq \ln (1+\dl B+a)-\ln (1+a) \no\\
&&+\int_0^a ds \ \Big(\dl \f{1}{(1+s)^2}B-\dl^2 \f{1}{(1+s)^3}B^2+\cdots \Big) \no\\
\aeq \ln (1+\dl B+a)-\ln (1+a) \no\\
&&+\int_0^a ds \ \Big(\dl \f{1}{(1+s)^2}B-\dl^2 \f{1}{(1+s)^3}B^2+\cdots \Big) \no\\
\aeq \ln\Big(1+\f{\dl B}{a+1}\Big)+\sum_{n=1}^\infty \f{(-1)^{n-1}}{n} \dl^n B^n \Big(1-\f{1}{(1+a)^n} \Big).
\eea
Using this equation for $a \to \infty$, we have
\bea
\ln(1+\dl B)\aeq \sum_{n=1}^\infty \f{(-1)^{n-1}}{n} \dl^n B^n,
\eea
which leads to \re{ln A+B+a_2}.

\newpage

\section{Definition of entropy production of the Markov jump process} \la{Appendix B}

Except \re{GMEn}, this chapter is based on Ref.\cite{Komatsu15}. 
We consider the Markov jump process on the states $n=1,2,\cdots,\mN$:
\bea
n(t)= n_k \hs{3mm} (t_k \le t<t_{k+1}), \ t_0= 0<t_1<t_2 \cdots <t_n<t_{N+1}=\tau . \la{MJ}
\eea
where $N=0,1,2,\cdots$ is the total number of jumps.
We denote the above path by 
\bea
\hat{n}= (N,(n_0,n_1,\cdots,n_N),(t_1,t_2,\cdots,t_N)) . \la{Hat_n}
\eea
The probability to find the system in a state $n$ is $p_n(t)$ and it obeys the master equation \re{M}. 
We suppose the trajectory of the control $\hat{\al}=\big( \al(t) \big)_{t=0}^{\tau}$ is smooth. 
Now we introduce
\bea
\theta_{nm}(\al) \defe  \left \{ \begin{array}{ll}
- \ln\f{K_{nm}(\al)}{K_{mn}(\al)} \hs{3mm} K_{nm}(\al) \ne 0 \\
0 \hs{20mm} K_{nm}(\al) = 0
\end{array} \right. .
\eea
If $n\ne m$, this is entropy production of process $m \to n$. 
The entropy production of process \re{Hat_n} is defined by 
\bea
\Theta^{\hat{\al}}[\hat{n}] = \sum_{k=1}^N \theta_{n_kn_{k-1}}(\al_{t_k}). \la{def_EP}
\eea
Then the weight (the transition probability density) associated with a path $\hat{n}$ is 
\bea
\mT^{\hat{\al}}[\hat{n}] = \prod_{k=1}^N K_{n_kn_{k-1}}(\al_{t_k}) \exp \Big[\sum_{k=0}^N \int_{t_k}^{t_{k+1}}dt \ K_{n_kn_k}(\al_t) \Big] .
\eea
The integral over all the paths is defined by
\bea
\int \mD \hn \ Y[\hn] \defe \sum_{N=0}^\infty \sum_{n_0,n_1,\cdots,n_N}^{n_{k-1}\ne n_k}\int_0^\tau dt_1 \int_{t_1}^\tau dt_2 \int_{t_3}^\tau dt_3 \cdots
 \int_{t_{N-1}}^\tau dt_N \ Y[\hn],
\eea
and the expectation value of $X[\hn]$ is defined by
\bea
\bra X \ket^\ha \defe \int \mD \hn \ X[\hn]p_{n_0}^\st(\al_0) \mT^\ha[\hn] .
\eea
Here, $p_{n}^\st(\al)$ is the instantaneous stationary probability distribution characterized by $\sum_m K_{nm}(\al)p_m^\st(\al) = 0$. 
We introduce a matrix $K^\lm(\al)$ by
\bea
[K^\lm(\al)]_{nm} \defe K_{nm}(\al)e^{i\lm \theta_{nm}(\al)} .
\eea
Then, the $k$-th order moment of the entropy production is given by 
\bea
\bra (\Theta^\ha[\hn])^k \ket^\ha
= \f{\partial^k}{\partial (i\lm)^k}\Big \vert_{\lm=0} \sum_{n,m} \Big[ {\rm{T}} \exp \big[ \int_0^\tau dt \ K^\lm(\al_t) \big] \Big]_{nm} p_m^\st(\al_0) . 
\la{GMEn}
\eea
In particular, the average is given by 
\bea
\sig^\RM{C} \defe \bra \Theta^\ha[\hn] \ket^\ha = \int_0^\tau dt \ \sum_{n,m} \sig_{nm}^\RM{C}(\al_t)p_m(t), \la{sig^C0}
\eea
where
\bea
\sig_{nm}^\RM{C}(\al) \defe K_{nm}(\al) \theta_{nm}(\al) =  -K_{nm}(\al) \ln \f{K_{nm}(\al)}{K_{mn}(\al)}.
\eea
According to Ref.\cite{Komatsu15}, for a quasistatic operation, 
\bea
\sig_\RM{ex}^\RM{C}= S_\RM{Sh}[p^\st(\al_\tau)]-S_\RM{Sh}[p^\st(\al_0)]+\mO(\ep^2\dl), \la{K15}
\eea
holds where 
\bea
\sig_\RM{ex}^\RM{C}\defe \sig^\RM{C}-\int_0^\tau dt \ \sum_{n,m} \sig_{nm}^\RM{C}(\al_t)p_m^\st(\al_t) , \la{sig_ex^C}
\eea 
and $ S_\RM{Sh}[p]\defe -\sum_n p_n\ln p_n$.


\begin{thebibliography}{99}
\bibitem{FN}Here, we supposed $\f{d}{dt}\bra O \ket_t \approx i^O(t)$ for $O=H_b,N_b$. 
However, because  the thermodynamic parameters $\be_b$ and $\mu_b$ are modulated, $\f{d}{dt}\bra H_b \ket_t$ and $\f{d}{dt}\bra N_b \ket_t$ also include 
the currents from the outside of the total system to the bath $b$.
\end{thebibliography}
\end{document}